\documentclass[
 article,
 twocolumn,
 groupedaddress,
 showpacs,
 preprintnumbers,
 amsmath,
 amssymb,
 aps,
 prx,
 prx,
 longbibliography,
 floatfix,
]{revtex4-1}

\usepackage{graphicx}					% Graphics
\usepackage{verbatim}					% Graphics
\usepackage{color}
\usepackage{hyperref}                   % Add hypertext capabilities
\usepackage{orcidlink}                     % I.M. orcid link

\newtheorem{proposition}{Proposition}[subsection]

\newcommand{\kk}{\kappa}                % Elliptic modulus
\newcommand{\sn}{\mathrm{sn}}           % Jacobi elliptic functions
\newcommand{\cn}{\mathrm{cn}}
\newcommand{\dn}{\mathrm{dn}}
\newcommand{\sd}{\mathrm{sd}}
\newcommand{\nd}{\mathrm{nd}}
\newcommand{\asn}{\mathrm{arcsn}}       % Inverse Jacobi elliptic functions
\newcommand{\acn}{\mathrm{arccn}}
\newcommand{\adn}{\mathrm{arcdn}}
\newcommand{\ads}{\mathrm{arcds}}
\newcommand{\acs}{\mathrm{arccs}}
\newcommand{\ans}{\mathrm{arcns}}

\newcommand{\J}{\mathrm{J}}             % Jacobian
\newcommand{\Tr}{\mathrm{Tr}}           % Trace
\newcommand{\const}{\mathrm{const}}     % const

\newcommand{\h}{\mathcal{H}}            % Hamiltonian
\newcommand{\K}{\mathcal{K}}            % Invariant of motion

\newcommand{\ds}{\displaystyle}         % Display style
\newcommand{\dd}{\mathrm{d}}            % General derivative
\newcommand{\pd}{\partial}              % Partial derivative

\begin{document}

%===============================================================================
\title{Dynamics of McMillan mappings I. McMillan multipoles}

\author{T.~Zolkin\,\orcidlink{0000-0002-2274-396X}}
\email{iguanodyn@gmail.com}
\affiliation{Fermilab, PO Box 500, Batavia, IL 60510-5011}

\author{S.~Nagaitsev\,\orcidlink{0000-0001-6088-4854}}
\email{snagaitsev@bnl.gov}
\affiliation{Brookhaven National Laboratory, Upton, NY 11973}
\affiliation{Old Dominion University, Norfolk, VA 23529}

% affiliation detox

\author{I.~Morozov\,\orcidlink{0000-0002-1821-7051}}
\email{i.morozov@corp.nstu.ru}
% \affiliation{Budker Institute of Nuclear Physics SB RAS, Novosibirsk 630090, Russia}
% this one should be ok
\affiliation{SRF Siberian Circular Photon Source "SKIF" Boreskov Institute of Catalysis SB RAS, Koltsovo 630559, Russia}
\affiliation{Novosibirsk State Technical University, Novosibirsk 630073, Russia}

\date{\today}

%===============================================================%
%===============================================================%
\begin{abstract}

%----------------------------------------------------------------
In this article, we consider two dynamical systems: the McMillan
sextupole and octupole integrable mappings, originally proposed
by Edwin McMillan.
Both represent the simplest symmetric McMillan maps, characterized
by a single intrinsic parameter.
While these systems find numerous applications across various
domains of mathematics and physics, some of their dynamical
properties remain unexplored.
We aim to bridge this gap by providing a comprehensive description
of all stable trajectories, including the parametrization of
invariant curves, Poincar\'e rotation numbers, and canonical
action-angle variables.

%----------------------------------------------------------------
In the second part, we establish connections between these maps
and general chaotic maps in standard form.
Our investigation reveals that the McMillan sextupole and octupole
serve as first-order approximations of the dynamics around the fixed
point, akin to the linear map and quadratic invariant
(known as the Courant-Snyder invariant in accelerator physics),
which represents zeroth-order approximations
(referred to as linearization).
Furthermore, we propose a novel formalism for nonlinear Twiss
parameters, which accounts for the dependence of rotation number
on amplitude.
This stands in contrast to conventional betatron phase advance used
in accelerator physics, which remains independent of amplitude.
Notably, in the context of accelerator physics, this new formalism
demonstrates its capability in predicting dynamical aperture around
low-order resonances for flat beams, a critical aspect in beam
injection/extraction scenarios.
\end{abstract}
%\pacs{00.00.Aa ,
%      00.00.Aa ,
%      00.00.Aa ,
%      00.00.Aa }% PACS, the Physics and Astronomy Classification Scheme.
%\keywords{Suggested keywords}% Use showkeys class option if keyword
                             % display desired

%===============================================================%
\maketitle
%\tableofcontents
%===============================================================%

%===============================================================%
%===============================================================%
%===============================================================%
\section{Introduction}

%----------------------------------------------------------------
A generic Hamiltonian dynamical system with $D \geq 2$ degrees of
freedom or a symplectic map with $D \geq 1$ often exhibits {\it
chaotic} behavior due to an insufficient number of {\it integrals
of motion}, also known as {\it invariants} or {\it conserved
quantities}.
In such cases, the system is highly sensitive to initial
conditions, rounding errors, and external perturbations, making
long-term predictions computationally demanding.

%----------------------------------------------------------------
Conversely, when a Hamiltonian system or symplectic map possesses
a sufficient number of independent first integrals in involution,
it is termed {\it integrable}, leading to regular, predictable
motion.
Completely integrable systems have historically played a crucial
role in understanding more complex dynamical scenarios.
For instance:
(i) Near stable equilibria, small-amplitude oscillations in
nonlinear Hamiltonian systems can often be approximated by a
multi-dimensional harmonic oscillator, as dictated by linearization
around elliptic fixed points.
(ii) The Kepler problem provides insight into celestial mechanics,
such as Earth's orbit, despite the chaotic nature of the entire
solar system.
(iii) Planetary motion around two fixed centers has applications
in satellite trajectory calculations
\cite{alexeev1965generalized,marchal1966calcul,marchal1986quasi},
electron acceleration in atomic collisions
\cite{jakas1995trapping,jakas1996production},
and determining energy levels in ${\mathrm{H}_2}^+$
\cite{strand1979semiclassical}.

%----------------------------------------------------------------
In this article, we focus on another renowned example known as the
McMillan integrable map~\cite{mcmillan1971problem}.
Originally proposed by Edwin McMillan as a simplified accelerator
lattice with one degree of freedom, consisting of linear optics
(corresponding to simple linear transformation in phase space) and
a special thin nonlinear lens (nonlinear vertical shear
transformation).
In the context of this article, we use the term ``accelerator
lattice'' to refer to the combined mapping that represents the
structured arrangement of magnetic and accelerating elements
in a particle accelerator, which guide and focus the beam along
its trajectory;
since our primary focus is on planar systems derived from
classical Hamiltonians describing a point-like particle or
on mappings corresponding to a single revolution of a particle
in an accelerator, the term ``particle’s trajectory'' can be
understood as the ``orbit'' of a dynamical system, while the term
``beam of particles'' refers to a collection of distinct initial
conditions.
As later demonstrated by Suris~\cite{suris1989integrable}, the
McMillan mapping is among the few possible integrable symplectic
transformations of the plane with an analytic integral of motion.
Specifically, for maps in the form~(\ref{math:McMsym}), the
invariant must be a regular, exponential, or trigonometric polynomial
of degree two in coordinate and momentum.

\newpage
%----------------------------------------------------------------
We show that the McMillan map is not merely a model
accelerator lattice but is also related to general symplectic
mappings of the same form, featuring chaotic dynamics as its
first and second-order approximations (Section~\ref{sec:PertTh}).
This insight aids in ``integrating-out'' additional nonlinear
features of small amplitude oscillations, such as detuning or
the frequency's dependence on the action variable.
As a consequence, we propose the concept of nonlinear Twiss
parameters for accelerator physics, serving as a natural extension
to the existing Courant-Snyder formalism.
This new framework enables a deeper understanding of
amplitude-dependent shifts in the betatron tune and facilitates
predicting the dynamical aperture around isolated low order
resonances, thereby representing a nonlinear integrable model
for slow resonant beam extraction for a flat beam.

%----------------------------------------------------------------
Finally, in the first part of this article, we address an important
gap in analytical results for the simplest symmetric McMillan maps,
which possess a single intrinsic parameter and quadratic (sextupole)
or cubic (octupole) nonlinearities.
Specifically, we provide a canonical transformation to action-angle
variables.
The octupole and, more generally, asymmetric McMillan maps have been
extensively studied, particularly in the work of Iatrou and Roberts
\cite{IR2002II}, who parametrized individual invariant curves.
However, explicit expressions for the rotation number --- and more
importantly, the canonical action integral --- have been missing until
now, see Section~\ref{sec:OctupoleMap} and Eqs.~(\ref{math:JOCMcM1},
\ref{math:JOCMcM2}).
As we demonstrate in Section~\ref{sec:PertTh}, canonical McMillan
mappings can be used to derive the leading twist coefficient for a
broad class of smooth mappings of the same form.
However, this requires explicit expressions for the rotation number
and action variable in the sextupole McMillan map, which we obtain
in Eqs.~(\ref{math:nuSXMcM}) and (\ref{math:JSXMcM1},\ref{math:JSXMcM2}),
respectively.
These results are essential for understanding nonlinear dynamics and
designing integrable accelerator lattices with desirable properties.
Additionally, they have practical implications for more realistic
integrable accelerator lattices based on the McMillan map, such as
those describing a particle with zero angular momentum in an integrable
(4D phase space) accelerator ring with an axially symmetric McMillan
electron lens~\cite{cathey2021calculations}.
Appendix~\ref{secAp:Stability} provides complementary bifurcation
diagrams for the McMillan sextupole (Fig.~\ref{fig:BifDSbig}) and
octupole (Fig.~\ref{fig:BifDObig}) mappings.

%----------------------------------------------------------------
The article is structured as follows: Section~\ref{sec:McMultipoles}
defines McMillan multipoles and auxiliary classical Hamiltonians,
which aid in building a qualitative understanding.
While these systems represent opposite extremes --- infinitely thin,
localized nonlinearity versus distributed nonlinearity --- their
simultaneous consideration provides a unified picture of integrable
planar symplectic dynamics.
Section~\ref{sec:Dynamics} introduces a general method for dynamical
analysis based on Danilov’s theorem, using the McMillan sextupole as
an illustrative example.
This section also bridges the parametrization of invariant curves
\cite{IR2002II} with the construction of canonical action-angle
variables.
Sections~\ref{sec:AmplFr} and \ref{sec:Stability} delve into a
detailed analysis of dynamics by examining amplitude-frequency
dependencies and stability diagrams for different motion regimes.
Next, Section~\ref{sec:PertTh} introduces low orders of
perturbation theory that relates McMillan multipoles to general
chaotic systems of the same map form.
We investigate small and large amplitudes using the H\'enon
quadratic and cubic mappings as examples for study.
Additionally, we present a general result for a model accelerator
lattice with a thin nonlinear lens and compare our findings with
other techniques such as Lie algebra.
Finally, Section~\ref{sec:General} discusses possible
generalizations and extensions of our results, including the
utilization of a more general form of the map.
Here, we apply perturbation theory to the horizontal dynamics
within the Fermilab delivery ring, which is employed for a
third-integer resonant extraction in the Mu2e experiment.

%----------------------------------------------------------------
At the conclusion of the article, we provide supplementary
materials.
Appendices~\ref{secAp:Symmetries} and \ref{secAp:Stability} offer
a detailed description of the symmetries of motion invariants and
the stability of fixed points and $n$-cycles for both, McMillan
sextupole and octupole mappings.
The last Appendix, \ref{secAp:Action}, contains a list of analytical
expressions for action integrals and their power series.

%===============================================================%
%===============================================================%
%===============================================================%
\section{\label{sec:McMultipoles}McMillan multipoles}

%------------------------------------------------------------- ---
The most general form of the {\it symmetric McMillan map} is
given by~\cite{mcmillan1971problem,quispel1988integrable,
quispel1989integrable,IR2002II}
\begin{equation}
\label{math:McMsym}
\begin{array}{l}
    q' =  p,       \\[0.25cm]
    p' = -q + f(p),
\end{array}
\end{equation}
where the prime ($'$) denotes the application of the map, and
$f(q)$ is the {\it force function} defined as
\[
    f(p) = -\,\frac{\mathrm{B}\,p^2 + 2\,\epsilon\,p + \Xi}
                 {\mathrm{A}\,p^2 + \mathrm{B}\,p + \Gamma}.
\]
This map is integrable for any set of parameters, with the
invariant of motion $\K[p,q]$ being a biquadratic function
of the coordinate $q$ and momenta $p$, that can be written
in matrix form as:
\begin{equation}
\label{math:McMinvM}
\K[p,q] =
\begin{bmatrix}
p^2 \\ p  \\ 1
\end{bmatrix}^\mathrm{T}
\left(
\mathrm{M}
\cdot
\begin{bmatrix}
q^2 \\ q  \\ 1
\end{bmatrix}
\right),
\quad
\mathrm{M} = 
\begin{bmatrix}
\mathrm{A}  & \mathrm{B} & \Gamma   \\
\mathrm{B}  & 2\,\epsilon& \Xi      \\
\Gamma      & \Xi        & \Lambda
\end{bmatrix}
.
\end{equation}

%----------------------------------------------------------------
In order to gain a deeper understanding of the physical meaning
behind the mapping's parameters, we introduce the auxiliary
Hamiltonian of a classical particle:
\begin{equation}
\label{math:HBoss}
\h[p,q;t] = \frac{p^2}{2\,m} + \lambda + \xi\,q + \gamma\,\frac{q^2}{2}
 + \beta\,\frac{q^3}{3}
 + \alpha\,\frac{q^4}{4}.
\end{equation}
%----------------------------------------------------------------
It is well known~\cite{LL_mechanics} that quartic and cubic terms
are the only contributors to detuning, i.e., the linear dependence
of frequency on the canonical action variable.
However, they differ fundamentally, producing qualitatively distinct
contributions $\propto \alpha$ or $\propto \beta^2$, respectively.
As we will see, the terms in the invariant proportional to 
$\mathrm{A}$ and $\mathrm{B}$ play similar qualitative roles, though
with notable differences, as discussed in Section~\ref{sec:AmplFr}.
Beyond this qualitative similarity, the mapping (\ref{math:McMsym})
and the Hamiltonian (\ref{math:HBoss}) represent two contrasting
models of horizontal motion in an accelerator: one describing motion
in the presence of an infinitely thin, localized nonlinear lens, and
the other modeling a nonlinear magnet uniformly distributed along the
machine circumference.
This contrast further motivates a detailed study of their differences.

%----------------------------------------------------------------
Rewriting Eq.~(\ref{math:McMinvM}) explicitly
\[
\begin{array}{l}
\K[p,q] = 
\underbrace{\mathrm{A}\,p^2q^2}_{\text{cubic nonlin.}}
    + \,
\underbrace{\mathrm{B}\,(p^2q + p\,q^2)}_{\text{quadratic nonlin.}}
    \, + \\[0.7cm]
    \qquad\qquad\qquad + \,
\underbrace{\Gamma\,(p^2 + q^2) + 2\,\epsilon\,p\,q}_{\text{harmonic
oscillator}}
    \,\, + \!
\underbrace{\Xi\,(p+q)}_\text{defines origin} + \,\,\Lambda
\end{array}
\]
allows us to establish a meaningful analogy between the terms.
%----------------------------------------------------------------
Foremost, adding a constant term to the Hamiltonian ($\lambda$)
or to the invariant of the map ($\Lambda$) only results in a shift
of the energy scale.
This shift doesn't affect the relative positions or momenta of the
particles in the system, and therefore, it doesn't alter the
physical trajectories or dynamics of the system.
Next, assuming an oscillatory regime (i.e., where $\h$ and $\K$
have a local extremum), we can set $\xi$ and $\Xi$ to zero by
positioning the origin at the stationary point of the Hamiltonian
or the fixed point of the map.
Further, when examining quadratic components:
\[
\frac{p^2}{2\,m} + \gamma\,\frac{q^2}{2}
\qquad\text{and}\qquad
\Gamma\,(p^2 + q^2) + 2\,\epsilon\,p\,q
\]
it is apparent that both contribute to inducing harmonic linear
oscillations in the system.
However, while we can normalize units of time and coordinate to
set $m = \gamma = 1$ in the Hamiltonian system, the mapping
introduces an intrinsic irreducible parameter
$a = -2\,\epsilon/\Gamma$ associated with the discrete nature of
time (see Appendix~\ref{secAp:Symmetries}).
Finally, the pair of parameters $\mathrm{A}$ and $\mathrm{B}$ or
$\alpha$ and $\beta$ contributes to higher-order nonlinear
effects, playing the roles of cubic and quadratic nonlinearities
respectively.

%----------------------------------------------------------------
If only one of nonlinear terms is present, after performing
non-dimensionalization of Hamiltonian~(\ref{math:HBoss}), we obtain
4 possible situations with stable trajectories:
\begin{equation}
\label{math:HAll}
\begin{array}{lc}
\ds\h_\mathrm{sxt\,}[p,q;t] =
    \frac{p^2}{2} + \frac{q^2}{2} + \frac{q^3}{3},  & \mathrm{(SX)} \\[0.25cm]
\ds\h^-_\mathrm{oct\,}[p,q;t] =
    \frac{p^2}{2} + \frac{q^2}{2} - \frac{q^4}{4},  & \mathrm{(DO)} \\[0.25cm]
\ds\h^+_\mathrm{oct\,}[p,q;t] =
    \frac{p^2}{2} + \frac{q^2}{2} + \frac{q^4}{4},  & \mathrm{(FO)} \\[0.25cm]
\ds\h_\mathrm{Duf}[p,q;t] =
    \frac{p^2}{2} - \frac{q^2}{2} + \frac{q^4}{4}.  & \mathrm{(Df)} \\[0.25cm]
\end{array}
\end{equation}
%----------------------------------------------------------------
These Hamiltonians serve as universal models for ``typical''
(i.e., those with at least nonzero cubic or quartic terms in
the potential energy) 1D nonlinear oscillators associated with
classical Hamiltonians and find applications across various
fields.
For instance, the first system serves as a prototype for the
1D H\'enon-Heiles potential~\cite{henon1964applicability} or
represents horizontal dynamics inside long sextupole
magnet combined with linear focusing (SX).
The subsequent two models describe planar defocusing (DO) and
focusing octupoles (FO), while the final one is associated with
the well-known unforced undamped Duffing oscillator (Df)
\cite{duffing1918erzwungene}.
Throughout this article, we will use lowercase subscripts ``sxt''
and ``oct'' to describe the leading nonlinearity in the Hamiltonian
or invariant, while uppercase labels (SX), (DO), (FO), and (Df)
will be used to specify the underlying regime of motion.

%----------------------------------------------------------------
We further expand our analogy by introducing two dynamical
systems: the \textit{McMillan sextupole} and \textit{octupole}
integrable mappings, defined by corresponding forces
\[
f_\text{sxt}(p) =-p\,\frac{p+2\,\epsilon}{p+\Gamma},
\qquad\text{and}\qquad
f_\text{oct}(p) =-\frac{2\,\epsilon\,p}{p^2+\Gamma},
\]
along with matrices of coefficients in the form
(\ref{math:McMinvM})
\[
\mathrm{M}_\text{sxt} =
\begin{bmatrix}
0       & 1             & \Gamma    \\
1       & 2 \,\epsilon  & 0         \\
\Gamma  & 0             & 0
\end{bmatrix},
\qquad\qquad
\mathrm{M}_\text{oct} =
\begin{bmatrix}
1       & 0             & \Gamma    \\
0       & 2 \,\epsilon  & 0         \\
\Gamma  & 0             & 0
\end{bmatrix}.
\]
%----------------------------------------------------------------
After eliminating all dependent parameters (Appendix
\ref{secAp:Symmetries}), we establish the following correspondence:
\begin{equation}
\label{math:mapAll}
\begin{array}{llc}
\ds\K_\mathrm{sxt}[p,q] =
    \K_0 + p^2 q + q\,p^2,  &\!|a| \leq 2, & \!\mathrm{(SX)} \\[0.25cm]
\ds\K^-_\mathrm{oct}[p,q] =
    \K_0 - p^2 q^2,         &\!|a| \leq 2,\,\Gamma < 0,& \!\mathrm{(DO)} \\[0.25cm]
\ds\K^+_\mathrm{oct}[p,q] =
    \K_0 + p^2 q^2,         &\!|a| \leq 2,\,\Gamma > 0,& \!\mathrm{(FO)} \\[0.25cm]
\ds\K^+_\mathrm{oct}[p,q] =
    \K_0 + p^2 q^2,         &\!|a| > 2   ,\,\Gamma > 0,& \!\mathrm{(Df)}
\end{array}
\end{equation}
where %$\K_0[p,q] = p^2 - a\,p\,q + q^2$.
\[\K_0[p,q] = p^2 - a\,p\,q + q^2.\]
%and (Df) for $\K^+_\mathrm{oct}$ with $|a|>2$,
%where $\K_0[p,q] = p^2 - a\,p\,q + q^2$.

%----------------------------------------------------------------
While the systems above are not precisely identical to the
Hamiltonians~(\ref{math:HAll}), we will observe substantial
similarities between them, aiding in constructing a comprehensive
understanding of the symmetric McMillan map dynamics.
For instance, (I) both Hamiltonians and mappings exhibit similar
amplitude-frequency dependencies below the intrinsic resonance.
(II) The solutions for $q$ involve Jacobi elliptic functions:
$\mathrm{sn}$, $\mathrm{cn}$, and $\mathrm{dn}$ for (DO), (FO),
and (Df), respectively~\cite{IR2002II}, and rational function of
$\mathrm{sn^2}$ for the (SX) case~\cite{zolkin2022mcmillan}.
Most importantly, we demonstrate that, akin to Hamiltonian
(\ref{math:HBoss}) being an approximation for a more general
potentials $U(q)$, McMillan multipoles serve as first and
second-order approximations for the map in a {\it McMillan-H\'enon
form}~(\ref{math:McMsym}) with a smooth $f(p)$.
This is particularly useful for ``integrating out'' leading
nonlinear effects around main resonances for chaotic systems,
akin to H\'enon~\cite{henon1969numerical}
or Chirikov standard mappings
\cite{chirikov1969research,chirikov1979universal}.

\newpage
%===============================================================%
%===============================================================%
%===============================================================%
\section{\label{sec:Dynamics}Solving the Equations of Motion}

%===============================================================%
%===============================================================%
\subsection{\label{sec:McSX}Sextupole map}

%----------------------------------------------------------------
Here, we will illustrate the general method of obtaining a
parametrization of an individual curve and a set of action-angle
variables based on Danilov's theorem
\cite{zolkin2017rotation,nagaitsev2020betatron,mitchell2021extracting},
using McMillan sextupole as an example.
We limit ourselves to cases with stable trajectories around fixed
point at the origin, considering only $-2 < a < 2$.

\noindent
%----------------------------------------------------------------
$\bullet$
We start by introducing a formal Hamiltonian
\[
    \mathrm{H}[p,q;t] \equiv \K_\mathrm{sxt}[p,q] =
    p^2q + p\,q^2 + p^2 -a\,p\,q + q^2
\]
which satisfies the system of Hamilton's equations
\begin{equation}
\label{math:dHsex}
\begin{array}{r}
\ds\qquad  \frac{\dd q}{\dd t} = \frac{\pd\mathrm{H}}{\pd p} =
    q^2 - a\,q + 2\,p + 2\,p\,q,       \\[0.35cm]
\ds-       \frac{\dd p}{\dd t} = \frac{\pd\mathrm{H}}{\pd q} =
    p^2 - a\,p + 2\,q + 2\,p\,q.
\end{array}
\end{equation}

\noindent
%----------------------------------------------------------------
$\bullet$
Since Hamiltonian function does not change along the trajectory,
we can solve for $p$ from $\mathrm{H}[p,q] = \mathrm{const}$:
\begin{equation}
\label{math:psex}
p = \frac{1}{2}\,\left[
    f_\text{sxt}(q) \pm \frac{\sqrt{\mathcal{P}(q)}}{q+1}
\right],
\end{equation}
where
\[
\mathcal{P}(q) =
    q^4 - 2\,(2 + a)\,q^3 + (a^2-4)\,q^2 + 4\,\mathrm{H}\,q +
    4\,\mathrm{H}.
\]
When the fixed point at the origin is stable ($|a|<2$),
for all closed invariant curves we have $\mathrm{H} > 0$, and,
polynomial $\mathcal{P}(q)$ has four distinct real roots $q_i$
such that
\[
    \sum_{i=1}^{4}\frac{1}{q_i} = -1,
    \qquad\qquad
    q_1 < q_2 \leq q \leq q_3 < q_4.
\]
Using Vieta's formulas for the quartic $\mathcal{P}(q)$,
one can express map parameter and energy level using $q_i$ as:
\[
    a  = \sum_{i=1}^{4}\frac{q_i}{2} - 2,
    \qquad\qquad\qquad
    \mathrm{H} = \frac{1}{4} \prod_{i=1}^{4} q_i.
\]

\noindent
%----------------------------------------------------------------
$\bullet$
Next, we should relate continuous flow given by Hamiltonian with
the mapping equations.
Substitution of (\ref{math:psex}) into the first Hamilton's
equation~(\ref{math:dHsex}) provides
\begin{equation}
\label{math:intRAW}
    \dd\,t = \pm\dd\,q/\sqrt{\mathcal{P}(q)}.
\end{equation}
Taking the integral of Eq.~(\ref{math:intRAW}) along the invariant
curve, we can find the period of motion
%----------------------------------------------------------------
\[
\mathrm{T} \equiv \oint \dd\,t =
    2\,\int_{q_2}^{q_3}\frac{\dd\,q}{\sqrt{\mathcal{P}(q)}} =
    \frac{4\,\mathrm{K}[\kappa]}{\sqrt{(q_4-q_2)(q_3-q_1)}}
\]
and the time of one-step of the map
\[
\ds \mathrm{T'} \equiv \int_0^\mathrm{T'} \dd\,t =
    \int_{q_0}^{q_0'}\frac{\dd\,q}{\sqrt{\mathcal{P}(q)}} =
    \int_{q_{2,3}}^{q_{2,3}'}\frac{\dd\,q}{\sqrt{\mathcal{P}(q)}}.
\]

%----------------------------------------------------------------
\begin{figure}[t!]
    \centering
    \includegraphics[width=\linewidth]{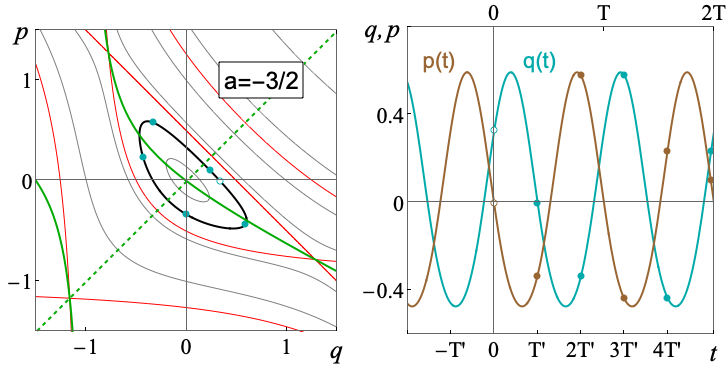}
    \caption{\label{fig:SxtParam}
    Phase space diagram for the McMillan sextupole map with
    $a=-3/2$ (left), and parametrization
    of invariant curve (\ref{math:param}) with initial conditions
    $(\{q_0\},\{p_0\})=(1/3,0)$ (right).
    }
\end{figure}
%----------------------------------------------------------------

The last integral is independent of choice of lower bound, so we
can choose one of the turning points as $q_0$, where with the
help of mapping equations we have
\[
q_{2,3}' = p(q_{2,3}) = f_\text{sxt}(q_{2,3})/2 =
    -q_{2,3}\,\frac{q_{2,3}-a}{q_{2,3}+1}.
\]
%----------------------------------------------------------------
$\text{T}'$ provides the discretization time interval and defines
the rotation number of the map as the ratio
\begin{equation}
\label{math:nuSXMcM}
    \nu = \frac{\mathrm{T}'}{\mathrm{T}} =
    \frac{\mathrm{F}\left[
        \arcsin\sqrt{
        \frac{q_3-q_1}{q_3-q_2}
        \frac{q_2(3\,q_2+2-a)}{2\,q_1(q_2+1) + q_2\,(q_2-a)}
        }
    ,\kappa\right]}
    {2\,\mathrm{K}[\kappa]}.
\end{equation}
Here $\mathrm{F}[\phi,\kappa]$ and $\mathrm{K}[\kappa]$ are
complete and incomplete elliptic integrals of the first kind
with elliptic modulus
\[
\kappa = \sqrt{\frac{(q_3-q_2)(q_4-q_1)}{(q_3-q_1)(q_4-q_2)}}.
\]
%----------------------------------------------------------------
$\bullet$
Taking integral~(\ref{math:intRAW}) from $q(0)=\{q_0\}$
to $q(t)$, we can obtain parametrization of the map
$\{q_n\},\{p_{n-1}\} = q(n\,\mathrm{T}')$:
\begin{equation}
\label{math:param}
q(t) = \frac{q_2 - q_1\,\frac{q_3-q_2}{q_3-q_1}\,\sn_1^2}
            {1-\frac{q_3-q_2}{q_3-q_1}\,\sn_1^2}
      = \frac{q_3 - q_4\,\frac{q_3-q_2}{q_4-q_2}\,\sn_2^2}
            {1-\frac{q_3-q_2}{q_4-q_2}\,\sn_2^2},
\end{equation}
where $\mathrm{sn}_{1,2}$ stands for one of two possible
Jacobi elliptic sine functions (see Fig.~\ref{fig:SxtParam})
\[
\mathrm{sn}_i = \mathrm{sn}\left[
    \sqrt{(q_4-q_2)(q_3-q_1)}\,\frac{t - t^*_i}{2},\kappa
\right]
\]
with initial phase shifts
\[
\ds t^*_{1,2} = \mp 2\,\frac{
    \mathrm{F}\left[
    \arcsin\sqrt{\frac{(q_{3,4}-q_{1,2})(\{q_0\}-q_{2,3})}
                          {(q_3-q_2)(\{q_0\}-q_{1,4})}}
      ,\kappa\right]
}{\sqrt{(q_4-q_2)(q_3-q_1)}}.
\]
%----------------------------------------------------------------
Here, we use roster notation to emphasize that $\{q_i\}$ (or any
other dynamical variable) is an element of the set representing
the orbit $\{q_0,q_1,q_2,\ldots\}$ defined by recurrence.
This distinction is crucial to clearly separate it from, say, the
roots $q_{1,2,3,4}$ of the characteristic polynomial $\mathcal{P}(q)$
or the initial point $q_0$, which appears as the lower limit in the
integral $\mathrm{T}'$.

%----------------------------------------------------------------
\begin{table*}[t!]
\begin{tabular}{p{4cm}p{1.75cm}p{3.5cm}p{3.75cm}p{3.75cm}}
                                                        \hline\hline\\[-0.25cm]
&   & DO ($\Gamma < 0$, $|a|<2$)
    & FO ($\Gamma > 0$)
    & Df ($\Gamma > 0$, $|a| > 2$)                                  \\[0.1cm]
Elliptic function   & ef & {\bf sn} & {\bf cn} & {\bf dn}           \\
                                                                    \hline
								                                \\[-0.2cm]
Domain of invariant & $\K\in$                                       & 
$\left[0; (|\frac{a}{2}|-1)^2\right] \geq 0$			            &
$[0;\infty) \geq 0$							                        &
$\left[-(|\frac{a}{2}|-1)^2;0\right] \leq 0$				        \\[0.4cm]
Amplitude & $h(\kappa)$							                    &
$\displaystyle \sqrt{k}$	                                        &
$\displaystyle \sqrt{k/k'}$		                                    &
$\displaystyle \sqrt{1/k'}$		                                    \\[0.4cm]
Phase advance ($a>0$) & $\eta_+$									&
$\displaystyle \ans\,\frac{\sqrt{k    }}{\sqrt[4]{\K}}$	        &
$\displaystyle \ads\,\frac{\sqrt{k\,k'}}{\sqrt[4]{\K}}$		    &
$\displaystyle \acs\,\frac{\sqrt{k'   }}{\sqrt[4]{|\K|}}$		    \\[0.4cm]
Initial phase & $\{u_0\}$									        &
$\displaystyle \asn\,\frac{\{q_0\}}{h(\kappa)\,\sqrt[4]{\K}}$     &
$\displaystyle \acn\,\frac{\{q_0\}}{h(\kappa)\,\sqrt[4]{\K}}$     &
$\displaystyle \adn\,\frac{\{q_0\}}{h(\kappa)\,\sqrt[4]{|\K|}}$     \\[0.4cm]
Elliptic modulus & $k(B)$									        &
$\displaystyle \frac{B-\sqrt{B^2-4}}{2}$			                &
$\displaystyle \frac{1}{\sqrt{2}}\sqrt{1+\frac{B}{\sqrt{B^2+4}}}$	&
$\displaystyle \frac{\sqrt{B+2}-\sqrt{B-2}}{2\,(B^2-4)^{-1/4}}$	    \\[0.4cm]
Complimentary modulus & $k'(B)$									    &
$\displaystyle \frac{\sqrt{B+2}-\sqrt{B-2}}{2\,(B^2-4)^{-1/4}}$	    &
$\displaystyle \frac{1}{\sqrt{2}}\sqrt{1-\frac{B}{\sqrt{B^2+4}}}$	&
$\displaystyle \frac{B-\sqrt{B^2-4}}{2}$			                \\[0.5cm]
$B$-function & $B(a,\K)$									        &
$\displaystyle-\frac{(a/2)^2-1-\K}{\sqrt{\K}}$                    &
$\displaystyle \frac{(a/2)^2-1+\K}{\sqrt{\K}}$                    &
$\displaystyle \frac{(a/2)^2-1+\K}{\sqrt{|\K|}}$                    \\[0.45cm]
&                                                                   &
$\displaystyle =\frac{1}{k} +k \geq 2$				                &
$\displaystyle =\frac{k}{k'}-\frac{k'}{k}\in \mathbb{R}$            &
$\displaystyle =\frac{1}{k'}+k'\geq 2$					            \\[0.25cm]
\hline
\hline
\end{tabular}
\caption{Elliptic parametrization of
        %constant level sets corresponding to
        stable trajectories for McMillan
        octupole map (after~\cite{IR2002II});
        for $a<0$ use $\eta_- = 2\,\mathrm{K}[\kappa] - \eta_+$.}
\label{tab:IR}
\vspace{-0.5cm}
\end{table*}
%----------------------------------------------------------------

\newpage

\noindent
%----------------------------------------------------------------
$\bullet$
Finally, we can rewrite the map in its canonical form
\[
\begin{array}{ll}
J' = J,                         &\qquad
\{J_n\} = \{J_0\},                      \\[0.25cm]
\psi' = \psi + 2\,\pi\,\nu(J),  &\qquad
\{\psi_n\} = \{\psi_0\} + 2\,\pi\,n\,\nu(\{J_0\}),
\end{array}
\]
where $J$ and $\psi$ are action-angle variables.
The action integral is given in Appendix~\ref{secAp:Action}, while
initial phase $\{\psi_0\}$ can be chosen e.g., as
$\arctan(\{p_0\}/\{q_0\})$ or $-2\,\pi\,t^*_{1,2}/\mathrm{T}$.

%===============================================================%
%===============================================================%
\subsection{\label{sec:OctupoleMap}Octupole map}

%----------------------------------------------------------------
The parametrization of individual curves for the octupole map was
obtained by Iatrou and Roberts in~\cite{IR2002II}.
Additionally, they provided two different methods demonstrating
how an arbitrary constant level set of an asymmetric McMillan map
can be transformed first into a symmetric, and then into the
octupole (canonical) form.
Consequently, the authors described the dynamics on each symmetric
biquadratic $\K^\mp_\mathrm{oct}[p,q] = \const$ using the elliptic
modulus $\kappa$ and the argument $\{u_n\} = \{u_0\} + n\,\eta$ of
the elliptic function as a pair of ``action-angle'' variables.
However, despite capturing the essence of the dynamics, the change
of variables from $(q,p)$ to $(\kappa,u)$ does not conserve the
phase space area.
Below, we summarize their results using our notations and
complement it with a form-invariant set of canonical action-angle
variables
\cite{kolmogorov1954conservation,moser1962invariant,arnol1963small},
thereby extending the description from an individual curve to a
continuum of stable trajectories.

%----------------------------------------------------------------
Due to additional symmetries of the invariant, the functional
dependence is much simpler compared to the sextupole case and all
trajectories can be parametrized as follows:
\begin{equation}
\label{math:OctParam}    
\left\{
\begin{array}{l}
\ds \{q_n\} = \sqrt[4]{|\K|}\,h(\kappa)\,\mathrm{ef}\left[
                \{u_0\} + \eta\,n,\kappa
              \right],          \\[0.25cm]
    \{p_n\} = \{q_{n+1}\},
\end{array}
\right.
\end{equation}
where $\mathrm{ef}$ is an appropriate elliptic function with
elliptic modulus $\kappa$,
$h(\kappa)$ is an amplitude function,
$\eta$ is the phase advance, and,
$\{u_0\}$ is the initial phase shift defined by initial condition
$\{q_0\}$.
%----------------------------------------------------------------
For defocusing octupole ($\Gamma < 0$), the proper Jacobi function
for stable trajectories is elliptic sine $\sn$, while for focusing
system ($\Gamma > 0$) one should use elliptic cosine $\cn$ for
trajectories rounding the origin, or delta amplitude $\dn$ for
closed orbits inside the figure-eight/lemniscate shape separatrix
(case $|a|>2$), akin regimes with finite motion in
classical Hamiltonians
\[
\h[p,q;t] =
    \frac{p^2}{2} + k\,\frac{q^2}{2} \mp \frac{q^4}{4}.
\]
Expressions for all parameters are listed in Table~\ref{tab:IR}
with $\eta_- = 2\mathrm{K}[\kappa] - \eta_+$.

%----------------------------------------------------------------
Using parametrization~(\ref{math:OctParam}), we can evaluate the
integral for {\it canonical action} as in
\cite{zolkin2016analytical}:
\begin{equation}
\label{math:JefSC}
\ds J_\text{ef} = \frac{1}{2\,\pi}\,\oint p\,\dd q
        = \frac{\sqrt{|\K|}}{2\,\pi} \times\left\{
        \begin{array}{l}
        \ds    \qquad\kappa  \,\,S_\sn,         \\[0.2cm]
        \ds    (\kappa/\kappa') \,S_\cn,        \\[0.2cm]
        \ds    (1/\kappa')      \,S_\dn,
        \end{array}
        \right.
\end{equation}
where $S_\mathrm{ef}$ is the area of corresponding {\it elliptic
Lissajous curve} with matching frequencies and phase difference
$\eta$.

%----------------------------------------------------------------
\begin{figure*}[t!]
    \centering
    \includegraphics[width=\linewidth]{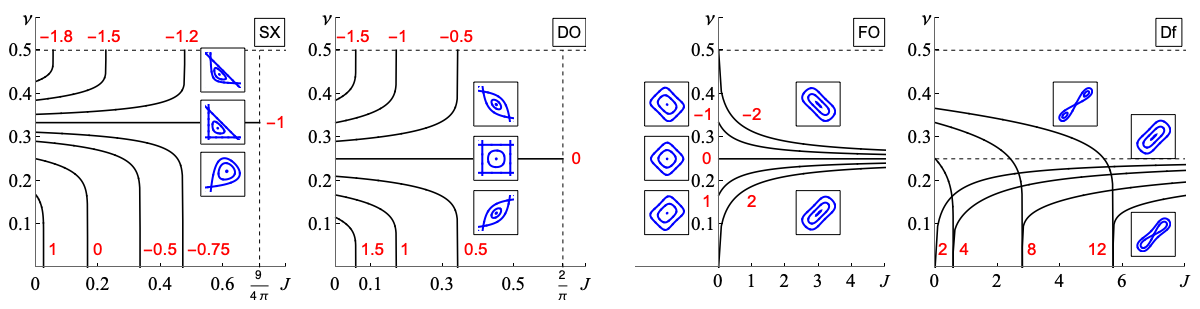}
    \caption{\label{fig:SpectraAll}
    Rotation number as a function of the action variable $\nu(J)$
    for McMillan sextupole (SX), defocusing (DO) and focusing (FO)
    octupoles, along with a focusing octupole in the Duffing regime
    (Df) with $a \geq 2$.
    The blue diagrams schematically depict phase space portraits
    corresponding to different curves, with the values of the map
    parameter $a$ indicated in red.
    The action variable for trajectories inside the figure-eight
    separatrix (case Df) is multiplied by a factor of 2 to ensure
    continuity of the graph with large amplitude trajectories rounding
    the origin.
    }
\end{figure*}
%----------------------------------------------------------------

%----------------------------------------------------------------
Finally, by rescaling the argument
$u \rightarrow \psi = 2\,\pi\,u/\mathrm{T}_\text{ef}$, we define
the {\it canonical angle variable}
\[
    \{\psi_n\} = \{\psi_0\} + 2\,\pi\,\nu\,n,
\]
where corresponding period of elliptic functions
$\mathrm{T}_\text{ef}$ is equal to $4\,\mathrm{K}[\kappa]$ for
oscillations rounding the origin ($\mathrm{sn}$ and $\mathrm{cn}$),
or, $2\,\mathrm{K}[\kappa]$ for trajectories inside the
figure-eight curve (dn).
The Poincar\'e rotation number again given by
$\mathrm{T'}/\mathrm{T_\text{ef}}$ where one-step time interval
$\mathrm{T'} \equiv \eta$ such that
\[
\eta =
\left\{
\begin{array}{ll}
\eta_+                         & \quad\text{for } a > 0,       \\[0.25cm]
2\,\mathrm{K}[\kappa] - \eta_+ & \quad\text{for } a < 0.
\end{array}
\right.
\]
%\[
%    a>0:\,\eta = \eta_+,
%    \qquad
%    a<0:\,\eta = \eta_- = 2\,\mathrm{K}[\kappa] - \eta_+.
%\]
Explicit expressions for $J_\text{ef}$, $S_\text{ef}$ and power
series $\nu(J_\text{ef})$ are given in Appendix~\ref{secAp:Action}.

%===============================================================%
%===============================================================%
%===============================================================%
\section{\label{sec:AmplFr}Amplitude-frequency dependence}

\begin{figure}[h!]
    \centering
    \includegraphics[width=\columnwidth]{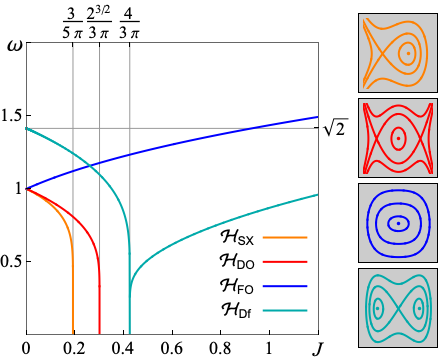}
    \caption{\label{fig:SpectraH}
    Frequency as a function of the action variable, $\omega(J)$,
    for all Hamiltonians~(\ref{math:HAll}).
    The corresponding phase space diagrams are displayed to the
    right, color-coded consistently.
    The action variable for trajectories inside the figure-eight
    separatrix (case Df) is multiplied by a factor of 2 to ensure
    continuity of the graph with large amplitude trajectories
    rounding the origin.
    }\vspace{-0.3cm}
\end{figure}

%----------------------------------------------------------------
In this section, we explore intrinsic dynamical properties that
remain invariant regardless of the mapping's representation and
examine limiting cases of both small and large amplitudes.
Figure~\ref{fig:SpectraAll} illustrates the dependence of the
rotation number on the action variable $\nu(J)$ for all the cases
under consideration.
While we postpone the detailed discussion on the stability of
critical points of the invariant until Appendix~\ref{secAp:Stability},
we encourage the reader to refer to Figure~\ref{fig:BifAllSmall},
which illustrates linear stability and the associated bifurcations,
as well as Figures~\ref{fig:BifDSbig} and \ref{fig:BifDObig},
which schematically depict phase space diagrams arranged in the
plane of map parameters.
To further deepen our understanding, we compare the qualitative
behavior of mappings with our reference classical Hamiltonians
(\ref{math:HAll}) by setting Fig.~\ref{fig:SpectraAll} side by
side with corresponding dependencies of frequency on action
$\omega(J)$, as depicted in Figure \ref{fig:SpectraH}.
For the readers' convenience, all exact expressions along with
their power series are listed at the end of the article in
Appendix~\ref{secAp:Action}.

%----------------------------------------------------------------
%----------------------------------------------------------------

\noindent
%----------------------------------------------------------------
$\bullet$ Small amplitude oscillations can be analyzed using a
linearization approach, where the system's behavior near an
equilibrium or fixed point is approximated.
In all cases

\newpage\noindent
with stable orbits around the origin (SX, DO, and
FO), such motion is determined by quadratic terms:
\[
    \mathcal{H}_0[p,q;t] = \frac{p^2}{2} + \frac{q^2}{2},
    \qquad\quad
    \K_0[p,q] = p^2 - a\,p\,q + q^2,
\]
for Hamiltonians or mappings, respectively.
This yields the corresponding frequency and rotation number
evaluated at $J=0$:
\[
    \omega_0 = 1,
    \qquad\qquad\qquad\qquad
    \nu_0 = \frac{1}{2\,\pi}\,\arccos\frac{a}{2}.
\]

\newpage
\noindent
For the Duffing oscillator, linearization around stable symmetric
stationary/fixed points provides:
\[
    \omega_0 = \sqrt{2},
    \qquad\qquad
    \nu_0 = \frac{1}{2\,\pi}\,\arccos\left(\frac{4}{a}-1\right).
\]
Although we can always normalize $\omega_0$ to unity by adjusting
units of time for Hamiltonians, mappings correspond to systems
with $1\frac{1}{2}$ degrees of freedom, resulting in an irreducible
intrinsic parameter even for the case of linear oscillations.

\noindent
%----------------------------------------------------------------
$\bullet$ When all trajectories are stable (cases FO and Df), the
frequency of oscillations for large amplitudes in Hamiltonian
systems experiences unbounded growth, as seen in the blue and cyan
curves in Fig.~\ref{fig:SpectraH}.
Conversely, with the rotation number constrained to the range
$\nu\in[0, 1/2]$, the behavior of $\nu(J)$ remains monotonic but may
increase or decrease towards a value of {\it intrinsic resonance}
$\nu_r = 1/4$, contingent upon whether the linear tune is below
$\nu_0 < \nu_r$ or above $\nu_0 > \nu_r$ with
\[
\lim_{J\rightarrow\infty} \omega(J) = \infty,
\qquad\qquad\qquad\qquad
\lim_{J\rightarrow\infty} \nu(J) = \nu_r,
\]
as depicted in the two plots on the right in
Fig.~\ref{fig:SpectraAll}.

%----------------------------------------------------------------
In this context, the term intrinsic resonance should be
interpreted carefully.
Unlike driven resonances, intrinsic resonance typically arises
from the interplay of the system's own frequencies, leading to
the excitation of oscillation amplitudes.
Here, the term {\it intrinsic} emphasizes the absence of any
externally applied force.
Next, note that for all canonical McMillan mappings, at the
resonant values of the parameter $a_r = 2\cos(2\,\pi\,\nu_r)$, the
mapping degenerates to linear with force function $f = a_r\,q$
and no dependence of rotation number on amplitude:
\[
    \nu(J) = \nu_r,\qquad\qquad\qquad
    (\nu_0 = \nu_r),
\]
refer to the two left plots in Fig.~\ref{fig:SpectraAll}.
In this regime, the map is not only linear but also periodic,
with rational rotation numbers.
%----------------------------------------------------------------
This leads to superdegeneracy, where the system possesses more
than one functionally independent invariant~\cite{ZKN2023PolI,
ZKN2024PolII}.
For instance, in the octupole map at the resonant value $a=0$,
all three functions below remain conserved under iteration
\[
p^2 + q^2,          \qquad
p^2 + q^2 + p^2 q^2,\qquad
p^2 + q^2 - p^2 q^2,\qquad
\]
as do any other functions with four-fold symmetry.
Although the function $p^2 + q^2 - p^2 q^2$ may initially seem
inadequate for describing a linear oscillator due to the presence
of a separatrix, it has the correct limiting behavior for
nonlinear McMillan mappings with small deviations from resonance,
$\nu_r+\delta\nu$.
Furthermore, as we will see, canonical McMillan maps approximate
quadratic ($f(p) = a\,p + p^2$) and cubic ($f(p) = a\,p \pm p^3$)
H\'enon mappings up to a special scaling~(Eq.~\ref{math:Kapprox}),
which results in denominators vanishing at resonance
$\nu_0 = \nu_r$.
This same scaling causes the area enclosed by the square (or
triangle, in the case of the sextupole) separatrix to shrink to
zero, leading to the disappearance of all closed level sets of
the invariant.
This effectively models nonlinear {\it resonance} in the mentioned
chaotic systems; see~\cite{zolkin2024dynamicsIII} for details.

%----------------------------------------------------------------
In all other instances (cases SX, DO, and small oscillations for
Df), the limiting amplitude is determined by the separatrix.
In Hamiltonian dynamics, this results in a vanishing frequency as
the period on the separatrix tends to infinity, while for mappings,
the limiting rotation number tends to $\frac{1}{2}$ for systems
above the intrinsic resonance ($\nu_0 > \nu_r$) due to the presence
of 2-cycles:
\[
    \omega(J_\text{sep}) = 0,
    \qquad
    \nu(J_\text{sep}) = \frac{1+\mathrm{sign}\,[\nu_0-\nu_r]}{4},
    \,\,(\nu_0\neq\nu_r)
\]
where $\nu_r = 1/3$ for the (SX) and $1/2$ for the (DO) cases.

\noindent
%----------------------------------------------------------------
$\bullet$ Within Hamiltonian systems~(\ref{math:HAll}), focusing
and defocusing octupoles display constant detuning values, assessed
at the equilibrium point at the origin, albeit with opposing signs:
\[
\left.\frac{\dd\omega_\mathrm{FO}}{\dd J_\mathrm{FO}}\right|_{J=0} = -
\left.\frac{\dd\omega_\mathrm{DO}}{\dd J_\mathrm{DO}}\right|_{J=0} =
\frac{3}{4}.
\]
The terms ``focusing'' and ``defocusing'' in the systems' names
correspond to their optical properties.
However, for McMillan octupoles, the detuning is a function of the
parameter $a$, with a reversal of optical properties upon crossing
the intrinsic resonance:
\begin{equation}
\label{math:detMcOCT}
\left.
    \frac{\dd\nu_\mathrm{FO,DO}}
         {\dd J_\mathrm{FO,DO}}
\right|_{J=0} =
    \pm \frac{3}{2\,\pi}\frac{a}{4-a^2} =
    \pm \frac{3}{4\,\pi}\frac{\cot(2\,\pi\,\nu_0)}
                             {\sin(2\,\pi\,\nu_0)}.
\end{equation}

\begin{figure}[h]
    \centering
    \includegraphics[width=\columnwidth]{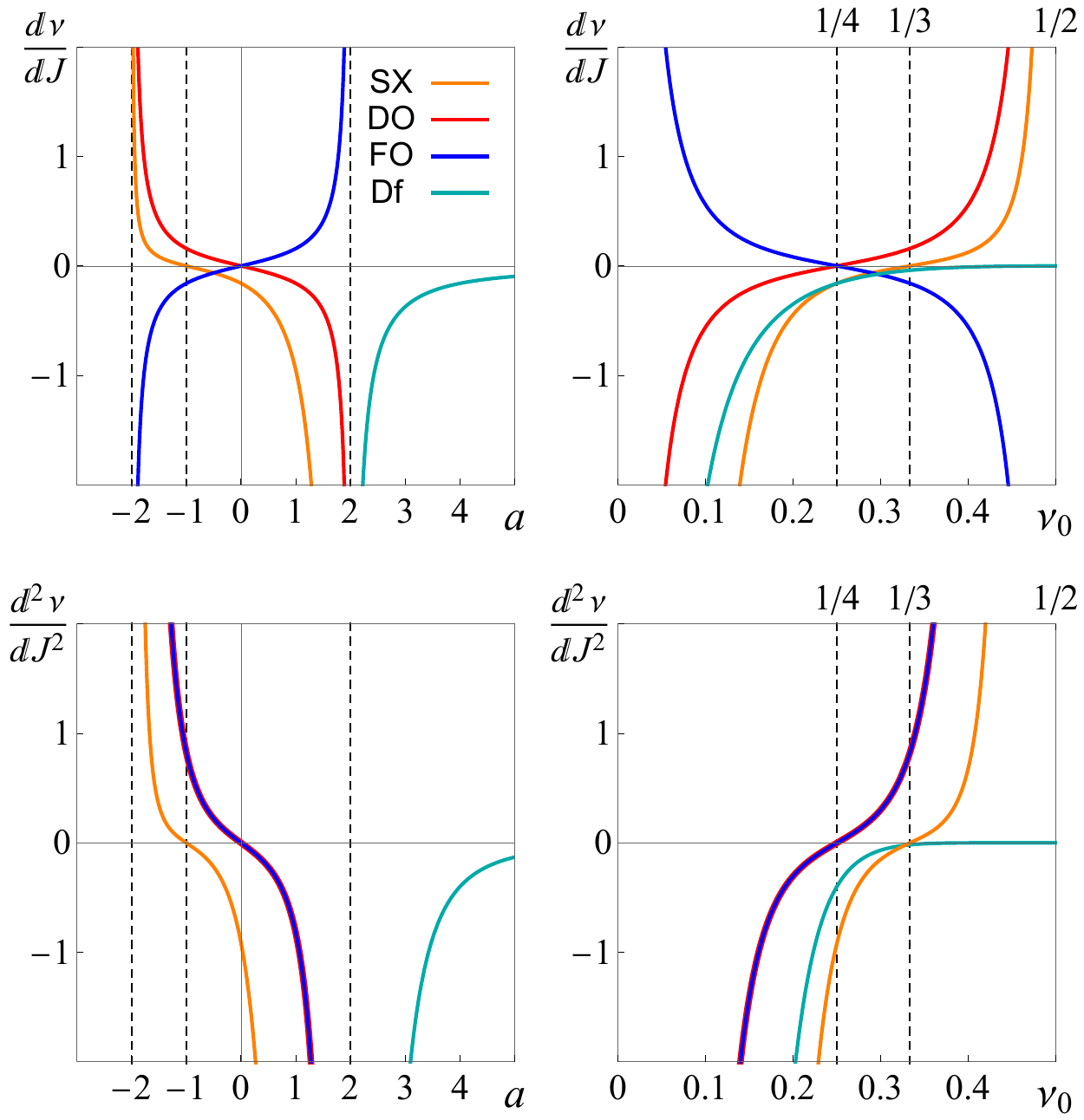}
    \caption{\label{fig:DetuningAll}
    Detuning $\dd\nu/\dd J$ and the second derivative
    $\dd^2\nu/\dd J^2$ evaluated at the fixed point ($J=0$)
    for all McMillan multipoles.
    The left and right columns show these quantities as
    functions of the map parameter $a$ and $\nu_0$ respectively.
    }
\end{figure}

%----------------------------------------------------------------
Similarly, the detuning for McMillan sextupole becomes positive
for $\nu_0 > 1/3$
\begin{equation}
\label{math:detMcSXT}
\begin{array}{l}
\ds\left.
    \frac{\dd\nu_\mathrm{SX}}
         {\dd J_\mathrm{SX}}
\right|_{J=0} = \ds
    -\frac{1}{2\,\pi}\frac{(1+a)(8+a)}{(2+a)(2-a)^2}        \\[0.4cm]
\ds\qquad =
-\frac{1}{32\,\pi}
 \frac{8\,\sin(2\,\pi\,\nu_0)+\sin(4\,\pi\,\nu_0)}
      {\sin^3(\pi\,\nu_0)    \,   \sin^3(2\,\pi\,\nu_0)}\,
      \sin(3\,\pi\,\nu_0)
\end{array}
\end{equation}
in contrast to the always negative value in its reference
Hamiltonian (SX):
\[
\left.
    \frac{\dd\omega_\mathrm{SX}}
         {\dd J_\mathrm{SX}}
\right|_{J=0} =-\frac{5}{6}.
\]
Additionally, above the resonance, the invariant
$\K_\text{sxt}[p,q]$ has 3 critical points, making its level sets
topologically different from $\h_\text{sxt}[p,q]$
(see Appendix~\ref{secAp:Stability} for details).
%----------------------------------------------------------------
Finally, considering the limit for the case (Df):
\begin{equation}
\label{math:detMcDF}
\ds\left.
    \frac{\dd\nu_\mathrm{Df}}
         {\dd J_\mathrm{Df}}
\right|_{J=0}  =
    \frac{1}{2\,\pi}\frac{4+a}{a\,(2-a)} =
    -\frac{1}{4\,\pi}
    \frac{2+\cos(2\,\pi\,\nu_0)}{\tan^2(\pi\,\nu_0)}.
\end{equation}
Figure~\ref{fig:DetuningAll} provides graphs for the detunings
(\ref{math:detMcOCT} -- \ref{math:detMcDF}) and the second
derivatives of the rotation number with respect to the action
variable at the fixed point as functions of $a$ and $\nu_0$.
%in the top and the bottom rows respectively.

%----------------------------------------------------------------
%----------------------------------------------------------------

%===============================================================%
%===============================================================%
%===============================================================%
\section{\label{sec:Stability}Stability diagrams}

\begin{figure}[b!]
    \centering
    \includegraphics[width=\columnwidth]{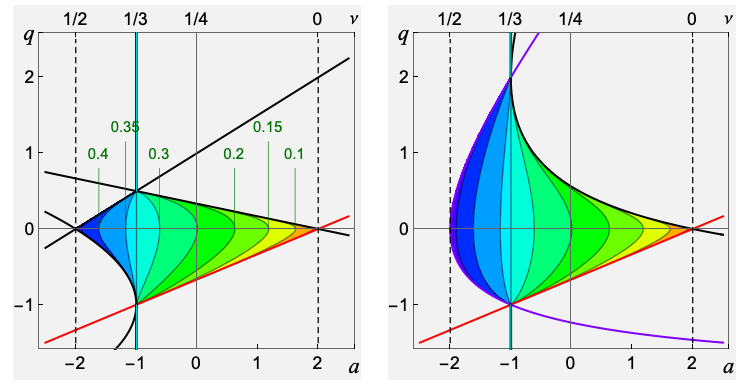}
    \caption{\label{fig:StabilitySXT}
    Stability diagrams for McMillan sextupole.
    The left and right plots correspond to initial conditions
    along the first or second symmetry lines.
    The color represents the rotation number for stable trajectories,
    ranging from 0 (red) to 1/2 (violet).
    Tick marks at the top provide resonant values of $\nu_0$.
    }
\end{figure}

\begin{figure}[t!]
    \centering
    \includegraphics[width=\columnwidth]{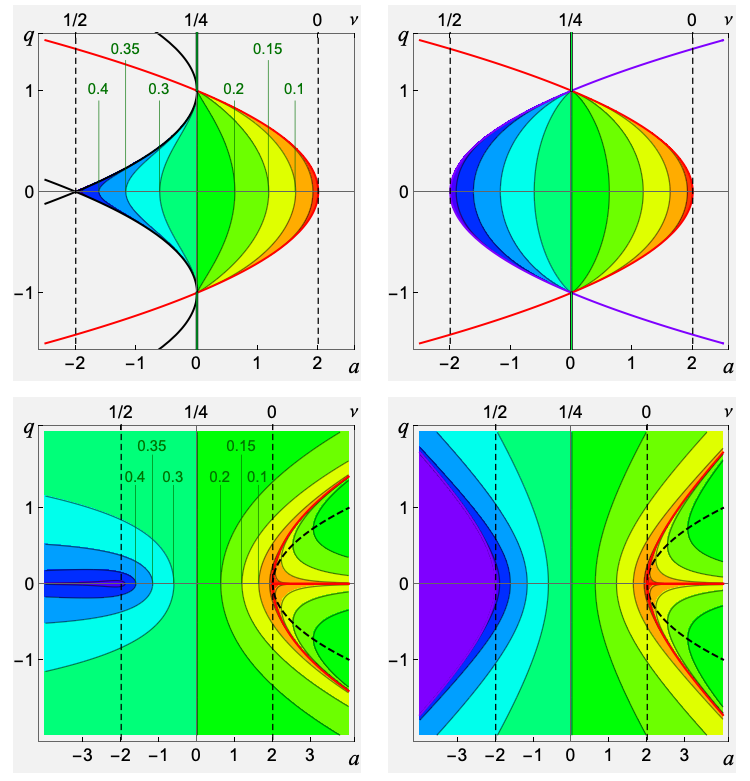}
    \caption{\label{fig:StabilityOCT}
    Same as in Fig.~\ref{fig:StabilitySXT} but for defocusing (top)
    and focusing (bottom) McMillan octupoles.
    }\vspace{-0.25cm}
\end{figure}

%----------------------------------------------------------------
To summarize the dynamical properties and establish their
connection to physical variables, we present stability diagrams,
Figs.~\ref{fig:StabilitySXT} and \ref{fig:StabilityOCT},
utilizing color to depict $\nu(q_0)$ for finite trajectories at
different parameter values $a$.
Each case includes two plots, corresponding to initial conditions
set up on the first (left column) and second (right column)
symmetry lines, respectively:
\[
    l_1:\,p_0 = q_0,
    \qquad\quad\text{and}\quad\qquad
    l_2:\,p_0 = f(q_0)/2.
\]

\vspace{-0.5cm}
%----------------------------------------------------------------
%----------------------------------------------------------------

\newpage
%----------------------------------------------------------------
%----------------------------------------------------------------

%----------------------------------------------------------------
These lines traverse through key features, and in the context of
accelerator physics, the second diagram defines the {\it dynamical
aperture} (the largest possible coordinate), delineating the phase
space available for the beam of particles.
Both symmetry lines intersect all fixed points, whereas the second
symmetry line and its inverse additionally pass through the 2-cycle.
For more details, please refer to \cite{mcmillan1971problem,
devogelaere1950,lewis1961reversible,ZKN2023} and 
\cite{roberts1992revers,zolkin2024HenonSet}.

%===============================================================%
%===============================================================%
%===============================================================%
\section{\label{sec:PertTh}McMillan map and perturbation theory}

%----------------------------------------------------------------
In Hamiltonian dynamics, a common technique involves expanding
the potential energy function around its minimum, especially when
examining small oscillations or perturbations.
This expansion enables us to approximate the potential and gain
insights into the particle's behavior in proximity to its
equilibrium position.
For a smooth potential, denoted as $U(q)$ with a minimum at the
origin and satisfying $U(0)=0$, its Taylor expansion is given by:
\[
    U(q) = \frac{U''(0)}{2!}\,q^2 +  \frac{U'''(0)}{3!}\,q^3 +
     \frac{U''''(0)}{4!}\,q^4 + \mathcal{O}(q^5).
\]
As a result, the Hamiltonian $\h_\mathrm{sxt}[p,q]$ serves as the
first-order approximation for a generic asymmetric potential,
while $\h^\pm_\mathrm{oct}[p,q]$ are the first-order
approximations for a symmetric $U(q)$, or more generally, for an
expansion with $U'''(0)=0$.
These model Hamiltonians play a crucial role in capturing
nonlinear effects, especially when higher-order terms in the
expansion come into play.

%----------------------------------------------------------------
Beyond its significance as one of the very few known exact
integrable models, the symmetric McMillan map holds deeper
meaning by serving as the first and second-order approximations
for a more general and even chaotic mappings in McMillan-H\'enon
form (\ref{math:McMsym}), with a differentiable but otherwise
arbitrary force function $f(q)$, resembling situation in
continuous dynamics.
%----------------------------------------------------------------
To demonstrate this, we introduce a small positive parameter
$\varepsilon$ representing the amplitude of oscillations.
This is achieved through a change of variables
$(q,p) \rightarrow \varepsilon\,(q,p)$:
\begin{equation}
\label{math:MapExp}
\!\!\!\!
\begin{array}{l}
q' = p							\\[0.25cm]
\ds	p' =-q + \frac{f(\varepsilon\,p)}{\varepsilon} =
	    -q + a\,p	+ \varepsilon\,b\,p^2
			+ \varepsilon^2 c\,p^3
			+ \ldots
\end{array}
\end{equation}
where the force function is expanded in a power series of
$(\varepsilon\,p)$
\[
	a \equiv \pd_p   f(0)/1!,\quad
	b \equiv \pd_p^2 f(0)/2!,\quad
	c \equiv \pd_p^3 f(0)/3!,\quad
	\ldots,
\]
and we assume the fixed point to be at the origin, necessitating
$f(0)=0$.
%----------------------------------------------------------------
Subsequently, we seek an \textit{approximated invariant} of
motion that is conserved with an accuracy of order
$\mathcal{O}(\varepsilon^{n+1})$:
\begin{equation}
\label{math:KK'}
\K^{(n)}[p',q'] - \K^{(n)}[p,q] = \mathcal{O}(\varepsilon^{n+1}).
\end{equation}
The invariant is sought in the form of a polynomial:
\[
\K^{(n)} =
    \K_0 + \varepsilon\,\K_1 + \varepsilon^2\,\K_2 + \ldots +
    \varepsilon^n\,\K_n,
\]
where $\K_m$ consists of homogeneous polynomials in $p$ and $q$
of $(m+2)$ degree
\[
\begin{array}{l}
	\K_0 =  C_{2,0}\,p^2    +
	        C_{1,1}\,p\,q\, +
	        C_{0,2}\,q^2,			    \\[0.2cm]
	\K_1 =  C_{3,0}\,p^3    +
	        C_{2,1}\,p^2 q  +
	        C_{1,2}\,p\,q^2 +
	        C_{0,3}\,q^3,               \\[0.1cm]
	\cdots,
\end{array}
\]
and $C_{i,j}$ are coefficients to be determined to satisfy 
Eq.~(\ref{math:KK'}).
%----------------------------------------------------------------
The reader can check that, in the first two orders of this
perturbation theory, a general result is provided
\begin{equation}
\label{math:Kapprox}
\begin{array}{l}
\ds \K^{(2)}[p,q] = \K_0[p,q]
    - \varepsilon\,\frac{b}{a+1}\,(p^2 q + p\,q^2)\,+   \\[0.35cm]
\qquad\ds +\,
    \varepsilon^2\left(\left[
        \frac{b^2}{a\,(a+1)} - \frac{c}{a}
    \right] p^2q^2 +
    C\,\K_0^2[p,q] \right)
\end{array}
\end{equation}
where $C$ is a coefficient such that Eq.~(\ref{math:KK'}) is
satisfied for any value it takes.

%----------------------------------------------------------------
Here, if we set $C=0$, one recognizes the invariant for the
symmetric McMillan map.
Thus, we learn that mappings~(\ref{math:mapAll}) are not just
integrable models.
%----------------------------------------------------------------
Similar to how the quadratic invariant $\K_0[p,q]$ with the
corresponding force $f(p) = a\,p$ allows us to ``integrate out''
linear dynamics around the fixed point (zeroth order), McMillan
sextupoles and octupoles provide the next order of approximation,
defining detuning and even approximating the dynamical aperture
around low-order resonances.
Below, we separately analyze cases of small
($\varepsilon \rightarrow 0$) and large amplitudes, providing
several examples and discussing applicability.
\newpage

%===============================================================%
\subsection{\label{sec:SmallAmp}Small amplitudes}
%===============================================================%

%===============================================================%
%===============================================================%
\subsubsection{\label{sec:DfRegime}McMillan mapping in Duffing regime}

%----------------------------------------------------------------
As our first example, let's consider the familiar case of the
McMillan octupole~(\ref{math:mapAll}) in the Duffing regime (Df).
Since the fixed point $\zeta^{(1-1)} = (0,0)$ is unstable, we can
shift the origin to one of the symmetric fixed points
(\ref{math:zeta1-23}):
\[
    (q,p) \rightarrow (\widetilde{q},\widetilde{p}) =
    (q,p)- \zeta^{(1-2)},
    \quad
    \zeta^{(1-2)} = \sqrt{\frac{a-2}{2}}\,(1,1).
\]
%----------------------------------------------------------------
This preserves the form of the map (\ref{math:McMsym}) and results
in the transformation of the force function:
\[
f(q) = \frac{a\,q}{1+q^2}
\,\,\rightarrow\,\,
\widetilde{f}(\,\!\widetilde{q}\,) =
    f\left(\widetilde{q} + q^{(1-2)}\right) - 2\,q^{(1-2)}.
\]
Expanding the new force in a series
\[
\begin{array}{l}
\ds \widetilde{f}(\,\!\widetilde{q}\,) =
    \alpha\,\widetilde{q} -
    \frac{(1+\alpha)\,\sqrt{4-\alpha^2}}{2}\,\widetilde{q}^{\,2} +
\\[0.35cm]
\ds \qquad\qquad\,\,\,\,\, + \,
    \frac{(2+\alpha)(2-\alpha^2)}{4}\,\widetilde{q}^{\,3} +
    \mathcal{O}(\widetilde{q}^{\,4}),
\end{array}
\]
where we introduced the parameter
\begin{equation}
\label{math:AAlpha}
\alpha = \frac{8}{a} - 2\,:\qquad
\nu_0 = \frac{\arccos(\alpha/2)}{2\,\pi},
\end{equation}
we can write an invariant in the form~(\ref{math:Kapprox})
\[
    \K[\widetilde{p},\widetilde{q}\,] =
    \K_0[\widetilde{p},\widetilde{q}\,] + 
    \frac{\sqrt{4-\alpha^2}}{2}\,(
        \widetilde{p}^{\,2}\widetilde{q} +
        \widetilde{p}\,\widetilde{q}^{\,2}
    ) +
    \frac{2+\alpha}{4}\,\widetilde{p}^{\,2}\widetilde{q}^{\,2}.
\]
%----------------------------------------------------------------
In this example, the second-order integral of motion from above
is an exact invariant (i.e., not approximated), which, when
written in new coordinates, provides coefficients in front of two
basic nonlinearities.
We see that oscillations inside the figure-eight separatrix are
equivalent to the symmetric McMillan map carrying both sextupole
(SX) and octupole (FO) components.
Thus, we can find detuning by adding Eqs.~(\ref{math:detMcSXT})
and (\ref{math:detMcOCT}) that are multiplied by appropriate
scaling coefficients from the invariant:
\[
\begin{array}{l}
\ds\left.
    \frac{\dd\nu_\mathrm{Df}}
         {\dd J_\mathrm{Df}}
\right|_{J=0}  =
    -\frac{1}{2\,\pi}\frac{(1+\alpha)(8+\alpha)}
                          {(2+\alpha)(2-\alpha)^2}
    \times\left(\frac{\sqrt{4-\alpha^2}}{2}\right)^2 + \\[0.5cm]
\ds \qquad +\,
    \frac{3}{2\,\pi}\frac{\alpha}{4-\alpha^2}
    \times\left(\frac{2+\alpha}{4}\right) =
    -\frac{1}{8\,\pi}\frac{(2+\alpha)(4+\alpha)}{2-\alpha}.
\end{array}
\]
%----------------------------------------------------------------
Using Eq.~(\ref{math:AAlpha}), we can convert this to the form
independent of the mapping's representation:
\[
\ds\left.
    \frac{\dd\nu_\mathrm{Df}}
         {\dd J_\mathrm{Df}}
\right|_{J=0}  =
    -\frac{1}{4\,\pi}
    \frac{2+\cos(2\,\pi\,\nu_0)}{\tan^2(\pi\,\nu_0)}
\]
which matches the previously obtained result~(\ref{math:detMcDF}).
Notice that this time we didn't solve for action-angle variables,
then take a limit of corresponding elliptic functions, but rather
used our knowledge of underlying nonlinearities.

%===============================================================%
%===============================================================%
\subsubsection{\label{sec:HNRegime}H\'enon quadratic and qubic mappings}

%----------------------------------------------------------------
It is important to recognize that for an arbitrary $f(p)$, the
global dynamics of equation (\ref{math:McMsym}) is known to be
chaotic
\cite{henon1969numerical,chirikov1969research,chirikov1979universal},
except for a few very special integrable cases
\cite{veselov1991integrable,mcmillan1971problem,quispel1988integrable,
quispel1989integrable,suris1989integrable,brown1993,cairns2016conewise,
ZKN2023,ZKN2024arxiv}.
For a brief review on the integrability of mappings in the form
(\ref{math:McMsym}), please refer to \cite{ZKN2023}.

%----------------------------------------------------------------
As a case study for chaotic dynamics, we use the aforementioned
{\it area-preserving H\'enon mappings} (H) with quadratic
\cite{henon1969numerical} and cubic force functions:
\begin{equation}
\label{math:fHenon}
f^\text{(H)}_\text{sxt}(q) = a\,q + q^2,
\qquad\qquad\qquad
f^\text{(H)}_\text{oct}(q) = a\,q \pm q^3.
\end{equation}

\noindent
%----------------------------------------------------------------
$\bullet$
Starting with the octupole forces, by comparing
$f^\text{(H)}_\text{oct}$ with Eq.~(\ref{math:MapExp}), we have
$b=0$ and $c=\pm1$, providing an approximated McMillan-H\'enon
(MH) invariant of the second order:
\[
    \K_\text{DO,FO}^\text{(MH)}[p,q] = \K_0[p,q] \mp \frac{1}{a}\,p^2 q^2
\]
corresponding to symmetric McMillan mappings with forces
\[
f_\text{DO,FO}^\text{(MH)}(q) = \frac{a^2 q}{a \mp q^2} =
    a\,q \pm q^3 + \mathcal{O}(q^5).
\]
Therefore, with the use of the scaling, from
Eq.~(\ref{math:detMcOCT}) we have
\begin{equation}
\label{math:HNOctDet}    
\ds\left.
    \frac{\dd\nu^\text{(MH)}_\text{DO,FO}}
         {\dd  J^\text{(MH)}_\text{DO,FO}}
\right|_{J=0}  =
\frac{1}{a}\times
\ds\left.
    \frac{\dd\nu_\text{DO,FO}}
         {\dd  J_\text{DO,FO}}
\right|_{J=0}  =
    \mp\frac{3}{8\,\pi}\frac{1}{\sin^2(2\,\pi\,\nu_0)}.
\end{equation}

\noindent
%----------------------------------------------------------------
$\bullet$ For the quadratic H\'enon map, we can also go up to the
second order of perturbation theory by first matching the quadratic
term in the force function ($b=1$)
\[
f^\text{(MH)}_\text{SX-1}(q) =
    \frac{a\,(a+1)+q}{(a+1)-q}\,q =
    a\,q + q^2 + \frac{q^3}{a+1} + \mathcal{O}(q^4),
\]
and then removing the cubic term from the expansion ($c=0$) by
using
\[
f^\text{(MH)}_\text{SX-2}(q) =
    \frac{a\,(a+1)+q}{(a+1)-q+\frac{1}{a}\,q^2}\,q =
    a\,q + q^2 + \mathcal{O}(q^4).
\]
While the first order is simply a rescaled McMillan sextupole
(SX), the second order represents a specific mixture~\cite{zolkin2024dynamicsIII} of both
McMillan sextupole and focusing octupole (FO), as can be seen
from the invariants:
\[
\begin{array}{l}
\ds     \K_\text{SX-1}^\text{(MH)}[p,q] = \K_0[p,q] - 
        \frac{p^2 q + p\,q^2}{a+1},                         \\[0.4cm]
\ds    \K_\text{SX-2}^\text{(MH)}[p,q] = \K_0[p,q] - 
        \frac{p^2 q + p\,q^2}{a+1} + \frac{p^2\,q^2}{a\,(a+1)}.
\end{array}
\]
%----------------------------------------------------------------
Applying the same logic, we evaluate the detuning for the quadratic
H\'enon map as
\[
\ds\left.
    \frac{\dd\nu^\text{(MH)}_\text{SX-1}}
         {\dd  J^\text{(MH)}_\text{SX-1}}
\right|_{J=0}  =
\left(\frac{-1}{1+a}\right)^2\times
\ds\left.
    \frac{\dd\nu_\text{SX}}
         {\dd  J_\text{SX}}
\right|_{J=0}
\]
and
\[
\ds\left.
    \frac{\dd\nu^\text{(MH)}_\text{SX-2}}
         {\dd  J^\text{(MH)}_\text{SX-2}}
\right|_{J=0}  =
\ds\left.
    \frac{\dd\nu^\text{(MH)}_\text{SX-1}}
         {\dd  J^\text{(MH)}_\text{SX-1}}
\right|_{J=0}  +
\frac{1}{a\,(1+a)}\times
\ds\left.
    \frac{\dd\nu_\text{FO}}
         {\dd  J_\text{FO}}
\right|_{J=0}
\]
resulting in
\begin{equation}
\label{math:HNSxtDet}
\begin{array}{l}
%\ds \ds\left.
%    \frac{\dd\nu^\text{(MH)}_\text{SX-1}}
%         {\dd  J^\text{(MH)}_\text{SX-1}}
%\right|_{J=0}  =
%-\frac{1}{16\,\pi}
% \frac{4+\cos(2\,\pi\,\nu_0)}
%      {\sin(\pi\,\nu_0)\,\sin^2(2\,\pi\,\nu_0)\,\sin(3\,\pi\,\nu_0)},
%         \\[0.55cm]
\ds \ds\left.
    \frac{\dd\nu^\text{(MH)}_\text{SX-1}}
         {\dd  J^\text{(MH)}_\text{SX-1}}
\right|_{J=0}  =
-\frac{1}{16\,\pi}
 \frac{9\,\cos(\pi\,\nu_0)+\cos(3\,\pi\,\nu_0)}
      {\sin^3(2\,\pi\,\nu_0)\,\sin(3\,\pi\,\nu_0)},
         \\[0.55cm]
\ds \ds\left.
    \frac{\dd\nu^\text{(MH)}_\text{SX-2}}
         {\dd  J^\text{(MH)}_\text{SX-2}}
\right|_{J=0}  =
    -\frac{1}{16\,\pi}
    \frac{3\,\cot(\pi\,\nu_0)+\cot(3\,\pi\,\nu_0)}
         {\sin^3(2\,\pi\,\nu_0)}.
\end{array}
\end{equation}

%----------------------------------------------------------------
To check the accuracy of Eqs.~(\ref{math:HNSxtDet}) and
(\ref{math:HNOctDet}) in describing actual dynamics, we compare
them with the dependencies obtained numerically from tracking.
Despite the chaotic nature of the global dynamics with forces
(\ref{math:fHenon}), the KAM theorem assures the existence of
action-angle variables in the vicinity of stable fixed points
\cite{kolmogorov1954conservation,moser1962invariant,arnol1963small}.
%----------------------------------------------------------------
We introduce a test particle that is initially placed, for example,
on one of the symmetry lines with a small deviation from the origin,
$q_0 = \delta q$.
First, we perform tracking for $N$ iterations:
\[
    \{q_0,p_0\} \rightarrow 
    \{q_1,p_1\} \rightarrow 
    \ldots      \rightarrow 
    \{q_N,p_N\},
\]
and then, we rewrite it in polar phase space coordinates
\[
    \{\rho_0,\phi_0\} \rightarrow
    \{\rho_1,\phi_1\} \rightarrow
    \ldots            \rightarrow
    \{\rho_N,\phi_N\}
\]
where
\[
    q = \rho\,\cos\phi,
    \qquad\qquad\qquad
    p = \rho\,\sin\phi.
\]
%----------------------------------------------------------------
From the sub-array of polar angles $\{\phi_0,\phi_q,\ldots,\phi_N\}$,
we evaluate the rotation number as:
\[
    \nu^*(\delta q) =
    \frac{1}{2\,\pi}\,\sum_{j=0}^{N-1} \frac{\phi_{j+1}-\phi_{j}}{N}.
\]
By sorting tuples $\{\rho_j,\phi_j\}$ such that $\forall\,j\in[1,N]$
we have $\phi_j>\phi_{j-1}$, we estimate the action integral
according to
\[
J^*(\delta q) =
    \frac{1}{2\,\pi}\,
    \sum_{j=0}^{N-1}\,\frac{\rho_j^2}{2}\,(\phi_{j+1} - \phi_j).
\]
Then, detuning can be numerically evaluated as:
\[
\left. \frac{\dd\nu}{\dd J}\right|_{J = 0} =
    \lim_{\substack{\delta q\rightarrow 0\,\,\, \\ N\rightarrow\infty}}
    \frac{\nu^*(\delta q) - \nu_0}{J^*(\delta q)}.
\]
By choosing a sufficiently large $N$ and sufficiently small
$\delta q$, we ensure convergence.

%----------------------------------------------------------------
\begin{figure}[t]
    \centering
    \includegraphics[width=\columnwidth]{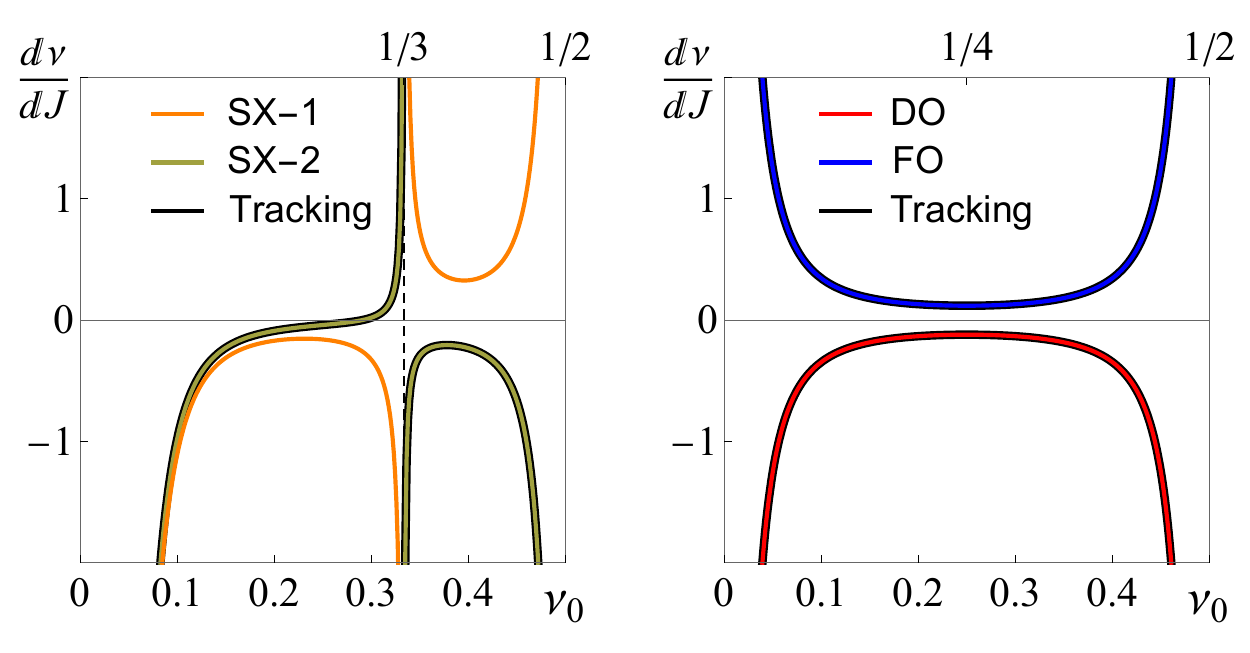}\vspace{-0.2cm}
    \caption{\label{fig:DetuningHN}
    Detuning for the H\'enon quadratic (left) and cubic (right)
    mappings~(\ref{math:fHenon}), assessed from tracking data
    (black curves), juxtaposed with analytical predictions from
    perturbation theory (illustrated with colors).
    }\vspace{-0.25cm}
\end{figure}
%----------------------------------------------------------------

%----------------------------------------------------------------
Figure~\ref{fig:DetuningHN} presents the comparison of tracking
(black curve) along with analytical results (shown in colors).
As observed, the first order of perturbation theory (SX-1) for the
H\'enon quadratic map is not sufficient and provides an accurate
approximation of detuning only in the range $0 \leq \nu_0 \lesssim 0.2$.
However, the second order for both quadratic and cubic mappings
(SX-2, DO and FO respectively) provides coincidence between both
approaches up to a machine-order accuracy.

%===============================================================%
%===============================================================%
\subsubsection{\label{sec:Twiss}Thin nonlinear lens}

%----------------------------------------------------------------
Let's now consider a more practical example of a simple accelerator
lattice with one degree of freedom consisting of linear optics
elements (drift spaces, dipoles, and quadrupoles) and a single thin
nonlinear lens~\cite{SYLee4}:
\[
\mathrm{F}:\quad\,\,
    \begin{bmatrix}
      x \\ \dot{x}
    \end{bmatrix}' =
    \begin{bmatrix}
      x \\ \dot{x}
    \end{bmatrix} +
    \begin{bmatrix}
      0 \\ F(x)
    \end{bmatrix}.
\]
The effect on a test particle from all linear elements can be
represented using a matrix with Courant-Snyder parametrization
\cite{courant1958theory}:
\[
\mathrm{M}:\quad
    \begin{bmatrix}
      x \\ \dot{x}
    \end{bmatrix}' =
    \begin{bmatrix}
      \cos \Phi + \alpha\,\sin \Phi	& \beta\,\sin \Phi		\\
      -\gamma\,\sin\Phi		& \cos \Phi - \alpha\,\sin \Phi
    \end{bmatrix}
    \begin{bmatrix}
      x \\ \dot{x}
    \end{bmatrix},
\]
where $\alpha$, $\beta$ and $\gamma$ are {\it Twiss parameters}
(also known as {\it Courant-Snyder parameters}) at the thin
lens location, and $\Phi$ is the {\it betatron phase advance}
over the linear optics insert
\[
    \Phi = \int \frac{\dd s}{\beta(s)}.
\]
%----------------------------------------------------------------
Without the nonlinear lens, the Twiss parameters are functions of
the longitudinal coordinate $s$ with $\beta(s)$ referred to as the
$\beta$-function, $\alpha(s) \equiv -\dot{\beta}(s)/2$, and
$\gamma(s) \equiv [1+\alpha^2(s)]/\beta(s)$.
At any location, the {\it Courant-Snyder invariant} is defined as:
\[
\gamma(s)\,x^2(s) + 
2\,\alpha(s)\,x(s)\,\dot{x}(s) + 
\beta(s)\,\dot{x}^2(s) = \const.
\]
The rotation number (or {\it betatron tune} in accelerator
physics) is independent of amplitude and given by:
\[
    \nu_0 = \frac{1}{2\,\pi}\,\oint\frac{\dd s}{\beta(s)}.
\]

%----------------------------------------------------------------
When the nonlinear lens is introduced, the combined one-turn map
$\mathrm{M}\circ\mathrm{F}$ can be rewritten in the form
(\ref{math:McMsym}) using a change of variables
\begin{equation}
\label{math:Thin2HN}
\left\{
\begin{array}{l}
	q = x,                                      \\[0.2cm]
	p = x\,(\cos \Phi + \alpha\,\sin \Phi) + 
            \dot{x}\,\beta\,\sin \Phi,
\end{array}\right.
\end{equation}
with the force function given by
\[
f(q) = 2\,q\,\cos\Phi + \beta\,\sin\Phi\,F(q).
\]
Using results from the previous subsection, we know that we can
define an approximated integral of motion~(\ref{math:Kapprox})
with
\begin{equation}
\label{math:Scaling}
\begin{array}{l}
\ds a = 2\,\cos\Phi + \beta\,\sin\Phi\,\pd_q\,F(0), \\[0.25cm]
\ds b = \beta\,\sin\Phi\,\pd_{qq }\,F(0),           \\[0.25cm]
\ds c = \beta\,\sin\Phi\,\pd_{qqq}\,F(0).
\end{array}
\end{equation}
This can be seen as a nonlinear analog of the Courant-Snyder
invariant that includes higher-order terms and can be easily
propagated through the linear part of the lattice, thus defined
for any azimuth, $\K[p,q;s]$.

%----------------------------------------------------------------
\begin{figure}[t]
    \centering
    \includegraphics[width=0.79\columnwidth]{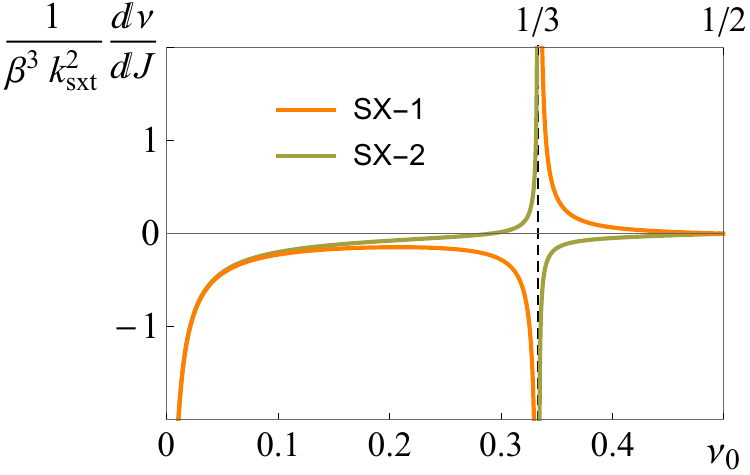}\vspace{-0.045cm}
    \caption{\label{fig:DetuningSX}
    Detuning for the thin sextupole lens as a function of the betatron
    tune $\nu_0$.
    The green curve corresponds to the value (SX-2) that matches
    numerical simulation, while the orange curve represents the
    first-order approximation (SX-1).
    }\vspace{-0.2cm}
\end{figure}
%----------------------------------------------------------------

%----------------------------------------------------------------
If necessary, $(q,p)$ can be inverted back to $(x,\dot{x})$,
providing results in terms of physical variables.
Using thin (Th) sextupole and thin octupole lenses as an example
\cite{SYLee4}:
\[
F_\mathrm{sxt}(x) = k_\text{sxt}\,x^2,
\qquad
F_\mathrm{oct}(x) =\pm k_\text{oct}\,x^3,
\]
we obtain first and second order detuning values:
\[
\begin{array}{l}
\ds \ds\left.
    \frac{\dd\nu^\text{(Th)}_\text{SX-1}}
         {\dd  J^\text{(Th)}_\text{SX-1}}
\right|_{J=0}  = -\frac{1}{16\,\pi}\,\,
    \frac{9\,\cos(\pi\,\nu_0)+\cos(3\,\pi\,\nu_0)}
         {\sin(3\,\pi\,\nu_0)}\,\,\,
    \beta^3\,k_\text{sxt}^2,
         \\[0.55cm]
%\ds \ds\left.
%    \frac{\dd\nu^\text{(Th)}_\text{SX-1}}
%         {\dd  J^\text{(Th)}_\text{SX-1}}
%\right|_{J=0}  = -\frac{1}{32\,\pi}\,
%    \frac{8\,\sin(2\,\pi\,\nu_0)+\sin(4\,\pi\,\nu_0)}{\sin(\pi\,\nu_0)\,\sin(3\,\pi\,\nu_0)}\,\,
%    \beta^3\,k_\text{sxt}^2,
%         \\[0.55cm]
\ds \ds\left.
    \frac{\dd\nu^\text{(Th)}_\text{SX-2}}
         {\dd  J^\text{(Th)}_\text{SX-2}}
\right|_{J=0}  =
   -\frac{3}{16\,\pi}\left[
        3\,\cot(\pi\,\nu_0)+\cot(3\,\pi\,\nu_0)
   \right]\,\beta^3\,k_\text{sxt}^2,
         \\[0.55cm]
\ds \ds\left.
    \frac{\dd\nu^\text{(Th)}_\text{DO,FO}}
         {\dd  J^\text{(Th)}_\text{DO,FO}}
\right|_{J=0}  =
   \mp\frac{3}{8\,\pi}\,\beta^2\,k_\text{oct},
\end{array}
\]
see Fig.~\ref{fig:DetuningSX}.
%----------------------------------------------------------------
The formulas above are consistent with other derivations using
various perturbation theories including the Deprit perturbation theory~\cite{michelotti1995intermediate} and the Lie algebra treatment~\cite{michelotti1984moser,bengtsson1997, morozov2017dynamical}.
Notice that in addition to the scaling provided by
Eqs.~(\ref{math:Scaling}), an additional factor equal to the
Jacobian of the transformation (\ref{math:Thin2HN}),
$\mathbf{J} = \beta\,\sin\Phi$, must be taken into account to
obtain the equations above from the McMillan-H\'enon detunings
(\ref{math:HNOctDet}) and (\ref{math:HNSxtDet}).

%===============================================================%
\subsection{\label{sec:LargeAmp}Large amplitudes}
%===============================================================%

%----------------------------------------------------------------
While the dependencies related to small amplitudes, such as the
rotation number or detuning around the fixed point, are smooth
functions of the map parameter, the distance to the largest stable
trajectory (dynamical aperture) in the case of general chaotic
dynamics forms a fractal curve.
Mathematically, the boundary of the area of stability in the
parameter space is defined by lines associated with $n$-cycles
of the system, and is known to be a non-trivial task since the
early works attributed to P.~Fatou, G.~Julia and B.~Mandelbrot
\cite{Fatou:1917A,Fatou:1917B,GastonJulia1918,BrooksMatelski,
Mandelbrot1980FRACTALAO}.

%----------------------------------------------------------------
Another nontrivial dependence in the case of large amplitudes
is the behavior of the rotation number.
Considering $\nu(q_0)$ for a fixed value of the map parameters
is an example of a singular (continuous, but not absolutely
continuous) function similar to a Cantor function.
When plotted against the initial condition $q_0$, the graph
reveals a ``devil's staircase'' resulting from mode-locking
(chaotic islands) that occurs for every rational $\nu$.

%----------------------------------------------------------------
The applicability of any perturbation theory in a particular
order greatly depends on the ``leading'' nonlinearity.
As previously noted, the area of stability exhibits a complex
shape, yet McMillan multipoles offer a reliable estimate for
the separatrix $q_\mathrm{sep}$ and $\nu(q_0)$ near low-order
(integer, half-integer, and third-integer) resonances.
While we do note a substantial alignment between our perturbation
theory and other methods in addressing dynamics around the fixed
point, it's important to highlight that the dependence of
$\nu(q_0)$ on large amplitudes through elliptic functions
inherently differs from the typical power series of $q_0$
often obtained in methods such as Lie algebra.
In particular, analytical expressions for the rotation number
of McMillan multipoles experience very rapid change around the
limiting $n$-cycle, providing a more realistic description of
behavior near the bounding separatrix.

\begin{figure}[t!]
    \centering
    \includegraphics[width=\columnwidth]{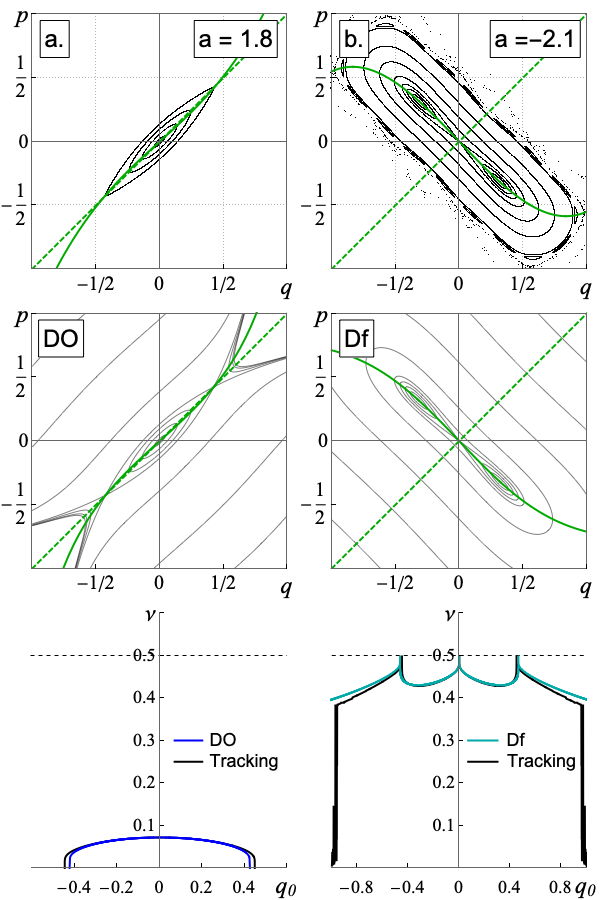}\vspace{0.45cm}
    \caption{\label{fig:GlobalOCT}
    The top row illustrates phase space diagrams for the H\'enon
    cubic map $f(q) = a\,q + q^3$ obtained through tracking.
    The middle row displays level sets for the corresponding
    approximated McMillan-H\'enon invariant of the second order.
    Dashed and solid green curves are the first ($p=q$) and
    second ($p=f(q)/2$) symmetry lines, respectively.
    The bottom row presents a comparison of the rotation number as
    a function of the initial coordinate along the second symmetry
    line $\nu(q_0)$, evaluated from tracking (black curve) and the
    analytical approximation (shown in color).
    }
\end{figure}

%----------------------------------------------------------------
To illustrate the concept, we first employ the H\'enon cubic map 
$f^\text{(H)}_\text{oct}(q) = a\,q + q^3$ in the defocusing
octupole regime ($a=1.8$) above the integer resonance
\[
    \delta\nu = \nu_0 - 0 \approx 0.072,
\]
and in the Duffing regime ($a=-2.1$) with an unstable orbit at
the origin and a stable 2-cycle slightly below the half-integer
resonance
\[
    \delta\nu = \nu_0 - \frac{1}{2} \approx-0.07.
\]
Both case studies are depicted in Fig.~\ref{fig:GlobalOCT},
showcasing tracking for stable trajectories in the top row, the
corresponding approximated invariant of the second order in the
middle row, and a comparison of rotation numbers along the second
symmetry line assessed from tracking versus perturbation theory
in the bottom.
In both situations, we observe a good estimate for the location of
the limiting separatrix (or figure-eight trajectory), as well as
qualitative and quantitative agreement of rotation numbers from
numerical experiments and analytical predictions.
When considering dynamics outside of the lemniscate (plots b.),
perturbation theory fails to predict the next isolating invariant
caused by a high-order resonance, while providing an accurate
estimate for $\nu(q_0)$ in the appropriate range.

%----------------------------------------------------------------
%----------------------------------------------------------------

%----------------------------------------------------------------
\begin{figure}[t!]
    \centering
    \includegraphics[width=\linewidth]{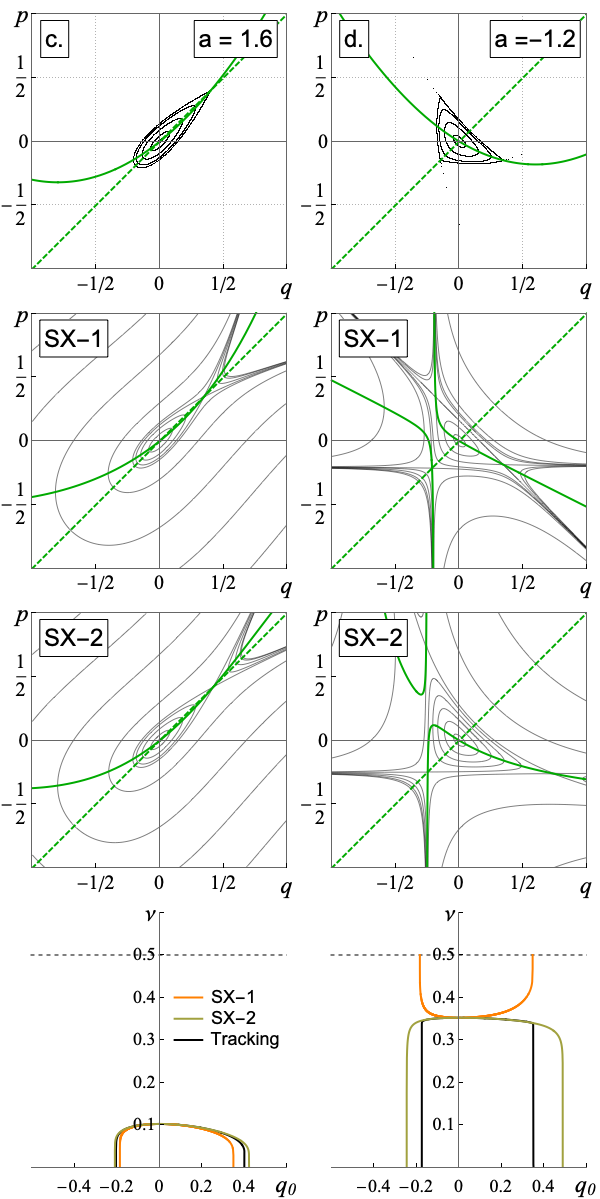}
    \caption{\label{fig:GlobalSXT}
    Same as Fig.~\ref{fig:GlobalOCT}, but for the H\'enon
    quadratic map $f(q) = a\,q + q^2$.
    An additional row (SX-1) and an orange curve in the bottom
    plots represent the first order of perturbation theory.
    }\vspace{-0.4cm}
\end{figure}
%----------------------------------------------------------------

\newpage
%----------------------------------------------------------------
Next, we examine the H\'enon quadratic map (\ref{math:fHenon}),
once again above the integer resonance ($a=1.6$)
\[
    \delta\nu = \nu_0 - 0 \approx 0.1,
\]
and then above the third-integer resonance ($a=-1.2$)
\[
    \delta\nu = \nu_0 - \frac{1}{3} \approx 0.02,
\]
as shown in Fig.~\ref{fig:GlobalSXT}.
This time, the two rows in the middle show the first and second
order approximated invariants, including corresponding
dependencies $\nu(q_0)$ in the plot at the bottom. 
For the third-order resonance (plots d.), the approximation SX-1
fails to predict the proper sign of detuning, as expected, but in
the next order SX-2 provides a quite accurate estimate of $\nu$
within the range of stable trajectories obtained by tracking.
Despite the discrepancy, both orders provide useful information
regarding the general shape and orientation of the phase space
trajectories, which is valuable in practical applications such
as resonant beam extraction, as demonstrated in the example in
the next section.

%===============================================================%
\section{\label{sec:General}Extensions and Generalizations}
%===============================================================%

%----------------------------------------------------------------
Numerous avenues exist for further generalizing the obtained results
and their application.
While each aspect merits its own dedicated publication, here we
provide a brief overview of the possibilities.

%----------------------------------------------------------------
$\bullet$ {\bf Higher dimensions}. The symmetric McMillan map can
be extended to an axially symmetric map with two degrees of freedom
(4D phase space), which is integrable and allows for the separation
of variables in polar coordinates.
This map holds potential for implementation in accelerator physics,
as it can be realized by incorporating a short electron lens into
existing linear lattice structures.

%----------------------------------------------------------------
This problem is addressed in the second part of this manuscript,
\cite{zolkin2024dynamicsII} while here, we provide a summary of
two key points.
Firstly, the McMillan octupole serves as an exact solution to
radial motion with zero angular momentum $p_\theta$, offering
insights into general dynamics.
Secondly, of particular significance is the observation that,
similar to perturbation theory for one degree of freedom, the
axially symmetric McMillan electron lens can be regarded as a
second-order approximation for a generic axially symmetric thin
lens.

%----------------------------------------------------------------
$\bullet$ {\bf Basis functions.} In constructing the perturbation
theory, the integral of motion was sought in the form of a
polynomial.
As previously mentioned, only in the first two orders (in addition
to the zeroth order) do we obtain an approximated invariant that is
an exact invariant for another system (symmetric McMillan map)
sharing the same map form.
This observation stems from Suris' theorem \cite{suris1989integrable},
which establishes that for mappings in the form~(\ref{math:McMsym}),
integrable systems with smooth invariants are limited to regular,
exponential, and trigonometric polynomials of degree two.
This limitation not only excludes polynomials of degree higher than
biquadratic in $q$ and $p$, but also severely restricts the class
of functions where integrable approximations can be sought.

%----------------------------------------------------------------
On the other hand, it should be noted that if trigonometric
polynomials are employed instead of regular ones, integrable
approximations via Suris mappings are still achieved in the
first two orders.
This alternative approach is more suitable for systems akin to
the Chirikov map~\cite{chirikov1969research,chirikov1979universal},
which feature periodic force functions $f(q)$ with specific
regularity properties \cite{ZKN2024arxiv}.

%----------------------------------------------------------------
$\bullet$ {\bf Higher orders}. Another question that remains beyond
the scope of this article is how to proceed with higher orders of
perturbation theory.
In an upcoming manuscript, we will describe the application of
Danilov's theorem to extract action-angle variables from an
approximated invariant, and, an averaging procedure that facilitates
the minimization of (\ref{math:KK'}) in higher orders by selecting
$C \neq 0$ in (\ref{math:Kapprox}).

%----------------------------------------------------------------
$\bullet$ {\bf Form of the map}. 
%----------------------------------------------------------------
Throughout this article, we have utilized a model form of the map
given by Eq.~(\ref{math:McMsym}).
While this may initially seem restrictive, we have demonstrated
that it encompasses numerous well-known dynamical systems, ranging
from the integrable symmetric McMillan map to chaotic systems like
H\'enon and Chirikov, as well as an accelerator lattice with a
single thin nonlinear lens.
Furthermore, according to Turaev theorem~\cite{turaev2002polynomial},
almost every symplectic map of the plane (and even higher-dimensional)
can be approximated by iterations of mappings in this form.

%----------------------------------------------------------------
However, what approach should be adopted for a general symplectic
transformation that represents more realistic systems, such as an
accelerator with multiple nonlinear lenses (not necessarily thin)
located at different positions along the lattice?

%----------------------------------------------------------------
In such cases, one approach is to decompose the mapping into the
involution of two consecutive mappings~(\ref{math:McMsym}), and,
then seek an approximated invariant of motion corresponding to the
asymmetric McMillan map.
Alternatively, one can employ physical variables and search for a
general polynomial.
It is worth noting that ``thick'' transformations or lenses can
always be incorporated into the analysis.
If an exact transformation is known for each element of the lattice,
the full one-turn map can be expanded into a power series up to the
third order.
If some transformations are unknown, a symplectic integrator
\cite{yoshida1990construction} consisting of drifts and thin kicks,
or the Dirac interaction picture along with the Magnus expansion
\cite{morozov2017dynamical}, can be used.

\begin{figure}[t!]
    \centering
    \includegraphics[width=0.9\linewidth]{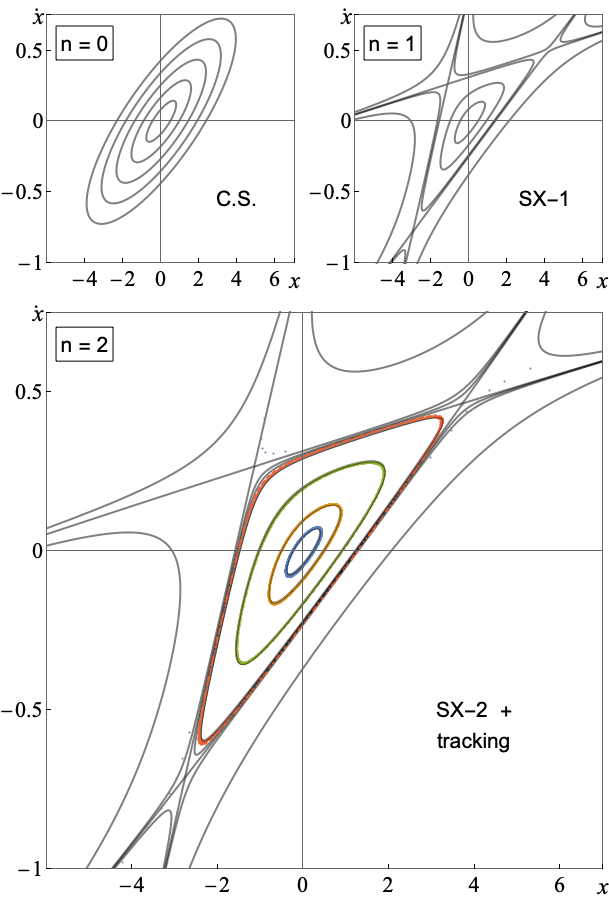}
    \caption{\label{fig:Mu2e}
    Third-integer resonant extraction for the Fermilab Mu2e
    experiment.
    The top plots present level sets for the zeroth (linear) and
    first-order approximated invariants (SX-1), while the plot at
    the bottom displays the second-order approximation (SX-2) along
    with tracking (colored dots).
    All plots are presented in physical coordinates $(x,\dot{x})$.
    }
\end{figure}

%----------------------------------------------------------------
In the first scenario, the dynamics of the asymmetric map can be
described with the help of parametrizations by Iatrou and Roberts
and Danilov theorem to provide canonical variables, similar to
what was demonstrated for the McMillan sextupole.
%----------------------------------------------------------------
To illustrate the second approach, let's consider the case of
third-integer resonant extraction for the Mu2e experiment at
Fermilab's muon campus.
The delivery ring consists of six sextupole magnets separated by
linear lattice inserts with a combined tune set up to
$\nu_0 = 0.35$, slightly above the third-integer resonance
\[
\delta\nu = \nu_0 - \frac{1}{3} \approx 0.017
\]
for this specific example.
After collecting a one-turn map (that includes all elements) and
expanding it into a power series, we apply perturbation theory to
dynamics in the horizontal plane $(x,\dot{x})$, since the beam is
flat.
Seeking approximated invariants from zeroth to the second order,
we obtain:
%----------------------------------------------------------------
\[
\begin{array}{l}
\K^{(0)}(x,\dot{x}) =
  7.2998    \,\dot{x}^2
- 2.1672    \,\dot{x}\,x
+ 0.2504  x^2,                  \\[0.25cm]
\K^{(1)}(x,\dot{x}) =
\K^{(0)}(x,\dot{x}) 
- 7.2429    \,\dot{x}^3 
+ 5.8544    \,\dot{x}^2 x\,-     \\[0.25cm]
\qquad\qquad\qquad
-\,1.0091   \,\dot{x}\,x^2
+ 0.0339    \,x^3,                 \\[0.25cm]
\K^{(2)}(x,\dot{x}) =
\K^{(1)}(x,\dot{x}) 
- 0.034     \,\dot{x}^3 x 
+ 0.037     \,\dot{x}^2 x^2\,-     \\[0.25cm]
\qquad\qquad\qquad
-\,0.028    \,\dot{x} x^3
+ 1.5\times10^{-5}  \,x^4,
\end{array}
\]
%----------------------------------------------------------------
where $K^{(0)}(x,\dot{x})$ is proportional to the conventional
Courant-Snyder invariant $\mathrm{C.S.}$.
Figure~\ref{fig:Mu2e} provides corresponding level sets alongside
a comparison with tracking (dots in color) for the second order
SX-2.

%----------------------------------------------------------------
%----------------------------------------------------------------

%----------------------------------------------------------------
While the conventional Courant-Snyder formalism has been
successfully utilized by the accelerator community for years, it's
important to acknowledge its limitations as it solely focuses on
linearization, disregarding nonlinear effects and leaving room for
improvement (as evidenced by the disparity between large amplitude
tracking with zeroth order $n=0$ invariant).
Of particular interest are the linearly independent amplitude
linear tune $\nu_0$ and the estimation of dynamical aperture.
Upon comparing tracking results with higher orders, we observe that
key features such as the shape and orientation of stable phase space
trajectories align quite well, mirroring our observations with the
standard form of the map.

%----------------------------------------------------------------
The suggested extension is highly natural, given that the original
Courant-Snyder and approximated invariants both follow polynomial
forms, allowing for seamless integration with virtually any existing
tracking software.
This extension could prove invaluable for rapid (non-tracking)
phase space analysis and facilitating the incorporation of chromatic
effects from sextupoles and tune shifts due to octupoles.
Moreover, if asymmetric/symmetric McMillan mappings serve as the
basis, the newly extended invariant aligns once more with an exactly
integrable system familiar to the accelerator community.

%----------------------------------------------------------------
The results of perturbation theory and derived formulas provide
asymptotic results for higher-dimensional systems in the following
sense.
Assuming a linearly decoupled symplectic map in higher dimensions
with a thin nonlinear kick, the 1D detuning in each plane (e.g.,
for $x$, $\dd\nu_x/\dd J_x$ with $J_y = J_z = \ldots = 0$) should
match the derived expressions;
in the presence of linear coupling, transformation to the
eigenbasis should be applied first.
This is analogous to how well-known formulas for sextupole
detuning remain valid in the presence of linear betatron coupling,
though they must be interpreted within the appropriately
transformed uncoupled planes.

%===============================================================%
%===============================================================%
%===============================================================%
\section{Summary}

\vspace{-0.2cm}
%----------------------------------------------------------------
In this article, we revisited an integrable McMillan map and
demonstrated its central role in the general symplectic dynamics
of the plane.
Through perturbation theory, we showed that McMillan sextupole and
octupole mappings serve as the first and second-order approximations
for planar transformations in standard form~Eq.(\ref{math:McMsym}).
This

\newpage
\noindent
%----------------------------------------------------------------
framework allows for a natural extension of the optical
function formalism employed in accelerator physics, encompassing
chromatic effects and tune shifts induced by sextupole and octupole
magnets, along with dynamical aperture around low-order resonances.
As an illustration,
we applied these concepts to a real accelerator
lattice utilized for the third-order resonant extraction at Mu2e
experiment (Fermilab), revealing a strong correspondence between
tracking results and analytical predictions of perturbation theory.

%----------------------------------------------------------------
This manuscript marks the first in a series of publications
dedicated to the McMillan map and related integrable systems
in higher dimensions, with applications to accelerator physics.
Here, we offer a systematic exposition of fundamental concepts,
providing a comprehensive description of stable trajectories,
including bifurcation diagrams, the parametrization of invariant
curves (see~\cite{IR2002II}), sets of canonical action-angle
variables, global expression for Poincar\'e rotation numbers
(nonlinear betatron tune) and detuning evaluated at the origin.

\vspace{0.2cm}
%===============================================================%
%===============================================================%
%===============================================================%
\section{Acknowledgments}

%----------------------------------------------------------------
The authors would like to thank Eric Stern (FNAL) and Taylor Nchako
(Northwestern University) for carefully reading this manuscript and
for their helpful comments.
Moreover, we would like to extend our gratitude to Vladimir
Nagaslaev (FNAL) for multiple discussions and his generous
contributions in preparation of Fig.~\ref{fig:Mu2e}.

%----------------------------------------------------------------
%----------------------------------------------------------------
This manuscript has been authored by Fermi Research Alliance, LLC
under Contract No. DE-AC02-07CH11359 with the U.S. Department of
Energy, Office of Science, Office of High Energy Physics. This work was also supported by Brookhaven
Science Associates, LLC under Contract No. DESC0012704 with the U.S. Department of Energy. I.M. acknowledges this work was partially supported by  the Ministry of Science and Higher Education of the Russian Federation (project FWUR-2024-0041).

%===============================================================%
%===============================================================%
%===============================================================%
%===============================================================%
\appendix
%===============================================================%
%===============================================================%
%===============================================================%
%===============================================================%

\newpage
%===============================================================%
%===============================================================%
%===============================================================%
%===============================================================%
%===============================================================%

\section{\label{secAp:Symmetries} Symmetries of invariants of motion}

%----------------------------------------------------------------
In this section, we discuss the symmetries of the invariants of
motion with respect to system parameters, beginning with the
McMillan sextupole
\[
\K_\mathrm{sxt}[p,q] =
   p^2 q + p\,q^2 + \Gamma\,(p^2 + q^2) + 2\,\epsilon\,p\,q.
\]
The following propositions are straightforward to verify.
%----------------------------------------------------------------
\begin{proposition}
\label{prop:Sex1}
Simultaneous change of map parameters ($\varepsilon \neq 0$)
\[
    \Gamma \rightarrow \varepsilon\,\Gamma,
    \qquad\qquad\qquad\qquad\quad
    \epsilon \rightarrow \varepsilon\,\epsilon,
\]
along with the scaling transformations
\[
(p,q) \rightarrow \varepsilon\,(p,q)
\qquad\text{and}\qquad
\K \rightarrow \varepsilon^3\,\K
\]
leaves the form of the map and biquadratic invariant.
\end{proposition}
%----------------------------------------------------------------
\begin{proposition}
\label{prop:Sex2}
Simultaneous change of map parameters
\[
    \Gamma \rightarrow \frac{\Gamma-2\,\epsilon}{3},
    \qquad\qquad
    \epsilon \rightarrow-\frac{4\,\Gamma + \epsilon}{3},
\]
along with the translation transformations
\[
(p,q) \rightarrow (p,q) + \frac{2}{3}\,(\Gamma+\epsilon)
\quad\mathrm{and}\quad
\K \rightarrow \K - \left[ \frac{2}{3}\,(\Gamma+\epsilon) \right]^3
\]
leaves the form of the map and biquadratic invariant.
\end{proposition}

%----------------------------------------------------------------
There are two consequences stemming from the first proposition.
Initially, in the parameter space of the map $(\Gamma, \epsilon)$,
all dynamical systems on a ray $\epsilon/\Gamma = \const$ that
starts at the origin are similar up to a scaling transformation.
This is the only intrinsic parameter corresponding to the trace of
the Jacobian evaluated at the fixed point at the origin, denoted
as $\zeta^{(1-1)}=(0,0)$:
\[
\Tr\,\J_\mathrm{sxt}(\zeta^{(1-1)})=-\frac{2\,\epsilon}{\Gamma}=a.
\]
Further, Proposition~\ref{prop:Sex1} tells us that if we know the
dynamics on a ray $\epsilon/\Gamma = \const$, the dynamics for the
systems on the opposite ray, with $\Gamma \rightarrow -\Gamma$ and
$\epsilon \rightarrow -\epsilon$, is given by the rotation of the
phase space by an angle of $\pi$ and inversion of $\K\rightarrow-\K$,
(case $\varepsilon =-1$).

\begin{figure}[t!]
    \centering
    \includegraphics[width=\columnwidth]{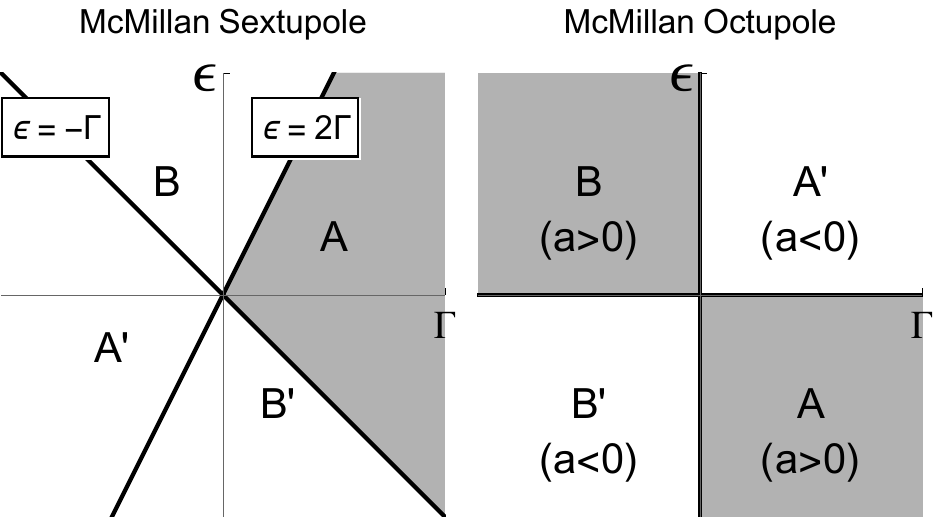}
    \caption{\label{fig:InvSymmetries}
    Left plot illustrates the $(\Gamma,\epsilon)$-plane of parameters
    for the McMillan sextupole map.
    The dynamics in sectors A and A$'$ (or B and B$'$) are interrelated
    through Proposition~\ref{prop:Sex1}, while the dynamics in sectors
    A and B (or A$'$ and B$'$) are connected through
    Proposition~\ref{prop:Sex2}.
    For the McMillan octupole (plot on the right), we observe focusing ($\Gamma > 0$) and defocusing ($\Gamma < 0$) regimes, where the
    dynamics in sectors A and A' or B and B' are linked by
    Proposition~\ref{prop:Oct3}.
    The filled regions denote the nonredundant parameter space.
    }
\end{figure}

%----------------------------------------------------------------
The second proposition arises from two potential choices of the
origin: the translation
\[
    (p,q) \rightarrow (p,q) - \zeta^{(1-2)}
\]
moves the origin to the second fixed point, denoted as
\[
    \zeta^{(1-2)} = -\frac{2}{3}\,(1,1)\,(\Gamma+\epsilon).
\]
The parameter transformation,
\[
\begin{bmatrix}
\Gamma \\[0.2cm] \epsilon
\end{bmatrix} \rightarrow
\begin{bmatrix}
 \frac{1}{3} &-\frac{2}{3} \\[0.2cm]
-\frac{4}{3} &-\frac{1}{3}
\end{bmatrix}
\begin{bmatrix}
\Gamma \\[0.2cm] \epsilon
\end{bmatrix},
\]
is an inversion ($\det = -1$) with two eigenvectors: the eigenvector
along the line $\epsilon = -\Gamma$ with a unit eigenvalue
\[
    v_1 = (-1,1)/\sqrt{2},
    \qquad
    \lambda_1 = 1,
\]
and another eigenvector with a minus unit egenvalue along the line
$\epsilon = 2\,\Gamma$,
\[
    v_2 = (1,2)/\sqrt{5},
    \qquad
    \lambda_2 =-1.
\]
%----------------------------------------------------------------
For all dynamical systems along the line $\epsilon=-\Gamma$, fixed
points $\zeta^{(1-1,2)}$ undergo transcritical bifurcation and
degenerate to a single point.
When $\epsilon = 2\,\Gamma$, the phase space portrait displays
mirror symmetry with respect to the line $p = -q - 1/2$,
with points $\zeta^{(1-1,2)}$ being mirror images of each other.
Without loss of generality, and based on Propositions
\ref{prop:Sex1}--\ref{prop:Sex2}, we will focus our analysis solely
on a specific region of the parameter space (see dashed sector A in
the left plot of the Fig.~\ref{fig:InvSymmetries})
\[
    (\epsilon >-\Gamma)
    \,\cap\,
    (\epsilon <2\,\Gamma)
    \qquad\mathrm{or}\qquad
    a \in [-4;2],\quad \Gamma \geq 0.
\]

%----------------------------------------------------------------
%----------------------------------------------------------------

%----------------------------------------------------------------
Next, for the invariant of the McMillan octupole,
\[
\K_\text{oct}[p,q] = p^2 q^2 + \Gamma\,(p^2 + q^2) + 2\,\epsilon\,p\,q,
\]
we have the following propositions:
%----------------------------------------------------------------
\begin{proposition}
\label{prop:Oct1}
Simultaneous change of map parameters ($\varepsilon > 0$)
\[
    \Gamma \rightarrow \varepsilon\,\Gamma,
    \qquad\qquad\qquad\qquad\quad
    \epsilon \rightarrow \varepsilon\,\epsilon,
\]
along with the scaling transformations
\[
(p,q) \rightarrow \sqrt{\varepsilon}\,(p,q)
\qquad\mathrm{and}\qquad
\K \rightarrow \varepsilon^2\,\K
\]
leaves the form of the map and biquadratic invariant.
\end{proposition}

%----------------------------------------------------------------
\begin{proposition}
\label{prop:Oct2}
Change of the map parameter
\[
    \epsilon \rightarrow -\epsilon,
\]
along with the reflection with respect to $p=0$ (or $q=0$),
\begin{equation}
\label{math:trans}
\begin{array}{l}
q' = q     \\[0.25cm]
p' =-p
\end{array}
\qquad\left(\text{or}\qquad
\begin{array}{l}
q' =-q     \\[0.25cm]
p' = p
\end{array}\right)
\end{equation}
leaves the biquadratic invariant.
\end{proposition}

%----------------------------------------------------------------
As a consequence of Proposition~\ref{prop:Oct1}, similar to the
sextupole case, the system still has one intrinsic parameter
($a=-2\,\epsilon/\Gamma$), and the dynamics on the ray
$\epsilon/\Gamma = \mathrm{const}$ with the initial point at the
origin is identical up to the scaling transformation.
However, since $\varepsilon$ is strictly positive, in contrast
to the previous situation, we have two possible forms of the
invariant corresponding to regimes with focusing and defocusing motion,
for $\Gamma \gtrless 0$.
The next Proposition~\ref{prop:Oct2} is less powerful, as it only
guarantees the conservation of the form of the invariant but not
the map.
In fact, it can be accompanied by another proposition:
%----------------------------------------------------------------
\begin{proposition}
\label{prop:Oct3}
Change of the map parameter
\[
    \epsilon \rightarrow -\epsilon,
\]
influences the dynamics of stable trajectory on a given energy
level $\K(p,q) \rightarrow \K(p,-q)=\K(-p,q)$, as follows:
\begin{itemize}
    \item Symmetric fixed points transform to 2-cycle, and vice versa
          2-cycle transforms to a pair of symmetric fixed points.
    \item Dynamics around fixed point at the origin transforms according
          to
    \[
    \begin{array}{l}
    J \rightarrow J,     \\[0.25cm]
    \nu \rightarrow \frac{1}{2} - \nu,
    \end{array}
    \]
    where $J$ is the action and $\nu$ is the rotation number.
\end{itemize}
\end{proposition}
%----------------------------------------------------------------
$\bigtriangleup$ Proof.
The transformation~(\ref{math:trans}) only alters the orientation
without affecting the area under the closed curve, implying
$J \rightarrow J$.
Due to the symmetries of the invariant, $\K(p,q) = \K(q,p)$
and $\K(p,q) = \K(-q,-p)$, we can represent the action as
$J = 2\,(J_\mathrm{I} + J_\mathrm{II})$, where $J_\mathrm{I}$
denotes the area over $2\pi$ in the I quadrant, and $J_\mathrm{II}$
represents the area in the II quadrant of the $(q,p)$-plane.
Utilizing Danilov's theorem
\cite{zolkin2017rotation,nagaitsev2020betatron}, we observe that
\[
    \nu = \frac{\dd J_\mathrm{II}}{\dd J}
    \qquad\rightarrow\qquad
    \frac{\dd J_\mathrm{I}}{\dd J} =
    \frac{\dd\left( \frac{J}{2}-J_\mathrm{II}\right)}{\dd J} = 
    \frac{1}{2} - \nu
\]
as $J_\mathrm{I}$ interchanges $J_\mathrm{II}$ under the action
of~(\ref{math:trans}). $\bigtriangledown$

%----------------------------------------------------------------
Consequently, we can focus solely on cases where $a > 0$ for both
the focusing ($\Gamma > 0$) and defocusing ($\Gamma < 0$) regimes.
Subsequently, with the aid of the propositions, we can deduce the
dynamics for $a < 0$.

\newpage
%===============================================================%
%===============================================================%
%===============================================================%
\section{\label{secAp:Stability}Stability of fixed points and
         2-cycles. \\
         Bifurcation diagrams.}

%----------------------------------------------------------------
Both fixed points ($n=1$) and $n$-cycles are fundamental concepts
that play significant roles in understanding the behavior of a
dynamical system.
For a mapping $\mathcal{M}$, they are respectively defined by
\[
    \zeta^{(n)}\equiv(q,p):\qquad\qquad
    \mathcal{M}^n (q,p) = (q,p)
\]
and are considered stable if the trace of the Jacobian is smaller
than two by its absolute value
\[
\left|
    \Tr\,\J_{\mathcal{M}^n}\left[ \zeta^{(n)} \right]
\right| < 2.
\]

%----------------------------------------------------------------
Iatrou and Roberts \cite{IR2002II} proved that for an integrable
map with a smooth integral, isolated critical points of the
integral belong to (isolated) cycles of the map, and the points
of isolated cycles of the map are (isolated) critical points of
the integral.
Applying this to a general asymmetric McMillan map, one can
verify that the invariant of motion can have up to 5 isolated
critical points.
Without loss of generality, we assume that one of them
corresponds to a fixed point at the origin.
The four other critical points have to be fixed points as well,
or two of them can form an isolated 2-cycle.
In an exceptional case, these fixed points and 2-cycles can
appear on the same level of the invariant and then form 3- or
4-cycles;
in this scenario, the map degenerates to a linear and, in fact,
all orbits (except the fixed point at the origin) become periodic
with corresponding periods of 3 and 4.

%----------------------------------------------------------------
Furthermore, for the mapping in the form~(\ref{math:McMsym}),
fixed points belong to the main diagonal and are given by the
intersection of two symmetry lines
\[
    p=q
    \qquad\mathrm{and}\qquad
    p = f(q)/2.
\]
2-cycles are given by the intersection of the second symmetry
line with its inverse
\[
    p = f(q)/2
    \qquad\mathrm{and}\qquad
    q = f(p)/2.
\]
If $f(q)$ is an odd function, then fixed points have to appear
in symmetric pairs with respect to the origin, and 2-cycle
belong to the intersection of the anti-diagonal $p=-q$ with
$p = f(q)/2$.

%----------------------------------------------------------------
Below, we summarize the stability of fixed points and $n$-cycles
for McMillan multipoles and provide associated bifurcation diagrams
of the invariant.

\vspace{0.5cm}

%===============================================================%
%===============================================================%
\subsection{McMillan sextupole}

\begin{figure}[t!]
    \centering
    \includegraphics[width=0.75\columnwidth]{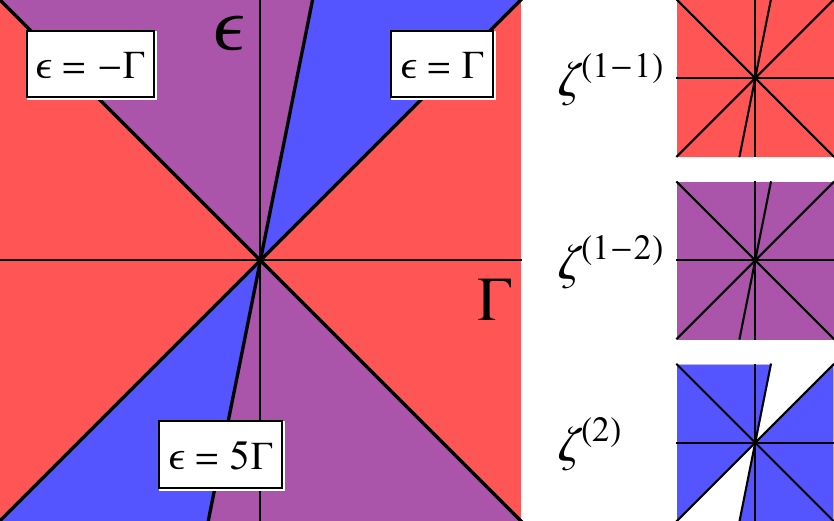}
    \caption{\label{fig:StabSXTsm}
    The left plot illustrates the domain of stability for fixed
    points and the 2-cycle in the $(\Gamma,\epsilon)$-plane of
    parameters for the McMillan sextupole.
    The corresponding real domain $\zeta\in\mathbb{R}$ for each
    case is depicted in the right column.
    If the fixed point/2-cycle is stable or real, then the
    corresponding area is filled with color.
    }
\end{figure}

%----------------------------------------------------------------
The sextupole map exhibits two fixed points defined for any
values of parameters on the $(\Gamma,\epsilon)$-plane: the
point at the origin,
\[
    \zeta^{(1-1)} = (0,0),
\]
and one on the main diagonal,
\[
    \zeta^{(1-2)} = -\frac{2}{3}\,(\Gamma+\epsilon)\,(1,1) =
    \frac{a-a_0}{3}\,(1,1)\,\Gamma.
\]
%----------------------------------------------------------------
Additionally, there exists a 2-cycle defined by
\[
\begin{array}{l}
    \zeta^{(2)} = (\Gamma-\epsilon)\,(1,1) \pm
    \sqrt{(\Gamma-\epsilon)(5\,\Gamma-\epsilon)}\,(1,-1)    \\[0.4cm]
\ds = \frac{1}{2}\,\left[
    (a-a_\frac{1}{2})\,(1,1) \pm
    \sqrt{ (a-a_\frac{1}{2}) (a-a^*_\frac{1}{2})}\,(1,-1)
    \right]\,\Gamma,
\end{array}
\]
which is restricted to a sector where
$(\Gamma-\epsilon)(5\,\Gamma-\epsilon)>0$, as shown in the
bottom plot of the right column in  Fig.~\ref{fig:StabSXTsm}.

%----------------------------------------------------------------
%----------------------------------------------------------------

%----------------------------------------------------------------
Stability in the space of parameters $(\Gamma,\epsilon)$ and
bifurcation diagram for fixed points/2-cycle as a function of
$a$ are presented in the main plot of Fig.~\ref{fig:StabSXTsm}
and left plot of Fig.~\ref{fig:BifAllSmall} respectively.
Analysis reveals that the fixed point at the origin is stable
for $a_{1/2} = -2 < a < a_0 = 2$, where the lower index
for a particular value of $a_\nu$ represents the corresponding
rotation number $\nu = \arccos(a_\nu/2)/(2\pi)$.
When $a=a_0$, $\zeta^{(1-1)}$ undergoes a transcritical
bifurcation [T], exchanging stability with $\zeta^{(1-2)}$.
Both points go through a sub-critical period doubling bifurcation
at $a=a_{1/2}$ [SBPD$_1$] or $a^*_{1/2}=-10$ [SBPD$_2$] for
$\zeta^{(1-1)}$ and $\zeta^{(1-2)}$ respectively.
Finally, for $a=a_{1/3}=-1$, the second fixed point and 2-cycle
approach the same energy level, causing the map to degenerate [D]
to linear with all orbits around the origin being period 3.
The 2-cycle is always unstable when defined on a real domain but
becomes stable if $\zeta^{(2)}\in \mathbb{C}^2$.

%----------------------------------------------------------------
Towards the conclusion of this section, Fig.~\ref{fig:BifDSbig}
offers an additional diagram providing insight into all potential
dynamical regimes.
It showcases phase space plots featuring different level sets of
the invariant $\K_\mathrm{sxt}[p,q]$.
These plots are organized along rays (or between them)
representing constant values of $a=-2\,\epsilon/\Gamma$, indicative
of the system's degeneracy or bifurcations.
The specific values of $a$ are highlighted on a scale delineated
with a dash-dotted circle.

\newpage
%===============================================================%
%===============================================================%
\subsection{McMillan octupole}

\begin{figure}[t!]
    \centering
    \includegraphics[width=0.75\columnwidth]{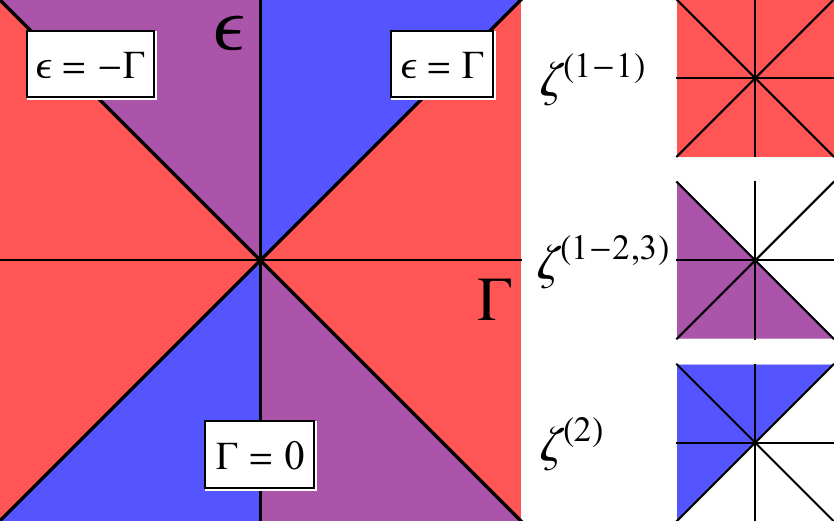}
    \caption{\label{fig:StabOCTsm}
    Same as Fig.~\ref{fig:StabSXTsm} but for McMillan octupole.
    }
\end{figure}

%----------------------------------------------------------------
Similar to the sextupole scenario, the octupole map also possesses
a fixed point at the origin:
\[
    \zeta^{(1-1)} = (0,0),
\]
which is defined for all $(\Gamma,\epsilon)\in \mathbb{R}^2$
and remains stable for $|\epsilon|<|\Gamma|$ (or $|a|<2$).
Additional symmetric fixed points and a 2-cycle are provided by:
\begin{equation}
\label{math:zeta1-23}
    \zeta^{(1-2,3)} = \pm\sqrt{-(\Gamma+\epsilon)}\,(1,1)
    = \pm\sqrt{\pm\frac{a-a_{0}}{2}}\,(1,1)\sqrt{|\Gamma|},
\end{equation}
\begin{equation}
\label{math:zeta2}
    \zeta^{(2)} =\pm\sqrt{-(\Gamma-\epsilon)}\,(1,-1)
    = \pm\sqrt{\mp\frac{a-a_\frac{1}{2}}{2}}\,(1,-1)\sqrt{|\Gamma|},
\end{equation}
where the $\pm$-sign under the square roots varies based on whether
$\Gamma$ is greater or less than 0;
note that, in contrast to the sextupole map, the ``natural'' unit
of distance shifts to $\sqrt{|\Gamma|}$ rather than $\Gamma$.
These points are real only for $\epsilon<-\Gamma$
(or $a \gtrless a_0$ for $\Gamma \gtrless 0$) and $\epsilon>\Gamma$
(or $a \lessgtr a_{1/2}$ for $\Gamma \gtrless 0$), as visualized
in the right column of Fig.~\ref{fig:StabOCTsm}.

%----------------------------------------------------------------
%----------------------------------------------------------------

\begin{figure*}[p!]
    \centering
    \includegraphics[width=\linewidth]{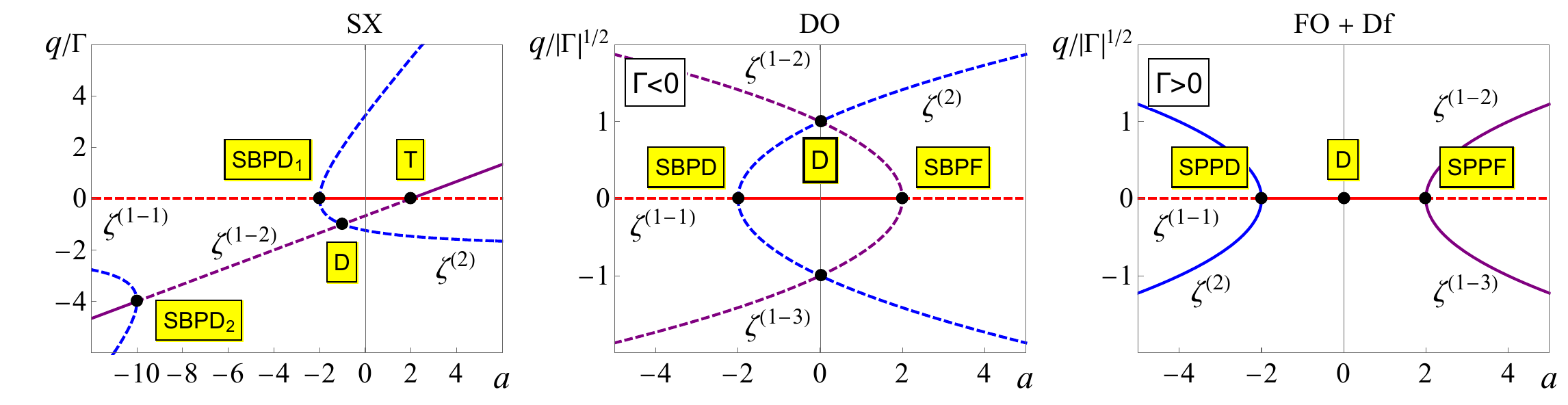}
    \caption{\label{fig:BifAllSmall}
    Bifurcation diagrams illustrating fixed points and 2-cycles of the
    McMillan sextupole (SX), octupole in defocusing (DO), focusing (FO),
    and Duffing (Df, $|a|>2$) regimes.
    Plots depict normalized coordinates as a function of the map
    parameter $a=-2\,\epsilon/\Gamma$.
    Stable fixed points/2-cycles are represented by solid lines, while
    unstable ones are dashed.
    Transcritical, pitchfork, and period doubling bifurcations are
    denoted by [T], [PF], and [PD] respectively;
    SB indicates sub-critical, and SP represents super-critical.
    Label [D] corresponds to system degeneracy.\vspace{-0.5cm}
}
\end{figure*}

\begin{figure*}[p!]
    \centering
    \includegraphics[width=0.6\linewidth]{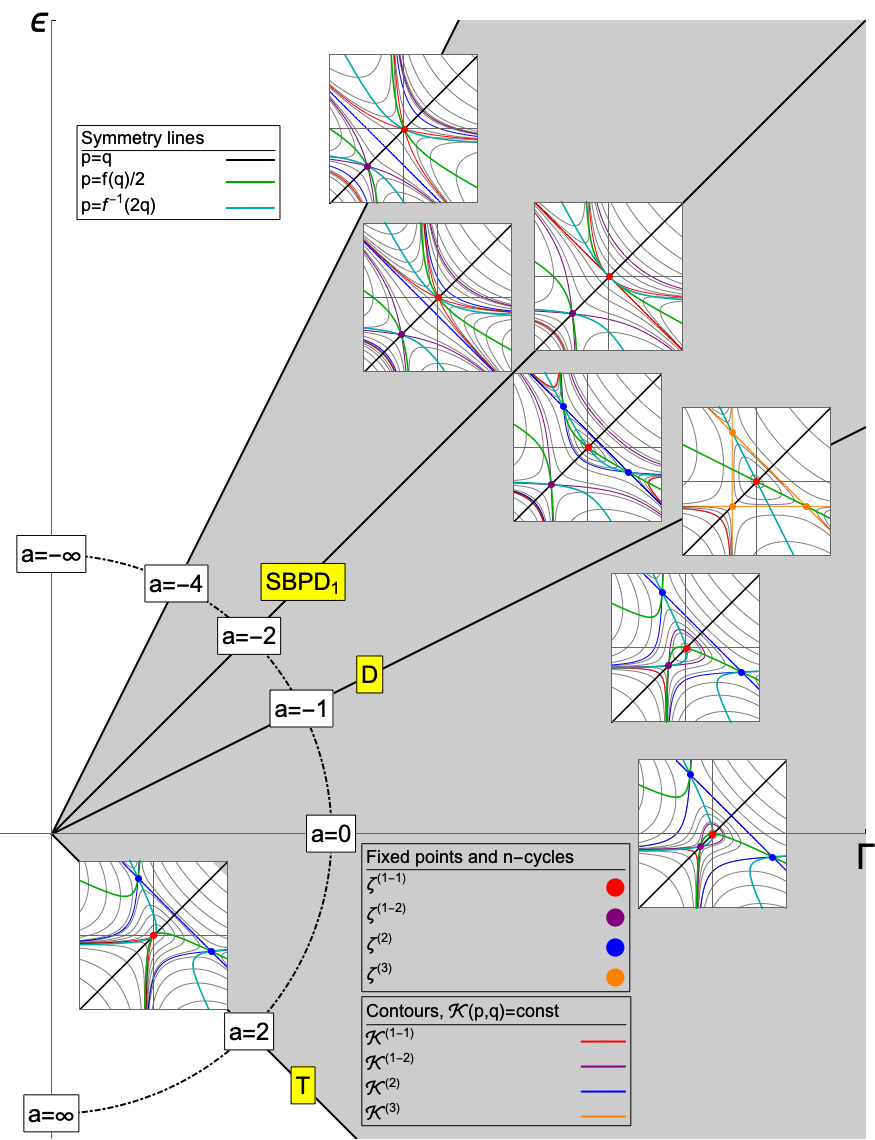}
    \caption{\label{fig:BifDSbig}
    Schematic phase space diagrams for the McMillan sextupole map
    arranged in the plane of map parameters $(\Gamma,\epsilon)$.
    Isolated fixed points, 2-cycles and degenerate 3-cycles, along
    with their corresponding level sets, are color-coded according
    to the legend;
    other level sets are represented in black. 
    The first ($p=q$), the second ($p=f(q)/2$) and the inverse
    ($q=f(p)/2$) symmetry lines are displayed in thick black,
    green, and cyan respectively.
    }
\end{figure*}

\begin{figure*}[p!]
    \centering
    \includegraphics[width=0.8\linewidth]{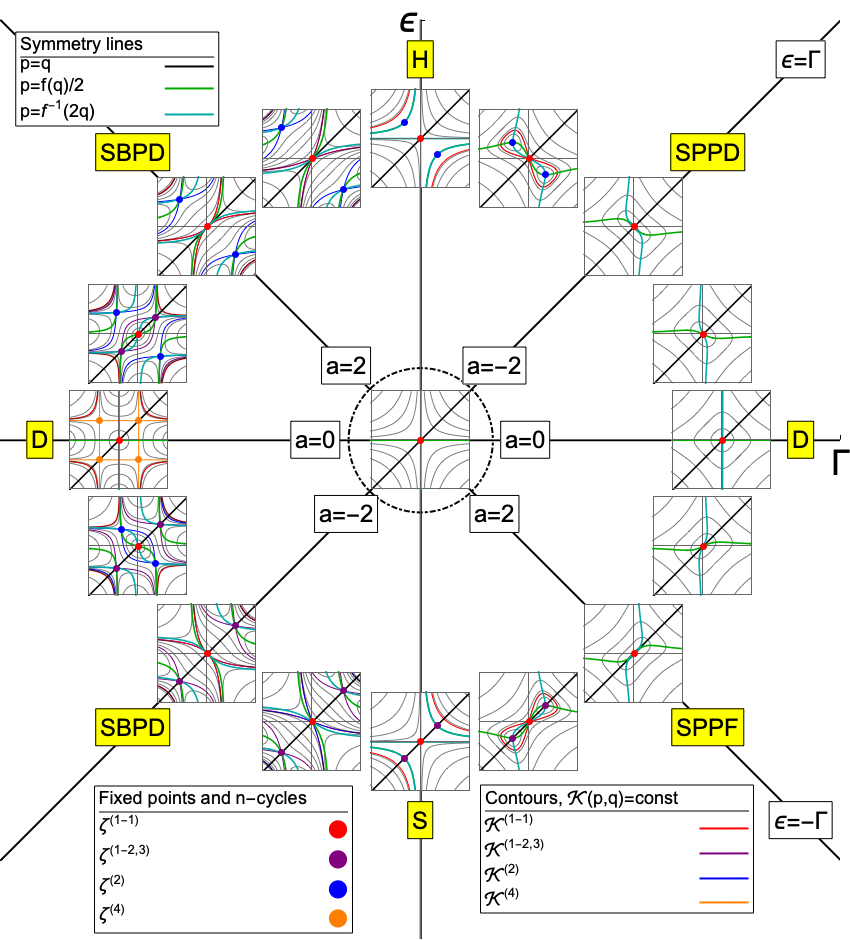}
    \caption{\label{fig:BifDObig}
    Schematic phase space diagrams for the McMillan octupole map
    arranged in the plane of map parameters $(\Gamma,\epsilon)$.
    Isolated fixed points, 2-cycles and degenerate 4-cycles, along
    with their corresponding level sets, are color-coded according
    to the legend;
    other level sets are represented in black. 
    The first ($p=q$), the second ($p=f(q)/2$) and the inverse
    ($q=f(p)/2$) symmetry lines are displayed in thick black,
    green, and cyan respectively.
    }
\end{figure*}

%----------------------------------------------------------------
The primary plot in Fig.~\ref{fig:StabOCTsm} alongside the
subsequent plots in Fig.~\ref{fig:BifAllSmall} illustrates the
stability and bifurcation diagrams across all conceivable
scenarios.
Now, let's break down the results:
(I) When $\Gamma < 0$, fixed points $\zeta^{(1-2,3)}$ and the
2-cycle are always unstable, and the only region with stable
trajectories is $a_{1/2}<a<a_0$.
At $a_{1/2,0}$, we observe that the fixed point at the origin
undergoes sub-critical period doubling [PD] or pitchfork [PF]
bifurcations, respectively.
(II) For $\Gamma > 0$, at $a_{1/2,0}$ the fixed point
$\zeta^{(1-1)}$ goes through super-critical [PD] or [PF]
bifurcations.
When it is stable ($|a|<2$), it is the only isolated cycle with
all trajectories being finite.
(III) In the Duffing regime, the point at the origin is locally
unstable, becoming a center of lemniscate separatrix.
If $a>a_0$, trajectories inside the figure-eight curve round
the centers of nearby ``eyes,'' while when $a<a_{1/2}$, the
orbit jumps from ``eye'' to ``eye'' since their centers form
a 2-cycle.
(IV) If $a = 0$, it degenerates into a linear system with a map
period of 4.
Fig.~\ref{fig:BifDObig} consolidates the aforementioned results,
depicting phase space diagrams on the plane of map parameters 
$(\Gamma,\epsilon)$, similar to Fig.~\ref{fig:BifDSbig}.

%----------------------------------------------------------------
%----------------------------------------------------------------

%----------------------------------------------------------------
%----------------------------------------------------------------

%----------------------------------------------------------------
%----------------------------------------------------------------

%===============================================================%
%===============================================================%
%===============================================================%
\section{\label{secAp:Action}Action integrals and power series}

%----------------------------------------------------------------
To derive action integrals for the McMillan octupole, we observe
that the normalized phase space trajectories (\ref{math:OctParam})
correspond to {\it elliptic Lissajous figures}
\[
\begin{array}{l}
    x = \mathrm{ef}\,(\omega_1\,t,\kappa),         \\[0.2cm]
    y = \mathrm{ef}\,(\omega_2\,t+\eta,\kappa).
\end{array}
\]
where the Jacobi elliptic functions $\mathrm{ef}$ are employed
instead of regular trigonometric functions for conventionally
defined Lissajous curves and a unit ratio of $\omega_1:\omega_2$.
%----------------------------------------------------------------
The area enclosed by such a curve can be determined using the
integral $S=\oint y\,\dd x$, which for ``major'' elliptic

\begin{figure}[t!]
    \centering
    \includegraphics[width=0.8\linewidth]{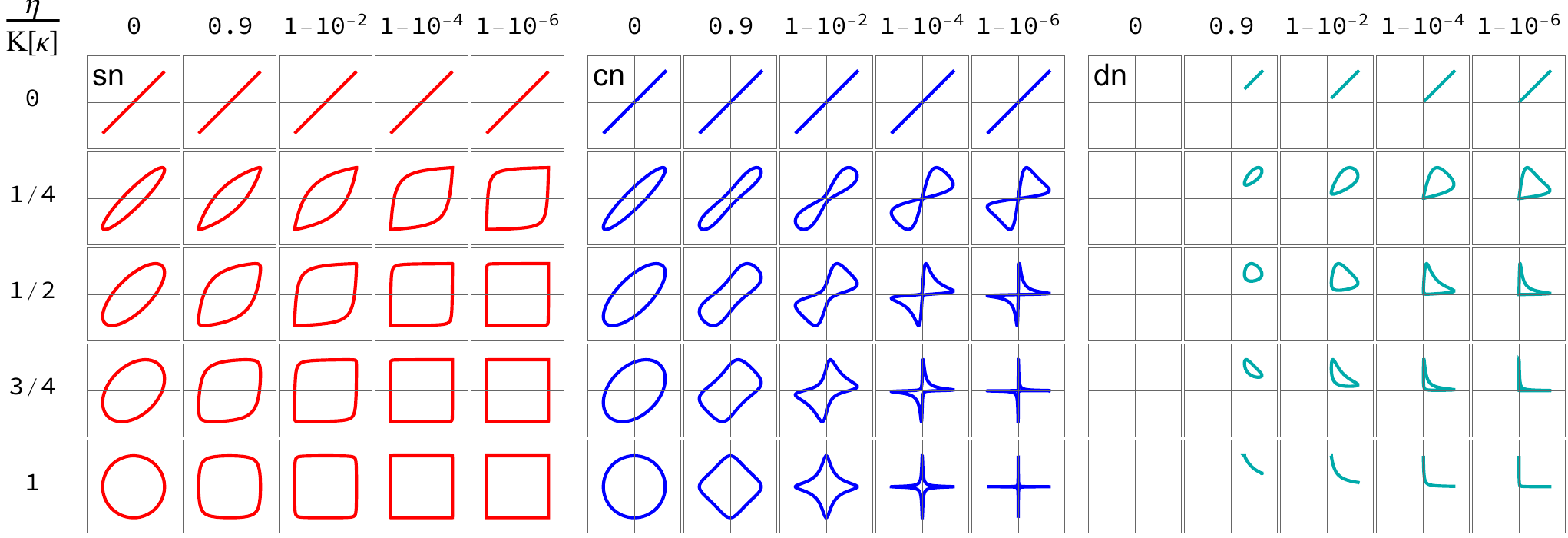}
    \includegraphics[width=0.8\linewidth]{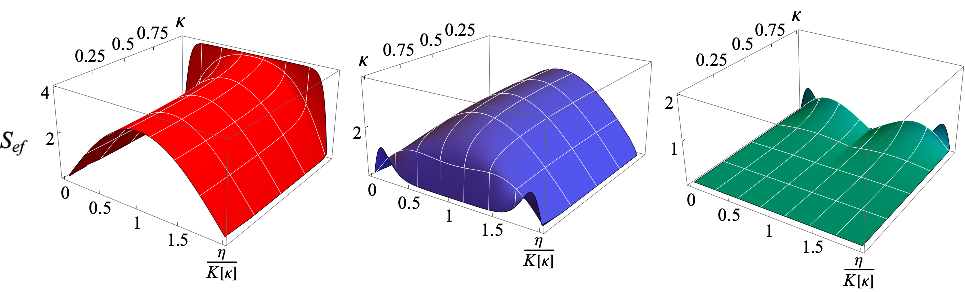}
    \caption{\label{fig:LssjArea}
    Sample elliptic Lissajous curves with a unit ratio of
    $\omega_1:\omega_2 = 1$ (top plots) and the corresponding
    area $S_\text{ef}$ enclosed under it (plots in the bottom)
    for major elliptic functions as a function of modulus
    $\kappa$ and phase difference $\eta$.
    Note that $\eta$ is measured in units of a quarter period
    $\mathrm{K}[\kappa]$.
    }
\end{figure}

\newpage
\noindent
%----------------------------------------------------------------
functions yields~\cite{zolkin2022mcmillan}:

\vspace{-0.2cm}
\[
\begin{array}{l}
\ds S_{\text{sn}} =
\frac{4}{\kappa^2}
    \frac{
    \sd^2 \eta\,\,\mathrm{E}[\kappa] - \mathrm{K}[\kappa] +
    \cn^2 \eta\,\,              \Pi[k^2\sn^2 \eta,\kappa]}
    {\sn\,\eta\,\,\sd^2 \eta},
\\[0.3cm]
\ds S_{\text{cn}} =
\frac{4}{\kappa^2}
    \frac{
    \mathrm{K}[\kappa] - \sn^2 \eta\,\,\mathrm{E}[\kappa] -
    \cn^2 \eta\,\,              \Pi[k^2\sn^2 \eta,\kappa]}
    {\sd\,\eta\,\,\sn^2 \eta},
\\[0.25cm]
\ds S_{\text{dn}} =
\frac{2}{\mathrm{sc}\,\eta\,\,\sn^2 \eta}\,\times\\[0.35cm]
\left\{
	  (\sn^2 \eta+\dn^2 \eta)\,\mathrm{K}[\kappa] -
	\sn^2 \eta\,\,\mathrm{E}[\kappa] -
	\dn^2 \eta\,           \Pi[k^2\sn^2 \eta,\kappa]
\right\}.
\end{array}
\]
%----------------------------------------------------------------
Figure~\ref{fig:LssjArea} illustrates the behavior of the area as
a function of $\eta$ and $\kappa$ along with several examples of
Lissajous-Jacobi orbits.
Utilizing the scaling outlined in~(\ref{math:JefSC}):

\onecolumngrid

\vspace{0.5cm}
%----------------------------------------------------------------
$\bullet$ For $\sn$ and $\cn$ cases, the action integrals and
their power series expressed in terms of elliptic modulus $\kk$
are:
%----------------------------------------------------------------
\begin{equation}
\label{math:JOCMcM1}
\begin{array}{rl}
J_\text{sn}[\kk] = &\ds\! \frac{2}{\pi}\,\frac{
    \sd^2\eta_+\,\mathrm{E}[\kk] - \mathrm{K}[\kk] +
    \cn^2\eta_+\,\Pi[k^2\,\sn^2 \eta_+,\kk]
}{\sd\,\eta_+\,\,\nd\,\eta_+}
\\[0.35cm]
= &\ds\! \frac{(4-a^2)^{\textstyle\frac{3}{2}}}{2^4}
\left[
    \kk^2 +
    \frac{2-7\,a^2}{{2^4}}\,\kk^4 +
    \frac{6-68\,a^2+33\,a^4}{2^7}\,\kk^6 +
    \frac{100-2398\,a^2+2838\,a^4-715\,a^6}{2^{12}}\,\kk^8 +
    \mathcal{O}(\kk^{10})
\right],
\\[0.4cm]
J_\text{cn}[\kk] = &\ds\! \frac{2}{\pi}\,\frac{
    \mathrm{K}[\kk] - \sn^2\eta_+\,\mathrm{E}[\kk] -
    \cn^2\eta_+\,\Pi[k^2\,\sn^2 \eta_+,\kk]
}{\sn\,\eta_+\,\,\dn\,\eta_+}
\\[0.35cm]
= &\ds\! \frac{(4-a^2)^{\textstyle\frac{3}{2}}}{2^4}
\left[
    \kk^2 +
    7\,\frac{2+a^2}{2^4}\,\kk^4 +
    \frac{102+44\,a^2+33\,a^4}{2^7}\,\kk^6 +
    \frac{3036+1246\,a^2+330\,a^4+715\,a^6}{2^{12}}\,\kk^8 +
    \mathcal{O}(\kk^{10})
\right],
\end{array}
\end{equation}
%----------------------------------------------------------------
or as a series of the invariant
\[
J_\text{sn,cn}[\K] = \frac{1}{\sqrt{4-a^2}}\,
\left[
    \K \pm
    \frac{2+a^2}{(4-a^2)^2}\,\K^2 +
    2\,\frac{6+12\,a^2+a^4}{(4-a^2)^4}\,\K^3 \pm
    5\,\frac{20+90\,a^2+30\,a^4+a^6}{(4-a^2)^6}\,\K^4 +
    \mathcal{O}(\K^5)
\right].
\]
%----------------------------------------------------------------
Furthermore, by expanding the rotation number in the same series
up to $\mathcal{O}(\kk^{10})$ or $\mathcal{O}(\K^{5})$
%----------------------------------------------------------------
\[
\begin{array}{rl}
\nu_\text{sn} \approx &\ds\! \nu_0 - \frac{a\,\sqrt{4-a^2}}{2^4\,2\,\pi}
\left[
    3\,\kk^2 +
    \frac{86-35\,a^2}{2^5}\,\kk^4 +
    7\,\frac{46-48\,a^2+11\,a^4}{2^7}\,\kk^6 +
    \frac{39340-74926\,a^2+40722\,a^4-6435\,a^6}{2^{14}}\,\kk^8
\right],
\\[0.35cm]
\nu_\text{cn} \approx &\ds\!  \nu_0 + \frac{a\,\sqrt{4-a^2}}{2^4\,2\,\pi}
\left[
    3\,\kk^2 +
    \frac{10+35\,a^2}{2^5}\,\kk^4 +
    \frac{18-56\,a^2+77\,a^4}{2^7}\,\kk^6 +
    \frac{1364-13\,(26+858\,a^2-495\,a^4)\,a^2}{2^{14}}\,\kk^8
\right],
\end{array}
\]

\vspace{-0.5cm}
\[
\begin{array}{ll}
\ds\text{or }\quad \nu_\text{sn,cn}[\K] = &\ds\! \nu_0 +
    \frac{a}{2\,\pi\,(4-a^2)^{3/2}}\,
\left[
    \mp\,3\,\K -
    \frac{1}{2}\,\frac{86+13\,a^2}{(4-a^2)^2}\,\K^2 \mp
    2\,\frac{322+200\,a^2+9\,a^4}{(4-a^2)^4}\,\K^3\, - \right.  \\[0.55cm]
&\ds \qquad\qquad\qquad\qquad\qquad\qquad\,\,\,\, - \left.
    \frac{39340+55506\,a^2+10386\,a^4+221\,a^6}{(4-a^2)^6}\,\K^4 +
    \mathcal{O}(\K^5)
\right],\qquad\qquad\qquad\qquad
\end{array}
\]
we derive series $\nu(J)$:
\[
\ds \nu_\text{sn,cn}[J] = \nu_0 +
\frac{a}{2\,\pi}\,\left[
    \mp
    \frac{3}{4-a^2}\,J -
    \frac{74+7\,a^2}{2\,(4-a^2)^{5/2}}\,J^2 \mp
    5\,\frac{(2+a^2)(46+a^2)}{(4-a^2)^4}\,J^3 +
    \mathcal{O}(J^4)
\right].
\]

%----------------------------------------------------------------
$\bullet$ For $\dn$ trajectories that orbit symmetric fixed points
in the Duffing regime, case (Df), we have:
\begin{equation}
\label{math:JOCMcM2}
\begin{array}{rl}
J_\text{dn}[\kk] = &\ds\! \frac{1}{\pi}\,\frac{
    (\sn^2\eta_++\dn^2\eta_+)\,\mathrm{K}[\kk] -
    \sn^2\eta_+\,\,\mathrm{E}[\kk] - 
    \dn^2\eta_+\,\,\Pi[k^2\,\sn^2 \eta_+,\kk]
}{\sn\,\eta_+\,\,\cn\,\eta_+}
\\[0.35cm]
= &\ds\! \frac{(a-2)^{3/2}}{2^4\sqrt{2}\,a}
\left[
    \kk^4 + \kk^6 +
    \left(\frac{119}{8} - \frac{4+3\,a}{a^2}\right)\,\frac{\kk^8   }{2^4} +
    \left(\frac{55 }{8} - \frac{4+3\,a}{a^2}\right)\,\frac{\kk^{10}}{2^3} +
    \mathcal{O}(\kk^{12})
\right],
\\[0.45cm]
J_\text{dn}[\widetilde{\K}] = &\ds\! \frac{2^{-3/2}}{\sqrt{a-2}}\,
\left[
    \widetilde{\K} +
    \frac{8-a}{2^4\,(a-2)^2}\,\widetilde{\K}^2 +
    \frac{192-32\,a+3\,a^2}{2^8\,(a-2)^4}\,\widetilde{\K}^3 +
    \frac{12800-5\,(480-72\,a+5\,a^2)\,a}{2^{13}\,(a-2)^6}\,\widetilde{\K}^4 +
    \mathcal{O}(\widetilde{\K}^5)
\right].
\end{array}
\end{equation}
and
%----------------------------------------------------------------
\[
\begin{array}{rl}
\nu_\text{dn} \approx &\ds\! \nu_0 - 
\frac{1}{2\,\pi}\frac{\sqrt{a-2}}{2^4\sqrt{2}\,a^2}
\left[
    (4+a)\,(\kk^4 + \kk^6) -
    \frac{384 + 96\,a - 964\,a^2 - 235\,a^3}{2^8\,a^2}\,\kk^8 -
    \frac{384 + 96\,a - 452\,a^2 - 107\,a^3}{2^7\,a^2}\,\kk^{10}
\right]
\\[0.35cm]
\approx &\ds\!  \nu_0 -
\frac{1}{2\,\pi}\frac{2^{-3/2}}{a\,(a-2)^{3/2}}
\left[
    (4+a)\,\widetilde{\K} -
    \frac{128-224\,a-68\,a^2+5\,a^3}{2^5\,a\,(a-2)^2}\,\widetilde{\K}^2\,+
\right.
\\[0.35cm]
&\ds\!\left. \qquad\qquad\qquad\qquad\qquad\qquad\quad\,\,\,\, +\,
    \frac{4096-11264\,a+12672\,a^2+4384\,a^3-476\,a^4 + 33\,a^5}
    {3 \times 2^8\,a^2\,(a-2)^4}\,\widetilde{\K}^3
\right]\\[0.35cm]
\approx &\ds\! \nu_0 - 
\frac{1}{2\,\pi}
\left[
    \frac{1}{a}\,\frac{4+a}{a-2}\,J -
    \frac{1}{a^2}\frac{128-160\,a-60\,a^2+3\,a^3}
    {2^{7/2}(a-2)^{5/2}}\,J^2
\right],
\qquad\quad\, \text{where }
\widetilde{\K} = \K + \left( \frac{a}{2}-1\right)^2.
\end{array}
\]

%----------------------------------------------------------------
%----------------------------------------------------------------

%----------------------------------------------------------------
$\bullet$ While the roots $q_{1,2,3,4}$ of the characteristic
polynomial $\mathcal{P}(q)$ for the sextupole map represent
turning points of the coordinate, they are not particularly
useful for obtaining expansions of $J$ or $\nu$.
(Indeed, this can be achieved using modern software.)
Therefore, we solely present an exact expression for the action
integral here:
\begin{equation}
\label{math:JSXMcM1}
\begin{array}{l}
\ds J_\mathrm{sxt} = \frac{1}{2\pi} \,\int_{q_2}^{q_3}
        \frac{\sqrt{\mathcal{P}(q)}}{q+1}\,\dd q =
        \frac{q_2-q_1}{\pi}\,\sqrt{\frac{q_3-q_1}{q_4-q_2}} \times 
\ds \left\{
        \frac{a+4+q_4-q_1}{4}\,\mathrm{K}[\kappa] +
        \frac{a+4}{4}\frac{q_4-q_2}{q_2-q_1}\,\mathrm{E}[\kappa]\,-
\right. \\[0.45cm]
\ds \qquad\qquad\qquad\qquad\qquad\qquad\qquad\left. -\,
        \frac{a+1}{q_3-q_1}
            \,\Pi\left[\frac{q_4-q_2}{q_4-q_1}\,\kappa^2,\kappa\right] -
        \frac{(1+q_3)(1+q_4)}{q_3-q_1}
            \,\Pi\left[\frac{1+q_1}{1+q_2}\frac{q_4-q_2}{q_4-q_1}\,\kappa^2,\kappa\right]
    \right\}.
\end{array}
\end{equation}

\newpage
%----------------------------------------------------------------
%----------------------------------------------------------------
\begin{figure*}[t!]
    \centering
    \includegraphics[width=0.8\linewidth]{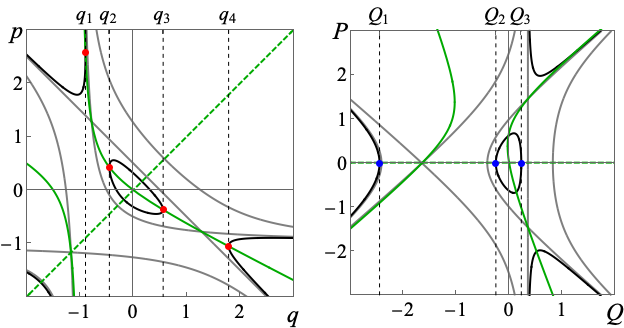}
    \caption{\label{fig:Rotation}
    Phase space portraits for the McMillan sextupole map with
    $a=-3/2$, illustrating level sets of the invariant in the
    original $(p,q)$ and new rotated coordinates $(P,Q)$.
    Turning points are identified by the intersection of the
    level set (depicted in black) with either the second
    (green curve) or first (dashed green line) symmetry lines,
    indicated by the red and blue dots respectively.
    }
\end{figure*}
%----------------------------------------------------------------

A more practical approach involves performing a rotation of the
phase space by an angle $\alpha = \pi/4$ clockwise:
\[
q\rightarrow Q\,\cos\alpha - P\,\sin\alpha,\qquad\qquad\qquad
p\rightarrow Q\,\sin\alpha + P\,\cos\alpha,
\]
which allows us to ``decouple'' one of the roots, as illustrated
in Fig.~\ref{fig:Rotation}.
After implementing this change of variables, the Hamiltonian
becomes:
\[
H[P,Q;t] \quad\rightarrow\quad
K[P,Q;t] = \frac{P^2\,(a + 2 - \sqrt{2}\,Q) -
                 Q^2\,(a - 2 - \sqrt{2}\,P)}{2}.
\]
Following the same logic as in Section~\ref{sec:McSX}, we again
solve for momentum
\[
P = \sqrt{\frac{\mathcal{G}(Q)}{Q_4 - Q}},
\]
which is now expressed through a third-order polynomial
$\mathcal{G}$ and a known root $Q_4$ that is independent of $K$:
\[
\mathcal{G}(Q) =
    \sqrt{2}\,\mathcal{K} - \frac{2-a}{\sqrt{2}}\,Q^2 - Q^3,
\qquad\qquad\qquad\qquad\qquad\qquad
Q_4 = \frac{2+a}{\sqrt{2}}.
\]
%----------------------------------------------------------------
For stable trajectories, we have $Q_1 < Q_2 < Q < Q_3 < Q_4$, such
that:
\[
\begin{array}{l}
\ds Q_1 = -\frac{2-a}{3\sqrt{2}}\left(
    1 + 2\,\cos\left[\frac{1}{3}\,\arccos
        \left[1-\frac{54\,K}{(2-a)^3}\right]
    \right]
\right),             \\[0.45cm]
\ds Q_2 = -\frac{2-a}{3\sqrt{2}}\left(
    1 - 2\,\sin\left[\frac{1}{3}\,\arcsin
        \left[1-\frac{54\,K}{(2-a)^3}\right]
    \right]
\right),             \\[0.45cm]
\ds Q_3 = -\frac{2-a}{3\sqrt{2}}\left(
    1 - 2\,\cos\left[\frac{1}{3}\,\arccos
        \left[\frac{54\,K}{(2-a)^3}-1\right]
    \right]
\right),
\end{array}
\]
allowing expansion in terms of $K$.
%----------------------------------------------------------------
Starting with the action integral:
\begin{equation}
\label{math:JSXMcM2}
\begin{array}{ll}
\ds J_\mathrm{sxt} &\ds\!\! = \frac{1}{2\,\pi}\,\oint P\,\dd Q =
\frac{1}{\pi}\,\int_{Q_2}^{Q_3} P\,\dd Q =
\frac{2}{\pi}\frac{1}{\sqrt{(Q_4-Q_2)(Q_3-Q_1)}}\,
\left\{
    (1+a)(Q_4-Q_3)\,\Pi\left[\frac{Q_3-Q_2}{Q_4-Q_2},\kk\right]
\right. +            \\[0.5cm]
    &\ds\!\! \qquad\qquad + \left.
\frac{(2+a)^2-K}{4}\,\left[\left(
    \frac{4+a}{Q_4-Q_1} - 3\sqrt{2}
\right)\,\mathrm{K}[\kk] +
\frac{(4+a)(Q_3-Q_1)}{(Q_4-Q_1)(Q_4-Q_3)}\,\mathrm{E}[\kk]
\right]
\right\}            \\[0.5cm]
    &\ds\!\! = \frac{(2-a)(4-a^2)^{3/2}}{2^8}\,\left[
\kk^4 + \kk^6 + \frac{97-27\,a-15\,a-a^3}{2^7}\,\kk^8 +
\frac{33-27\,a-15\,a-a^3}{2^6}\,\kk^{10} + \mathcal{O}(\kk^{12})
\right]            \\[0.5cm]
    &\ds\!\! = \frac{1}{\sqrt{4-a^2}}\,\left[
    K - 3\,\frac{2+2\,a+a^2}{(a-2)(a^2-4)^2}\,K^2
    + 10\,\frac{22+48\,a+42\,a^2+16\,a^3+3\,a^4}
        {(a-2)^2(a^2-4)^4}\,K^3
    + \mathcal{O}(K^4)
\right],
\end{array}
\end{equation}
%----------------------------------------------------------------
where the new elliptic modulus is given by:
\[
\kappa = \sqrt{\frac{(Q_3-Q_2)(Q_4-Q_1)}{(Q_3-Q_1)(Q_4-Q_2)}}.
\]
Next, substitution of $P$ into the equation of motion
\[
\dd t = \left(\frac{\pd K}{\pd P}\right)^{-1}\,\dd q=
\dd q/\sqrt{2\,(Q_4-Q)(Q_3-Q)(Q-Q_2)(Q-Q_1)}
\]
provides the period of motion and time of one step of the map:
\[
\mathrm{T} = \oint \dd t =
    \frac{2\,\sqrt{2}\,\mathrm{K}[\kk]}
                {\sqrt{(Q_4-Q_2)(Q_3-Q_1)}},
\qquad\qquad
\mathrm{T}' = \int_{Q_0}^{Q_0'} \dd t =
    \frac{\sqrt{2}\,\mathrm{F}[\arcsin\phi,\kk]}
                {\sqrt{(Q_4-Q_2)(Q_3-Q_1)}}.
\]
%----------------------------------------------------------------
Using new turning points as the initial point for the integral,
$Q_0 = Q_{2,3}$:
\[
\left\{Q_0,P_0\right\} = \left\{\sqrt{2}\,q_0,0\right\}
\qquad\rightarrow\qquad
\left\{Q_0',P_0'\right\} = 
\left\{\frac{a-q_0}{1+q_0}\frac{q_0}{\sqrt{2}},
       \frac{a-2-3\,q_0}{1+q_0}\frac{q_0}{\sqrt{2}}\right\}
\]
defines
\[
\phi = \sqrt{
    Q_2\,\frac{Q_3-Q_1}{Q_3-Q_2}\frac{2\,\sqrt{2}-\sqrt{2}\,a+3\,Q_2}
    {2\,Q_1\,(\sqrt{2}+Q_2)-Q_2\,(\sqrt{2}\,a-Q_2)}
} = \sqrt{
    Q_3\,\frac{(2\,\sqrt{2}+\sqrt{2}\,a-2\,Q_2)
               (2\,\sqrt{2}-\sqrt{2}\,a+3\,Q_3)}
    {2\,(Q_3-Q_2)(4+2\,a+2\,\sqrt{2}\,Q_3+Q_3^2)}
},
\]
and finally, the rotation number:
%----------------------------------------------------------------
\[
\begin{array}{rl}
\nu_\mathrm{sxt} = &\ds\!\!\frac{\mathrm{T}'}{\mathrm{T}} =
    \frac{\mathrm{F}[\arcsin\phi,\kk]}{2\,\mathrm{K}[\kk]}
\\ [0.5cm]
=&\ds\!\!
\nu_0 - \frac{1+a}{2\,\pi}\,\frac{\sqrt{4-a^2}}{2^8}\left[
 (8+a)\,(\kk^4+\kk^6-\kk^{10}) +
 5\,\frac{768+30\,a-78\,a^2-17\,a^3-a^4}{2^9}\,(\kk^8+2\,\kk^{10}) +
\mathcal{O}(\kk^{12})
\right]
\\ [0.5cm]
=&\ds\!\!
\nu_0 - \frac{1}{2\,\pi}\frac{1+a}{(2-a)(4-a^2)^{3/2}}\left[
 (8+a)\,K +
 \frac{832+734\,a+258\,a^2+13\,a^3-a^4}{2\,(2-a)(4-a^2)^2}\,K^2 +
\mathcal{O}(K^3)
\right]
\\ [0.5cm]
=&\ds\!\!
\nu_0 - \frac{1}{2\,\pi}\frac{1+a}{(2-a)(4-a^2)}\left[
 (8+a)\,J +
 \frac{736+626\,a+198\,a^2+7\,a^3-a^4}{2\,(2-a)(4-a^2)^{3/2}}\,J^2 +
\mathcal{O}(J^3)
\right].
\end{array}
\]
%----------------------------------------------------------------

\newpage
%\bibliography{papers}

\begin{thebibliography}{52}%
\makeatletter
\providecommand \@ifxundefined [1]{%
 \@ifx{#1\undefined}
}%
\providecommand \@ifnum [1]{%
 \ifnum #1\expandafter \@firstoftwo
 \else \expandafter \@secondoftwo
 \fi
}%
\providecommand \@ifx [1]{%
 \ifx #1\expandafter \@firstoftwo
 \else \expandafter \@secondoftwo
 \fi
}%
\providecommand \natexlab [1]{#1}%
\providecommand \enquote  [1]{``#1''}%
\providecommand \bibnamefont  [1]{#1}%
\providecommand \bibfnamefont [1]{#1}%
\providecommand \citenamefont [1]{#1}%
\providecommand \href@noop [0]{\@secondoftwo}%
\providecommand \href [0]{\begingroup \@sanitize@url \@href}%
\providecommand \@href[1]{\@@startlink{#1}\@@href}%
\providecommand \@@href[1]{\endgroup#1\@@endlink}%
\providecommand \@sanitize@url [0]{\catcode `\\12\catcode `\$12\catcode
  `\&12\catcode `\#12\catcode `\^12\catcode `\_12\catcode `\%12\relax}%
\providecommand \@@startlink[1]{}%
\providecommand \@@endlink[0]{}%
\providecommand \url  [0]{\begingroup\@sanitize@url \@url }%
\providecommand \@url [1]{\endgroup\@href {#1}{\urlprefix }}%
\providecommand \urlprefix  [0]{URL }%
\providecommand \Eprint [0]{\href }%
\providecommand \doibase [0]{http://dx.doi.org/}%
\providecommand \selectlanguage [0]{\@gobble}%
\providecommand \bibinfo  [0]{\@secondoftwo}%
\providecommand \bibfield  [0]{\@secondoftwo}%
\providecommand \translation [1]{[#1]}%
\providecommand \BibitemOpen [0]{}%
\providecommand \bibitemStop [0]{}%
\providecommand \bibitemNoStop [0]{.\EOS\space}%
\providecommand \EOS [0]{\spacefactor3000\relax}%
\providecommand \BibitemShut  [1]{\csname bibitem#1\endcsname}%
\let\auto@bib@innerbib\@empty
%</preamble>
\bibitem [{\citenamefont {Alexeev}(1965)}]{alexeev1965generalized}%
  \BibitemOpen
  \bibfield  {author} {\bibinfo {author} {\bibfnamefont {V.~M.}\ \bibnamefont
  {Alexeev}},\ }\bibfield  {title} {\enquote {\bibinfo {title} {Generalized
  three-dimensional problem of two fixed centers of gravitation --- a
  classification of movements},}\ }\href@noop {} {\bibfield  {journal}
  {\bibinfo  {journal} {Bull. Inst. Theoret. Astron.}\ }\textbf {\bibinfo
  {volume} {10}},\ \bibinfo {pages} {241--271} (\bibinfo {year}
  {1965})}\BibitemShut {NoStop}%
\bibitem [{\citenamefont {Marchal}(1966)}]{marchal1966calcul}%
  \BibitemOpen
  \bibfield  {author} {\bibinfo {author} {\bibfnamefont {Christian}\
  \bibnamefont {Marchal}},\ }\bibfield  {title} {\enquote {\bibinfo {title}
  {Calcul du mouvement des satellites artificiels {\`a} partir des solutions
  exactes du probl{\`e}me des deux centres fixes. ({F}rench). {C}alculation of
  the movement of artificial satellites from exact solutions of the problem of
  the two fixed centers},}\ }\href@noop {} {\bibfield  {journal} {\bibinfo
  {journal} {Bulletin Astronomique}\ }\textbf {\bibinfo {volume} {1}},\
  \bibinfo {pages} {189--213} (\bibinfo {year} {1966})},\ \bibinfo {note}
  {[{3$^e$} {S}{\'e}rie, {T}ome {I}, {F}ascicule 3]}\BibitemShut {NoStop}%
\bibitem [{\citenamefont {Marchal}(1986)}]{marchal1986quasi}%
  \BibitemOpen
  \bibfield  {author} {\bibinfo {author} {\bibfnamefont {Christian}\
  \bibnamefont {Marchal}},\ }\bibfield  {title} {\enquote {\bibinfo {title} {On
  quasi-integrable problems. {The} example of the artificial satellites
  perturbed by the {E}arth's zonal harmonics},}\ }\href@noop {} {\bibfield
  {journal} {\bibinfo  {journal} {Celestial Mechanics}\ }\textbf {\bibinfo
  {volume} {38}},\ \bibinfo {pages} {377--387} (\bibinfo {year}
  {1986})}\BibitemShut {NoStop}%
\bibitem [{\citenamefont {Jakas}(1995)}]{jakas1995trapping}%
  \BibitemOpen
  \bibfield  {author} {\bibinfo {author} {\bibfnamefont {Mario~M.}\
  \bibnamefont {Jakas}},\ }\bibfield  {title} {\enquote {\bibinfo {title}
  {Trapping of a classical electron between two heavy scattering centers},}\
  }\href {\doibase 10.1103/PhysRevA.52.866} {\bibfield  {journal} {\bibinfo
  {journal} {Phys. Rev. A}\ }\textbf {\bibinfo {volume} {52}},\ \bibinfo
  {pages} {866--869} (\bibinfo {year} {1995})}\BibitemShut {NoStop}%
\bibitem [{\citenamefont {Jakas}(1996)}]{jakas1996production}%
  \BibitemOpen
  \bibfield  {author} {\bibinfo {author} {\bibfnamefont {Mario~M.}\
  \bibnamefont {Jakas}},\ }\bibfield  {title} {\enquote {\bibinfo {title} {The
  production of high-energy electrons during low energy atomic collisions in
  solids},}\ }\href {\doibase https://doi.org/10.1016/0168-583X(95)01559-0}
  {\bibfield  {journal} {\bibinfo  {journal} {Nuclear Instruments and Methods
  in Physics Research Section B: Beam Interactions with Materials and Atoms}\
  }\textbf {\bibinfo {volume} {115}},\ \bibinfo {pages} {255--260} (\bibinfo
  {year} {1996})}\BibitemShut {NoStop}%
\bibitem [{\citenamefont {Strand}\ and\ \citenamefont
  {Reinhardta}(1979)}]{strand1979semiclassical}%
  \BibitemOpen
  \bibfield  {author} {\bibinfo {author} {\bibfnamefont {Michael~P.}\
  \bibnamefont {Strand}}\ and\ \bibinfo {author} {\bibfnamefont {William~P.}\
  \bibnamefont {Reinhardta}},\ }\bibfield  {title} {\enquote {\bibinfo {title}
  {Semiclassical quantization of the low lying electronic states of
  {$\mathrm{H}_2^+$}},}\ }\href@noop {} {\bibfield  {journal} {\bibinfo
  {journal} {The Journal of Chemical Physics}\ }\textbf {\bibinfo {volume}
  {70}},\ \bibinfo {pages} {3812--3827} (\bibinfo {year} {1979})}\BibitemShut
  {NoStop}%
\bibitem [{\citenamefont {McMillan}(1971)}]{mcmillan1971problem}%
  \BibitemOpen
  \bibfield  {author} {\bibinfo {author} {\bibfnamefont {Edwin~M.}\
  \bibnamefont {McMillan}},\ }\bibfield  {title} {\enquote {\bibinfo {title} {A
  problem in the stability of periodic systems},}\ }in\ \href@noop {} {\emph
  {\bibinfo {booktitle} {Topics in modern physics. A {T}ribute to {E}dward {U}.
  {C}ondon}}},\ \bibinfo {editor} {edited by\ \bibinfo {editor} {\bibfnamefont
  {W.~E.}\ \bibnamefont {Brittin}}\ and\ \bibinfo {editor} {\bibfnamefont
  {H.}~\bibnamefont {Odabasi}}}\ (\bibinfo  {publisher} {Colorado Associated
  University Press},\ \bibinfo {address} {Boulder, CO},\ \bibinfo {year}
  {1971})\ pp.\ \bibinfo {pages} {219--244}\BibitemShut {NoStop}%
\bibitem [{\citenamefont {Suris}(1989)}]{suris1989integrable}%
  \BibitemOpen
  \bibfield  {author} {\bibinfo {author} {\bibfnamefont {Yuri~B.}\ \bibnamefont
  {Suris}},\ }\bibfield  {title} {\enquote {\bibinfo {title} {Integrable
  mappings of the standard type},}\ }\href {\doibase
  https://doi.org/10.1007/BF01078586} {\bibfield  {journal} {\bibinfo
  {journal} {Functional Analysis and Its Applications}\ }\textbf {\bibinfo
  {volume} {23}},\ \bibinfo {pages} {74--76} (\bibinfo {year}
  {1989})}\BibitemShut {NoStop}%
\bibitem [{\citenamefont {Iatrou}\ and\ \citenamefont
  {Roberts}(2002)}]{IR2002II}%
  \BibitemOpen
  \bibfield  {author} {\bibinfo {author} {\bibfnamefont {Apostolos}\
  \bibnamefont {Iatrou}}\ and\ \bibinfo {author} {\bibfnamefont {John A.~G.}\
  \bibnamefont {Roberts}},\ }\bibfield  {title} {\enquote {\bibinfo {title}
  {Integrable mappings of the plane preserving biquadratic invariant curves
  {II}},}\ }\href {\doibase 10.1088/0951-7715/15/2/313} {\bibfield  {journal}
  {\bibinfo  {journal} {Nonlinearity}\ }\textbf {\bibinfo {volume} {15}},\
  \bibinfo {pages} {459--489} (\bibinfo {year} {2002})}\BibitemShut {NoStop}%
\bibitem [{\citenamefont {Cathey}\ \emph {et~al.}(2021)\citenamefont {Cathey},
  \citenamefont {Stancari}, \citenamefont {Valishev},\ and\ \citenamefont
  {Zolkin}}]{cathey2021calculations}%
  \BibitemOpen
  \bibfield  {author} {\bibinfo {author} {\bibfnamefont {Brandon}\ \bibnamefont
  {Cathey}}, \bibinfo {author} {\bibfnamefont {Giulio}\ \bibnamefont
  {Stancari}}, \bibinfo {author} {\bibfnamefont {Alexander}\ \bibnamefont
  {Valishev}}, \ and\ \bibinfo {author} {\bibfnamefont {Timofey}\ \bibnamefont
  {Zolkin}},\ }\bibfield  {title} {\enquote {\bibinfo {title} {Calculations of
  detuning with amplitude for the {McMillan} electron lens in the {F}ermilab
  {I}ntegrable {O}ptics {T}est {A}ccelerator ({IOTA})},}\ }\href {\doibase
  10.1088/1748-0221/16/03/P03041} {\bibfield  {journal} {\bibinfo  {journal}
  {Journal of Instrumentation}\ }\textbf {\bibinfo {volume} {16}},\ \bibinfo
  {pages} {P03041} (\bibinfo {year} {2021})}\BibitemShut {NoStop}%
\bibitem [{\citenamefont {Quispel}\ \emph {et~al.}(1988)\citenamefont
  {Quispel}, \citenamefont {Roberts},\ and\ \citenamefont
  {Thompson}}]{quispel1988integrable}%
  \BibitemOpen
  \bibfield  {author} {\bibinfo {author} {\bibfnamefont {Gilles
  Reinout~Willem}\ \bibnamefont {Quispel}}, \bibinfo {author} {\bibfnamefont
  {Jhon A.~G.}\ \bibnamefont {Roberts}}, \ and\ \bibinfo {author}
  {\bibfnamefont {Colin~J.}\ \bibnamefont {Thompson}},\ }\bibfield  {title}
  {\enquote {\bibinfo {title} {Integrable mappings and soliton equations},}\
  }\href {\doibase https://doi.org/10.1016/0375-9601(88)90803-1} {\bibfield
  {journal} {\bibinfo  {journal} {Physics Letters A}\ }\textbf {\bibinfo
  {volume} {126}},\ \bibinfo {pages} {419--421} (\bibinfo {year}
  {1988})}\BibitemShut {NoStop}%
\bibitem [{\citenamefont {Quispel}\ \emph {et~al.}(1989)\citenamefont
  {Quispel}, \citenamefont {Roberts},\ and\ \citenamefont
  {Thompson}}]{quispel1989integrable}%
  \BibitemOpen
  \bibfield  {author} {\bibinfo {author} {\bibfnamefont {Gilles
  Reinout~Willem}\ \bibnamefont {Quispel}}, \bibinfo {author} {\bibfnamefont
  {Jhon A.~G.}\ \bibnamefont {Roberts}}, \ and\ \bibinfo {author}
  {\bibfnamefont {Colin~J.}\ \bibnamefont {Thompson}},\ }\bibfield  {title}
  {\enquote {\bibinfo {title} {Integrable mappings and soliton equations
  {II}},}\ }\href {\doibase https://doi.org/10.1016/0167-2789(89)90233-9}
  {\bibfield  {journal} {\bibinfo  {journal} {Physica D: Nonlinear Phenomena}\
  }\textbf {\bibinfo {volume} {34}},\ \bibinfo {pages} {183--192} (\bibinfo
  {year} {1989})}\BibitemShut {NoStop}%
\bibitem [{\citenamefont {Landau}\ and\ \citenamefont
  {Lifshitz}(1976)}]{LL_mechanics}%
  \BibitemOpen
  \bibfield  {author} {\bibinfo {author} {\bibfnamefont {L.~D.}\ \bibnamefont
  {Landau}}\ and\ \bibinfo {author} {\bibfnamefont {E.~M.}\ \bibnamefont
  {Lifshitz}},\ }\href {\doibase https://doi.org/10.1016/C2009-0-25569-3}
  {\emph {\bibinfo {title} {Mechanics}}},\ \bibinfo {edition} {3rd}\ ed.,\
  \bibinfo {series} {Course of Theoretical Physics}, Vol.~\bibinfo {volume}
  {1}\ (\bibinfo  {publisher} {Butterworth-Heinemann},\ \bibinfo {address}
  {Oxford},\ \bibinfo {year} {1976})\BibitemShut {NoStop}%
\bibitem [{\citenamefont {H{\'e}non}\ and\ \citenamefont
  {Heiles}(1964)}]{henon1964applicability}%
  \BibitemOpen
  \bibfield  {author} {\bibinfo {author} {\bibfnamefont {Michel}\ \bibnamefont
  {H{\'e}non}}\ and\ \bibinfo {author} {\bibfnamefont {Carl}\ \bibnamefont
  {Heiles}},\ }\bibfield  {title} {\enquote {\bibinfo {title} {The
  applicability of the third integral of motion: some numerical experiments},}\
  }\href {\doibase 10.1086/109234} {\bibfield  {journal} {\bibinfo  {journal}
  {The Astronomical Journal}\ }\textbf {\bibinfo {volume} {69}},\ \bibinfo
  {pages} {73--79} (\bibinfo {year} {1964})}\BibitemShut {NoStop}%
\bibitem [{\citenamefont {Duffing}(1918)}]{duffing1918erzwungene}%
  \BibitemOpen
  \bibfield  {author} {\bibinfo {author} {\bibfnamefont {Georg}\ \bibnamefont
  {Duffing}},\ }\href@noop {} {\emph {\bibinfo {title} {Erzwungene
  {S}chwingungen bei ver{\"a}nderlicher {E}igenfrequenz und ihre technische
  {B}edeutung. ({G}erman). {F}orced oscillations with changing natural
  frequencies and their technical significance}}}\ (\bibinfo  {publisher}
  {Vieweg \& Sohn},\ \bibinfo {address} {Braunschweig},\ \bibinfo {year}
  {1918})\BibitemShut {NoStop}%
\bibitem [{\citenamefont {Zolkin}\ \emph {et~al.}(2022)\citenamefont {Zolkin},
  \citenamefont {Nagaitsev},\ and\ \citenamefont
  {Morozov}}]{zolkin2022mcmillan}%
  \BibitemOpen
  \bibfield  {author} {\bibinfo {author} {\bibfnamefont {Timofey}\ \bibnamefont
  {Zolkin}}, \bibinfo {author} {\bibfnamefont {Sergei}\ \bibnamefont
  {Nagaitsev}}, \ and\ \bibinfo {author} {\bibfnamefont {Ivan}\ \bibnamefont
  {Morozov}},\ }\href@noop {} {\enquote {\bibinfo {title} {Mcmillan map and
  nonlinear {T}wiss parameters},}\ } (\bibinfo {year} {2022}),\ \Eprint
  {http://arxiv.org/abs/2204.12691} {arXiv:2204.12691 [nlin.SI]} \BibitemShut
  {NoStop}%
\bibitem [{\citenamefont {H{\'e}non}(1969)}]{henon1969numerical}%
  \BibitemOpen
  \bibfield  {author} {\bibinfo {author} {\bibfnamefont {Michel}\ \bibnamefont
  {H{\'e}non}},\ }\bibfield  {title} {\enquote {\bibinfo {title} {Numerical
  study of quadratic area-preserving mappings},}\ }\href
  {http://www.jstor.org/stable/43635985} {\bibfield  {journal} {\bibinfo
  {journal} {Quarterly of Applied Mathematics}\ }\textbf {\bibinfo {volume}
  {27}},\ \bibinfo {pages} {291--312} (\bibinfo {year} {1969})}\BibitemShut
  {NoStop}%
\bibitem [{\citenamefont {Chirikov}(1971)}]{chirikov1969research}%
  \BibitemOpen
  \bibfield  {author} {\bibinfo {author} {\bibfnamefont {Boris~V.}\
  \bibnamefont {Chirikov}},\ }\href {https://cds.cern.ch/record/325497?ln=en}
  {\enquote {\bibinfo {title} {Research conserning the theory of non-linear
  resonance and stochasticity},}\ } (\bibinfo {year} {1971}),\ \bibinfo {note}
  {translated at CERN by A. T. Sanders from the Russian [CERN-Trans-71-40].
  Nuclear Physics Institute of the Siberian Section of the USSR Academy of
  Science, Report 267, Novosibirsk, 1969 [IYAF-267-TRANS-E]}\BibitemShut
  {NoStop}%
\bibitem [{\citenamefont {Chirikov}(1979)}]{chirikov1979universal}%
  \BibitemOpen
  \bibfield  {author} {\bibinfo {author} {\bibfnamefont {Boris~V.}\
  \bibnamefont {Chirikov}},\ }\bibfield  {title} {\enquote {\bibinfo {title} {A
  universal instability of many-dimensional oscillator systems},}\ }\href
  {\doibase https://doi.org/10.1016/0370-1573(79)90023-1} {\bibfield  {journal}
  {\bibinfo  {journal} {Physics Reports}\ }\textbf {\bibinfo {volume} {52}},\
  \bibinfo {pages} {263--379} (\bibinfo {year} {1979})}\BibitemShut {NoStop}%
\bibitem [{\citenamefont {Zolkin}\ \emph {et~al.}(2017)\citenamefont {Zolkin},
  \citenamefont {Nagaitsev},\ and\ \citenamefont
  {Danilov}}]{zolkin2017rotation}%
  \BibitemOpen
  \bibfield  {author} {\bibinfo {author} {\bibfnamefont {Timofey}\ \bibnamefont
  {Zolkin}}, \bibinfo {author} {\bibfnamefont {Sergei}\ \bibnamefont
  {Nagaitsev}}, \ and\ \bibinfo {author} {\bibfnamefont {Viatcheslav}\
  \bibnamefont {Danilov}},\ }\href@noop {} {\enquote {\bibinfo {title}
  {Rotation number of integrable symplectic mappings of the plane},}\ }
  (\bibinfo {year} {2017}),\ \Eprint {http://arxiv.org/abs/1704.03077}
  {arXiv:1704.03077 [nlin.SI]} \BibitemShut {NoStop}%
\bibitem [{\citenamefont {Nagaitsev}\ and\ \citenamefont
  {Zolkin}(2020)}]{nagaitsev2020betatron}%
  \BibitemOpen
  \bibfield  {author} {\bibinfo {author} {\bibfnamefont {Sergei}\ \bibnamefont
  {Nagaitsev}}\ and\ \bibinfo {author} {\bibfnamefont {Timofey}\ \bibnamefont
  {Zolkin}},\ }\bibfield  {title} {\enquote {\bibinfo {title} {Betatron
  frequency and the {P}oincar{\'e} rotation number},}\ }\href {\doibase
  10.1103/PhysRevAccelBeams.23.054001} {\bibfield  {journal} {\bibinfo
  {journal} {Phys. Rev. Accel. Beams}\ }\textbf {\bibinfo {volume} {23}},\
  \bibinfo {pages} {054001} (\bibinfo {year} {2020})}\BibitemShut {NoStop}%
\bibitem [{\citenamefont {Mitchell}\ \emph {et~al.}(2021)\citenamefont
  {Mitchell}, \citenamefont {Ryne}, \citenamefont {Hwang}, \citenamefont
  {Nagaitsev},\ and\ \citenamefont {Zolkin}}]{mitchell2021extracting}%
  \BibitemOpen
  \bibfield  {author} {\bibinfo {author} {\bibfnamefont {Chad~E.}\ \bibnamefont
  {Mitchell}}, \bibinfo {author} {\bibfnamefont {Robert~D.}\ \bibnamefont
  {Ryne}}, \bibinfo {author} {\bibfnamefont {Kilean}\ \bibnamefont {Hwang}},
  \bibinfo {author} {\bibfnamefont {Sergei}\ \bibnamefont {Nagaitsev}}, \ and\
  \bibinfo {author} {\bibfnamefont {Timofey}\ \bibnamefont {Zolkin}},\
  }\bibfield  {title} {\enquote {\bibinfo {title} {Extracting dynamical
  frequencies from invariants of motion in finite-dimensional nonlinear
  integrable systems},}\ }\href {\doibase 10.1103/PhysRevE.103.062216}
  {\bibfield  {journal} {\bibinfo  {journal} {Phys. Rev. E}\ }\textbf {\bibinfo
  {volume} {103}},\ \bibinfo {pages} {062216} (\bibinfo {year}
  {2021})}\BibitemShut {NoStop}%
\bibitem [{\citenamefont {Kolmogorov}(1954)}]{kolmogorov1954conservation}%
  \BibitemOpen
  \bibfield  {author} {\bibinfo {author} {\bibfnamefont {Andrey~Nikolaevich}\
  \bibnamefont {Kolmogorov}},\ }\bibfield  {title} {\enquote {\bibinfo {title}
  {On conservation of conditionally periodic motions for a small change in
  {H}amilton's function},}\ \ }(\bibinfo {year} {1954})\ pp.\ \bibinfo {pages}
  {527--530}\BibitemShut {NoStop}%
\bibitem [{\citenamefont {Moser}(1962)}]{moser1962invariant}%
  \BibitemOpen
  \bibfield  {author} {\bibinfo {author} {\bibfnamefont {J{\"u}rgen}\
  \bibnamefont {Moser}},\ }\bibfield  {title} {\enquote {\bibinfo {title} {On
  invariant curves of area-preserving mapping of an annulus},}\ }\href
  {https://www.mathnet.ru/links/ca5cbd9baef2265dd4d67e2c7e2539a8/mat236.pdf}
  {\bibfield  {journal} {\bibinfo  {journal} {Matematika}\ }\textbf {\bibinfo
  {volume} {6}},\ \bibinfo {pages} {51--68} (\bibinfo {year}
  {1962})}\BibitemShut {NoStop}%
\bibitem [{\citenamefont {Arnold}(1963)}]{arnol1963small}%
  \BibitemOpen
  \bibfield  {author} {\bibinfo {author} {\bibfnamefont {Vladimir~I.}\
  \bibnamefont {Arnold}},\ }\bibfield  {title} {\enquote {\bibinfo {title}
  {Proof of a theorem of {A}. {N}. {K}olmogorov on the preservation of
  conditionally periodic motions under a small perturbation of the
  {H}amiltonian},}\ }\href {\doibase 10.1070/RM1963v018n05ABEH004130}
  {\bibfield  {journal} {\bibinfo  {journal} {Russian Mathematical Surveys}\
  }\textbf {\bibinfo {volume} {18}},\ \bibinfo {pages} {9--36} (\bibinfo {year}
  {1963})}\BibitemShut {NoStop}%
\bibitem [{\citenamefont {Zolkin}\ and\ \citenamefont
  {Nagaitsev}(2016)}]{zolkin2016analytical}%
  \BibitemOpen
  \bibfield  {author} {\bibinfo {author} {\bibfnamefont {Timofey}\ \bibnamefont
  {Zolkin}}\ and\ \bibinfo {author} {\bibfnamefont {Sergei}\ \bibnamefont
  {Nagaitsev}},\ }\href@noop {} {\enquote {\bibinfo {title} {Analytical theory
  of {McMillan} map},}\ }\bibinfo {howpublished} {available at
  \url{https://accelconf.web.cern.ch/napac2016/talks/wea1co06_talk.pdf}}
  (\bibinfo {year} {2016}),\ \bibinfo {note} {presented at the North American
  Particle Accelerator Conference (NAPAC'16), Chicago, IL, USA, October 9--14,
  2016, {[WEA1CO06]}}\BibitemShut {NoStop}%
\bibitem [{\citenamefont {Zolkin}\ \emph
  {et~al.}(2023{\natexlab{a}})\citenamefont {Zolkin}, \citenamefont {Kharkov},\
  and\ \citenamefont {Nagaitsev}}]{ZKN2023PolI}%
  \BibitemOpen
  \bibfield  {author} {\bibinfo {author} {\bibfnamefont {T.}~\bibnamefont
  {Zolkin}}, \bibinfo {author} {\bibfnamefont {Y.}~\bibnamefont {Kharkov}}, \
  and\ \bibinfo {author} {\bibfnamefont {S.}~\bibnamefont {Nagaitsev}},\
  }\bibfield  {title} {\enquote {\bibinfo {title} {Machine-assisted discovery
  of integrable symplectic mappings},}\ }\href {\doibase
  10.1103/PhysRevResearch.5.043241} {\bibfield  {journal} {\bibinfo  {journal}
  {Phys. Rev. Res.}\ }\textbf {\bibinfo {volume} {5}},\ \bibinfo {pages}
  {043241} (\bibinfo {year} {2023}{\natexlab{a}})}\BibitemShut {NoStop}%
\bibitem [{\citenamefont {Zolkin}\ \emph
  {et~al.}(2024{\natexlab{a}})\citenamefont {Zolkin}, \citenamefont {Kharkov},\
  and\ \citenamefont {Nagaitsev}}]{ZKN2024PolII}%
  \BibitemOpen
  \bibfield  {author} {\bibinfo {author} {\bibfnamefont {T.}~\bibnamefont
  {Zolkin}}, \bibinfo {author} {\bibfnamefont {Y.}~\bibnamefont {Kharkov}}, \
  and\ \bibinfo {author} {\bibfnamefont {S.}~\bibnamefont {Nagaitsev}},\
  }\bibfield  {title} {\enquote {\bibinfo {title} {Integrable symplectic maps
  with a polygon tessellation},}\ }\href {\doibase
  10.1103/PhysRevResearch.6.023324} {\bibfield  {journal} {\bibinfo  {journal}
  {Phys. Rev. Res.}\ }\textbf {\bibinfo {volume} {6}},\ \bibinfo {pages}
  {023324} (\bibinfo {year} {2024}{\natexlab{a}})}\BibitemShut {NoStop}%
\bibitem [{\citenamefont {Zolkin}\ \emph
  {et~al.}(2024{\natexlab{b}})\citenamefont {Zolkin}, \citenamefont
  {Nagaitsev}, \citenamefont {Morozov}, \citenamefont {Kladov},\ and\
  \citenamefont {Kim}}]{zolkin2024dynamicsIII}%
  \BibitemOpen
  \bibfield  {author} {\bibinfo {author} {\bibfnamefont {Tim}\ \bibnamefont
  {Zolkin}}, \bibinfo {author} {\bibfnamefont {Sergei}\ \bibnamefont
  {Nagaitsev}}, \bibinfo {author} {\bibfnamefont {Ivan}\ \bibnamefont
  {Morozov}}, \bibinfo {author} {\bibfnamefont {Sergei}\ \bibnamefont
  {Kladov}}, \ and\ \bibinfo {author} {\bibfnamefont {Young-Kee}\ \bibnamefont
  {Kim}},\ }\href {https://arxiv.org/abs/2410.10380} {\enquote {\bibinfo
  {title} {Dynamics of {M}c{M}illan mappings {III}. {S}ymmetric map with mixed
  nonlinearity},}\ } (\bibinfo {year} {2024}{\natexlab{b}}),\ \Eprint
  {http://arxiv.org/abs/2410.10380} {arXiv:2410.10380 [nlin.SI]} \BibitemShut
  {NoStop}%
\bibitem [{\citenamefont {DeVogelaere}(1958)}]{devogelaere1950}%
  \BibitemOpen
  \bibfield  {author} {\bibinfo {author} {\bibfnamefont {Ren{\'e}}\
  \bibnamefont {DeVogelaere}},\ }\enquote {\bibinfo {title} {{IV}. {O}n the
  structure of symmetric periodic solutions of conservative systems, with
  applications},}\ in\ \href {\doibase doi:10.1515/9781400881758-005} {\emph
  {\bibinfo {booktitle} {Contributions to the Theory of Nonlinear Oscillations
  (AM-41), Volume IV}}},\ \bibinfo {editor} {edited by\ \bibinfo {editor}
  {\bibfnamefont {Solomon}\ \bibnamefont {Lefschetz}}}\ (\bibinfo  {publisher}
  {Princeton University Press},\ \bibinfo {address} {Princeton},\ \bibinfo
  {year} {1958})\ pp.\ \bibinfo {pages} {53--84}\BibitemShut {NoStop}%
\bibitem [{\citenamefont {Lewis~Jr.}(1961)}]{lewis1961reversible}%
  \BibitemOpen
  \bibfield  {author} {\bibinfo {author} {\bibfnamefont {Daniel~C.}\
  \bibnamefont {Lewis~Jr.}},\ }\bibfield  {title} {\enquote {\bibinfo {title}
  {Reversible transformations},}\ }\href@noop {} {\bibfield  {journal}
  {\bibinfo  {journal} {Pacific Journal of Mathematics}\ }\textbf {\bibinfo
  {volume} {11}},\ \bibinfo {pages} {1077--1087} (\bibinfo {year}
  {1961})}\BibitemShut {NoStop}%
\bibitem [{\citenamefont {Zolkin}\ \emph
  {et~al.}(2023{\natexlab{b}})\citenamefont {Zolkin}, \citenamefont {Kharkov},\
  and\ \citenamefont {Nagaitsev}}]{ZKN2023}%
  \BibitemOpen
  \bibfield  {author} {\bibinfo {author} {\bibfnamefont {Timofey}\ \bibnamefont
  {Zolkin}}, \bibinfo {author} {\bibfnamefont {Yaroslav}\ \bibnamefont
  {Kharkov}}, \ and\ \bibinfo {author} {\bibfnamefont {Sergei}\ \bibnamefont
  {Nagaitsev}},\ }\bibfield  {title} {\enquote {\bibinfo {title}
  {Machine-assisted discovery of integrable symplectic mappings},}\ }\href
  {\doibase 10.1103/PhysRevResearch.5.043241} {\bibfield  {journal} {\bibinfo
  {journal} {Phys. Rev. Res.}\ }\textbf {\bibinfo {volume} {5}},\ \bibinfo
  {pages} {043241} (\bibinfo {year} {2023}{\natexlab{b}})}\BibitemShut
  {NoStop}%
\bibitem [{\citenamefont {Roberts}\ and\ \citenamefont
  {Quispel}(1992)}]{roberts1992revers}%
  \BibitemOpen
  \bibfield  {author} {\bibinfo {author} {\bibfnamefont {J.A.G.}\ \bibnamefont
  {Roberts}}\ and\ \bibinfo {author} {\bibfnamefont {G.R.W.}\ \bibnamefont
  {Quispel}},\ }\bibfield  {title} {\enquote {\bibinfo {title} {Chaos and
  time-reversal symmetry. {O}rder and chaos in reversible dynamical systems},}\
  }\href {\doibase https://doi.org/10.1016/0370-1573(92)90163-T} {\bibfield
  {journal} {\bibinfo  {journal} {Physics Reports}\ }\textbf {\bibinfo {volume}
  {216}},\ \bibinfo {pages} {63--177} (\bibinfo {year} {1992})}\BibitemShut
  {NoStop}%
\bibitem [{\citenamefont {Zolkin}\ \emph
  {et~al.}(2024{\natexlab{c}})\citenamefont {Zolkin}, \citenamefont
  {Nagaitsev}, \citenamefont {Morozov}, \citenamefont {Kladov},\ and\
  \citenamefont {Kim}}]{zolkin2024HenonSet}%
  \BibitemOpen
  \bibfield  {author} {\bibinfo {author} {\bibfnamefont {Tim}\ \bibnamefont
  {Zolkin}}, \bibinfo {author} {\bibfnamefont {Sergei}\ \bibnamefont
  {Nagaitsev}}, \bibinfo {author} {\bibfnamefont {Ivan}\ \bibnamefont
  {Morozov}}, \bibinfo {author} {\bibfnamefont {Sergei}\ \bibnamefont
  {Kladov}}, \ and\ \bibinfo {author} {\bibfnamefont {Young-Kee}\ \bibnamefont
  {Kim}},\ }\href {https://arxiv.org/abs/2412.05541} {\enquote {\bibinfo
  {title} {Isochronous and period-doubling diagrams for symplectic maps of the
  plane},}\ } (\bibinfo {year} {2024}{\natexlab{c}}),\ \Eprint
  {http://arxiv.org/abs/2412.05541} {arXiv:2412.05541 [nlin.CD]} \BibitemShut
  {NoStop}%
\bibitem [{\citenamefont {Veselov}(1991)}]{veselov1991integrable}%
  \BibitemOpen
  \bibfield  {author} {\bibinfo {author} {\bibfnamefont {Aleksandr~Petrovich}\
  \bibnamefont {Veselov}},\ }\bibfield  {title} {\enquote {\bibinfo {title}
  {Integrable maps},}\ }\href {\doibase 10.1070/RM1991v046n05ABEH002856}
  {\bibfield  {journal} {\bibinfo  {journal} {Russian Mathematical Surveys}\
  }\textbf {\bibinfo {volume} {46}},\ \bibinfo {pages} {1--51} (\bibinfo {year}
  {1991})}\BibitemShut {NoStop}%
\bibitem [{\citenamefont {Brown}(1993)}]{brown1993}%
  \BibitemOpen
  \bibfield  {author} {\bibinfo {author} {\bibfnamefont {Morton}\ \bibnamefont
  {Brown}},\ }\bibfield  {title} {\enquote {\bibinfo {title} {A periodic
  homeomorphism of the plane},}\ }in\ \href@noop {} {\emph {\bibinfo
  {booktitle} {Continuum theory and dynamical systems}}},\ \bibinfo {series and
  number} {\bibinfo {series} {Lecture Notes in Pure and Appl. Math.}\ No.\
  \bibinfo {number} {149}}\ (\bibinfo  {publisher} {Marcel Dekker AG},\
  \bibinfo {address} {New York},\ \bibinfo {year} {1993})\ pp.\ \bibinfo
  {pages} {83--87}\BibitemShut {NoStop}%
\bibitem [{\citenamefont {Cairns}\ \emph {et~al.}(2016)\citenamefont {Cairns},
  \citenamefont {Nikolayevsky},\ and\ \citenamefont
  {Rossiter}}]{cairns2016conewise}%
  \BibitemOpen
  \bibfield  {author} {\bibinfo {author} {\bibfnamefont {Grant}\ \bibnamefont
  {Cairns}}, \bibinfo {author} {\bibfnamefont {Yuri}\ \bibnamefont
  {Nikolayevsky}}, \ and\ \bibinfo {author} {\bibfnamefont {Gavin}\
  \bibnamefont {Rossiter}},\ }\bibfield  {title} {\enquote {\bibinfo {title}
  {Conewise linear periodic maps of the plane with integer coefficients},}\
  }\href {https://www.jstor.org/stable/10.4169/amer.math.monthly.123.4.363}
  {\bibfield  {journal} {\bibinfo  {journal} {The American Mathematical
  Monthly}\ }\textbf {\bibinfo {volume} {123}},\ \bibinfo {pages} {363--375}
  (\bibinfo {year} {2016})}\BibitemShut {NoStop}%
\bibitem [{\citenamefont {Zolkin}\ \emph
  {et~al.}(2023{\natexlab{c}})\citenamefont {Zolkin}, \citenamefont {Kharkov},\
  and\ \citenamefont {Nagaitsev}}]{ZKN2024arxiv}%
  \BibitemOpen
  \bibfield  {author} {\bibinfo {author} {\bibfnamefont {Timofey}\ \bibnamefont
  {Zolkin}}, \bibinfo {author} {\bibfnamefont {Yaroslav}\ \bibnamefont
  {Kharkov}}, \ and\ \bibinfo {author} {\bibfnamefont {Sergei}\ \bibnamefont
  {Nagaitsev}},\ }\href@noop {} {\enquote {\bibinfo {title} {Integrable
  symplectic maps with a polygon tessellation},}\ } (\bibinfo {year}
  {2023}{\natexlab{c}}),\ \Eprint {http://arxiv.org/abs/2311.17616}
  {arXiv:2311.17616 [nlin.SI]} \BibitemShut {NoStop}%
\bibitem [{\citenamefont {Lee}(2018)}]{SYLee4}%
  \BibitemOpen
  \bibfield  {author} {\bibinfo {author} {\bibfnamefont {Shyh-Yuan}\
  \bibnamefont {Lee}},\ }\href {\doibase 10.1142/11111} {\emph {\bibinfo
  {title} {Accelerator Physics}}},\ \bibinfo {edition} {4th}\ ed.\ (\bibinfo
  {publisher} {World Scientific Publishing Company},\ \bibinfo {address}
  {Singapore},\ \bibinfo {year} {2018})\BibitemShut {NoStop}%
\bibitem [{\citenamefont {Courant}\ and\ \citenamefont
  {Snyder}(1958)}]{courant1958theory}%
  \BibitemOpen
  \bibfield  {author} {\bibinfo {author} {\bibfnamefont {Ernest~D.}\
  \bibnamefont {Courant}}\ and\ \bibinfo {author} {\bibfnamefont {Hartland~S.}\
  \bibnamefont {Snyder}},\ }\bibfield  {title} {\enquote {\bibinfo {title}
  {Theory of the alternating-gradient synchrotron},}\ }\href {\doibase
  https://doi.org/10.1016/0003-4916(58)90012-5} {\bibfield  {journal} {\bibinfo
   {journal} {Annals of Physics}\ }\textbf {\bibinfo {volume} {3}},\ \bibinfo
  {pages} {1--48} (\bibinfo {year} {1958})}\BibitemShut {NoStop}%
\bibitem [{\citenamefont {Michelotti}(1995)}]{michelotti1995intermediate}%
  \BibitemOpen
  \bibfield  {author} {\bibinfo {author} {\bibfnamefont {Leo~P.}\ \bibnamefont
  {Michelotti}},\ }\href {https://books.google.ru/books?id=DImTQgAACAAJ} {\emph
  {\bibinfo {title} {Intermediate Classical Dynamics with Applications to Beam
  Physics}}},\ Wiley Series in Beam Physics and Accelerator Technology\
  (\bibinfo  {publisher} {Wiley},\ \bibinfo {address} {New York},\ \bibinfo
  {year} {1995})\BibitemShut {NoStop}%
\bibitem [{\citenamefont {Michelotti}(1984)}]{michelotti1984moser}%
  \BibitemOpen
  \bibfield  {author} {\bibinfo {author} {\bibfnamefont {Leo~P.}\ \bibnamefont
  {Michelotti}},\ }\bibfield  {title} {\enquote {\bibinfo {title} {Moser like
  transformations using the lie transform},}\ }\href@noop {} {\bibfield
  {journal} {\bibinfo  {journal} {Part. Accel.}\ }\textbf {\bibinfo {volume}
  {16}},\ \bibinfo {pages} {233--252} (\bibinfo {year} {1984})}\BibitemShut
  {NoStop}%
\bibitem [{\citenamefont {Bengtsson}(1997)}]{bengtsson1997}%
  \BibitemOpen
  \bibfield  {author} {\bibinfo {author} {\bibfnamefont {Johan}\ \bibnamefont
  {Bengtsson}},\ }\href@noop {} {\emph {\bibinfo {title} {The sextupole scheme
  for the {S}wiss {L}ight {S}ource ({SLS}): an analytic approach}}},\ \bibinfo
  {type} {Tech. Rep.}\ \bibinfo {number} {SLS-TME-TA-1997-0009}\ (\bibinfo
  {institution} {Paul Scherrer Institut},\ \bibinfo {address} {Villigen,
  Switzerland},\ \bibinfo {year} {1997})\BibitemShut {NoStop}%
\bibitem [{\citenamefont {Morozov}\ and\ \citenamefont
  {Levichev}(2017)}]{morozov2017dynamical}%
  \BibitemOpen
  \bibfield  {author} {\bibinfo {author} {\bibfnamefont {Ivan}\ \bibnamefont
  {Morozov}}\ and\ \bibinfo {author} {\bibfnamefont {Evgeny}\ \bibnamefont
  {Levichev}},\ }\bibfield  {title} {\enquote {\bibinfo {title} {Dynamical
  aperture control in accelerator lattices with multipole potentials},}\ }in\
  \href {\doibase 10.23727/CERN-Proceedings-2017-001.195} {\emph {\bibinfo
  {booktitle} {Proceedings of the CERN-BINP Workshop for Young Scientists in
  {$e^+e^-$} Colliders, Geneva, Switzerland, 22 -- 25 August 2016}}},\ \bibinfo
  {series} {CERN Proceedings}, Vol.\ \bibinfo {volume} {1/2017},\ \bibinfo
  {editor} {edited by\ \bibinfo {editor} {\bibfnamefont {Valeria}\ \bibnamefont
  {Brancolini}}\ and\ \bibinfo {editor} {\bibfnamefont {Lucie}\ \bibnamefont
  {Linssen}}}\ (\bibinfo {organization} {CERN},\ \bibinfo {address} {Geneva},\
  \bibinfo {year} {2017})\ pp.\ \bibinfo {pages} {195--206},\ \bibinfo {note}
  {{CERN-Proceedings-2017-001}}\BibitemShut {NoStop}%
\bibitem [{\citenamefont {Fatou}(1917{\natexlab{a}})}]{Fatou:1917A}%
  \BibitemOpen
  \bibfield  {author} {\bibinfo {author} {\bibfnamefont {Pierre Joseph~Louis}\
  \bibnamefont {Fatou}},\ }\bibfield  {title} {\enquote {\bibinfo {title} {Sur
  les substitutions rationnelles. ({F}rench) {O}n rational substitutions},}\
  }\href@noop {} {\bibfield  {journal} {\bibinfo  {journal} {Comptes Rendus de
  l'Acad{\'e}mie des Sciences de Paris}\ }\textbf {\bibinfo {volume} {164}},\
  \bibinfo {pages} {806--808} (\bibinfo {year}
  {1917}{\natexlab{a}})}\BibitemShut {NoStop}%
\bibitem [{\citenamefont {Fatou}(1917{\natexlab{b}})}]{Fatou:1917B}%
  \BibitemOpen
  \bibfield  {author} {\bibinfo {author} {\bibfnamefont {Pierre Joseph~Louis}\
  \bibnamefont {Fatou}},\ }\bibfield  {title} {\enquote {\bibinfo {title} {Sur
  les substitutions rationnelles. ({F}rench). {O}n rational substitutions},}\
  }\href@noop {} {\bibfield  {journal} {\bibinfo  {journal} {Comptes Rendus de
  l'Acad{\'e}mie des Sciences de Paris}\ }\textbf {\bibinfo {volume} {165}},\
  \bibinfo {pages} {992--995} (\bibinfo {year}
  {1917}{\natexlab{b}})}\BibitemShut {NoStop}%
\bibitem [{\citenamefont {Julia}(1918)}]{GastonJulia1918}%
  \BibitemOpen
  \bibfield  {author} {\bibinfo {author} {\bibfnamefont {Gaston~Maurice}\
  \bibnamefont {Julia}},\ }\bibfield  {title} {\enquote {\bibinfo {title}
  {M{\'e}moire sur l'it{\'e}ration des fonctions rationnelles. {(French)}.
  {M}emoir on the iteration of rational functions},}\ }\href
  {http://eudml.org/doc/234994} {\bibfield  {journal} {\bibinfo  {journal}
  {Journal de Math{\'e}matiques Pures et Appliqu{\'e}es}\ }\textbf {\bibinfo
  {volume} {8}},\ \bibinfo {pages} {47--245} (\bibinfo {year} {1918})},\
  \bibinfo {note} {{[8{$^e$} {S}{\'e}rie, {T}ome 1]}}\BibitemShut {NoStop}%
\bibitem [{\citenamefont {Brooks}\ and\ \citenamefont
  {Matelski}(1981)}]{BrooksMatelski}%
  \BibitemOpen
  \bibfield  {author} {\bibinfo {author} {\bibfnamefont {Robert}\ \bibnamefont
  {Brooks}}\ and\ \bibinfo {author} {\bibfnamefont {J.~Peter}\ \bibnamefont
  {Matelski}},\ }\enquote {\bibinfo {title} {The dynamics of 2-generator
  subgroups of {PSL(2,$\mathbb{C}$)}},}\ in\ \href {\doibase
  doi:10.1515/9781400881550-007} {\emph {\bibinfo {booktitle} {Riemann Surfaces
  And Related Topics: Proceedings of the 1978 Stony Brook Conference. (AM-97),
  Volume 97}}},\ \bibinfo {editor} {edited by\ \bibinfo {editor} {\bibfnamefont
  {Irwin}\ \bibnamefont {Kra}}\ and\ \bibinfo {editor} {\bibfnamefont
  {Bernard}\ \bibnamefont {Maskit}}}\ (\bibinfo  {publisher} {Princeton
  University Press},\ \bibinfo {address} {Princeton},\ \bibinfo {year} {1981})\
  pp.\ \bibinfo {pages} {65--72}\BibitemShut {NoStop}%
\bibitem [{\citenamefont {Mandelbrot}(1980)}]{Mandelbrot1980FRACTALAO}%
  \BibitemOpen
  \bibfield  {author} {\bibinfo {author} {\bibfnamefont {Benoit~B.}\
  \bibnamefont {Mandelbrot}},\ }\bibfield  {title} {\enquote {\bibinfo {title}
  {Fractal aspects of the iteration of {$z \rightarrow \lambda\,z\,(1-z)$} for
  complex {$\lambda$} and {$z$}},}\ }\href {\doibase
  https://doi.org/10.1111/j.1749-6632.1980.tb29690.x} {\bibfield  {journal}
  {\bibinfo  {journal} {Annals of the New York Academy of Sciences}\ }\textbf
  {\bibinfo {volume} {357}},\ \bibinfo {pages} {249--259} (\bibinfo {year}
  {1980})}\BibitemShut {NoStop}%
\bibitem [{\citenamefont {Zolkin}\ \emph
  {et~al.}(2024{\natexlab{d}})\citenamefont {Zolkin}, \citenamefont {Cathey},\
  and\ \citenamefont {Nagaitsev}}]{zolkin2024dynamicsII}%
  \BibitemOpen
  \bibfield  {author} {\bibinfo {author} {\bibfnamefont {Tim}\ \bibnamefont
  {Zolkin}}, \bibinfo {author} {\bibfnamefont {Brandon}\ \bibnamefont
  {Cathey}}, \ and\ \bibinfo {author} {\bibfnamefont {Sergei}\ \bibnamefont
  {Nagaitsev}},\ }\href {https://arxiv.org/abs/2405.05657} {\enquote {\bibinfo
  {title} {Dynamics of {M}c{M}illan mappings {II}. {A}xially symmetric map},}\
  } (\bibinfo {year} {2024}{\natexlab{d}}),\ \Eprint
  {http://arxiv.org/abs/2405.05657} {arXiv:2405.05657 [nlin.SI]} \BibitemShut
  {NoStop}%
\bibitem [{\citenamefont {Turaev}(2002)}]{turaev2002polynomial}%
  \BibitemOpen
  \bibfield  {author} {\bibinfo {author} {\bibfnamefont {Dmitry}\ \bibnamefont
  {Turaev}},\ }\bibfield  {title} {\enquote {\bibinfo {title} {Polynomial
  approximations of symplectic dynamics and richness of chaos in non-hyperbolic
  area-preserving maps},}\ }\href {\doibase 10.1088/0951-7715/16/1/308}
  {\bibfield  {journal} {\bibinfo  {journal} {Nonlinearity}\ }\textbf {\bibinfo
  {volume} {16}},\ \bibinfo {pages} {123} (\bibinfo {year} {2002})}\BibitemShut
  {NoStop}%
\bibitem [{\citenamefont {Yoshida}(1990)}]{yoshida1990construction}%
  \BibitemOpen
  \bibfield  {author} {\bibinfo {author} {\bibfnamefont {Haruo}\ \bibnamefont
  {Yoshida}},\ }\bibfield  {title} {\enquote {\bibinfo {title} {Construction of
  higher order symplectic integrators},}\ }\href {\doibase
  https://doi.org/10.1016/0375-9601(90)90092-3} {\bibfield  {journal} {\bibinfo
   {journal} {Physics Letters A}\ }\textbf {\bibinfo {volume} {150}},\ \bibinfo
  {pages} {262--268} (\bibinfo {year} {1990})}\BibitemShut {NoStop}%
\end{thebibliography}

%

\end{document}